\documentstyle[12pt,epsfig]{article}

\def\be{\begin{equation}}
\def\ee{\end{equation}}
\def\la{\langle}
\def\ra{\rangle}
\def\IP{\hbox{\rm I\kern -1.6pt{\rm P}}}
\def\IC{{\hbox{\rm C\kern-.58em{\raise.53ex\hbox{$\scriptscriptstyle|$}}
    \kern-.55em{\raise.53ex\hbox{$\scriptscriptstyle|$}} }}}
\def\IN{\hbox{I\kern-.2em\hbox{N}}}
\def\IR{\hbox{\rm I\kern-.2em\hbox{\rm R}}}
\def\ZZ{\hbox{{\rm Z}\kern-.3em{\rm Z}}}
\def\IT{\hbox{\rm T\kern-.38em{\raise.415ex\hbox{$\scriptstyle|$}} }}

\newtheorem{theorem}{Theorem}[section]
\newtheorem{lemma}[theorem]{Lemma}

\newtheorem{proposition}[theorem]{Proposition}
\newtheorem{corollary}[theorem]{Corollary}

\begin{document}

\title{Dynamics of a Massive Piston in an Ideal Gas}
\author{N. Chernov$^{1,4}$,
J.~L.~Lebowitz$^{2,4}$, and Ya.~Sinai$^{3}$}
\date{\today}
\maketitle

\begin{abstract}
We study a dynamical system consisting of a massive piston in a
cubical container of large size $L$ filled with an ideal gas. The
piston has mass $M\sim L^2$ and undergoes elastic collisions with
$N\sim L^3$ non-interacting gas particles of mass $m=1$. We find
that, under suitable initial conditions, there is, in the limit $L
\to \infty$, a scaling regime with time and space scaled by $L$,
in which the motion of the piston and the one particle
distribution of the gas satisfy autonomous coupled equations
(hydrodynamical equations), so that the mechanical trajectory of
the piston converges, in probability, to the solution of the
hydrodynamical equations for a certain period of time. We also
discuss heuristically the dynamics of the system on longer
intervals of time.

\footnotetext[1]{Department of Mathematics,
University of Alabama at Birmingham, Alabama 35294}
\footnotetext[2]{Department of Mathematics, Rutgers
University, New Jersey 08854}
\footnotetext[3]{Department of Mathematics, Princeton University,
New Jersey 08544}
\footnotetext[4]{Current address: Institute for Advanced Study,
Princeton, NJ 08540}

\end{abstract}

\tableofcontents

\renewcommand{\theequation}{\arabic{section}.\arabic{equation}}

\section{Introduction}
\label{secI} \setcounter{equation}{0}

The evolution of a macroscopic system consisting of a gas in a
container divided by a massive movable wall (piston) is an old problem
in statistical physics with a colorful history. It was discussed by
Landau and Lifshitz \cite{LL} and later by Lebowitz \cite{L1}, Feynman
\cite{F}, Kubo \cite{Ku}, see recent surveys by Lieb \cite{Li}, Gruber
\cite{G} and others \cite{KBM}.

\begin{figure}[h]
\centering
\epsfig{figure=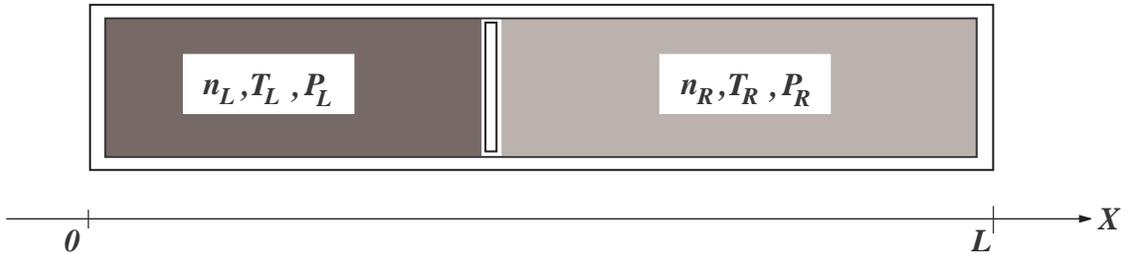}\caption{Piston in a cylinder filled
with gas.}
\end{figure}

In its simplest form, the model consists of an isolated cylinder filled
with gas and divided into two compartments by a large piston which is
free to move along the axis of the cylinder, see Fig.~1. Initially, the
piston is held fixed by a clamp and the gas in each compartment evolves
independently and is in equilibrium\footnote{For an isolated system of
$N$ atoms and total energy $E$, the equilibrium distribution is defined
in statistical mechanics by a uniform probability distribution
$\rho_{\rm eq}$ on the energy surface $E=\,{\rm const}$ in the phase
space ($\rho_{\rm eq}$ is called a microcanonical ensemble, it remains
invariant under the dynamics by the Liouville theorem). One says that
the system (gas) is ``in equilibrium'' if its states are ``typical''
for the measure $\rho_{\rm eq}$. In this state, a macroscopic gas will
have (approximately, for large $N$) a uniform spatial density and a
Maxwellian velocity distribution. The latter is defined so that the
$x,y,z$ components of the velocity vectors are independent normal
random variables $N(0,\sigma^2)$ with the same variance
$\sigma^2=k_BT/m$, where $k_B$ is Boltzmann's constant, $T$ the
temperature of the gas (which is a function of $E$, see below), and $m$
the mass of an atom.}. We denote the density and temperature in the
left and right compartments by $n_L,T_L$ and $n_R,T_R$, respectively.
The gas exerts pressure (= force per unit area) on the piston, which is
given by equilibrium statistical mechanics as a function of density and
temperature: $P_L=P(n_L,T_L)$ and $P_R=P(n_R,T_R)$ on the left and on
the right, respectively. At time $t=0$ the clamp is removed and the
piston is released. Now one wants to describe the evolution of the
system, especially its limit (final) state as $t\to\infty$.

Starting with $P_L\neq P_R$, the piston moves under the net pressure
difference and compresses the gas whose pressure is lower, until its
pressure builds up and it pushes the piston back. Depending on the
initial values of $n_L,T_L,n_R,T_R$ and the dynamical characteristics
of the gases, the piston may follow a complicated trajectory, sloshing
back and forth, but gradually it comes to rest at a place where the
pressures are equalized on both sides: $P_L=P_R$. At that time one also
expects that the gas in each compartment will again be in equilibrium.

We observe, however, that the equality of pressures $P_L=P_R$ and the
fact that the gas in each compartment separately is at equilibrium does
not guarantee that $T_L=T_R$. In particular, for dilute gases (which we
shall consider from now on), the pressure is related to the density and
temperature by $P=nk_BT$, where $k_B$ is Boltzmann's constant, and it
is possible that $T_L < T_R$ while $n_L > n_R$, so that the gas in the
left compartment is cooler but denser, and in the right one hotter but
more dilute (or vice versa). The exact values of the temperatures $T_L$
and $T_R$, at the time when there is no longer any pressure difference
between the left and right and the piston comes to rest, depend on the
initial conditions and other characteristics of the system, see
\cite{CPS,G}, for example. Therefore, we have two possibilities now. If
it happens that $T_L=T_R$ as well as $P_L=P_R$, then the system as a
whole will be in equilibrium, and we say that it came to a {\em thermal
equilibrium}. On the contrary, if $P_L=P_R$ but $T_L\neq T_R$, the
system is said to be at {\em mechanical equilibrium}, or {\em
quasi-equilibrium}.

One may now ask whether the mechanical equilibrium is stable in the
sense that it can last forever (assuming that the whole system remains
perfectly isolated from the outside world), or will the gases find ways
to exchange energy through the piston and eventually bring the system
to a thermal equilibrium? It was claimed in some textbooks, based on a
simplistic interpretation of the laws of thermodynamics, see below,
that indeed the mechanical equilibrium could persist ``forever'', cf.\
\cite{G,GF} for some history.

On the other hand, Landau and Lifshitz \cite{LL}, Feynman \cite{F} and
many others argued intuitively that the system should converge from the
mechanical equilibrium to a thermal equilibrium. They predicted that
the cooler compartment should gradually heat up and the hotter one cool
down, while the piston slowly moves from the cooler side to the hotter
side, so that the pressure balance is maintained until the temperatures
are equalized, and the piston makes its final stop.

The confusion about the evolution of the gas after the establishment of
mechanical equilibrium is due to the following: Heat conduction through
a wall is normally associated with the internal motion of the molecules
of the wall colliding with those of the gas and thus exchanging
momentum and energy. However, the piston and the walls in our idealized
model are supposed to be rigid, solid, and structureless bodies and the
gas atoms bounce off them elastically. This idealization is exactly the
reason why the gases in the different compartments could be in
equilibrium at different temperatures when the piston was clamped. The
unclamped piston, on the other hand, interacts with gas atoms as a
whole, i.e.\ as one huge and massive molecule. It then makes tiny
microscopic movements (vibrations) induced by collisions with atoms on
both sides. Hence, some microscopic exchange of momentum and energy
does take place. But these microscopic vibrations of the piston are not
part of macroscopic thermodynamics, in which the action of the piston
on the gas in each compartment is regarded as an external mechanical
force. Under this condition (and assuming that the piston has no
entropy of its own)  the second law of thermodynamics would predict
that the entropy of the gas, as it goes from some initial equilibrium
state to a final equilibrium state, could not decrease. When the gas in
each compartment is in equilibrium, its thermodynamic entropy is known
to be \cite{LL,Ca}
$$
        S_i = N_i\left [ -\log P_i+(1+3/2)\log T_i\right ] + f(N_i),
        \ \ \ \ \ i=L,R
$$
where $N_L=n_LV_L$ and $N_R=n_RV_R$ denote the number of atoms in the
gases, and the explicit form of $f(N_i)$ is irrelevant for us, since
its value does not change in time. Now, if our system does evolve from
mechanical equilibrium to thermal equilibrium, keeping the pressure
balance $P_L=P_R$ and the total kinetic energy $\frac 32
k_B(N_LT_L+N_RT_R)$ fixed, then one can easily compute (we leave this
as an exercise) that the pressure of the gases stays constant in time
and the total entropy of the system $S=S_L+S_R$ grows until it reaches
its maximum at the point of thermal equilibrium. At the same time, the
entropy $S_R$ decreases, while $S_L$ increases, since $T_R$ goes down,
$T_L$ goes up, and the pressure $P_R=P_L$ remains constant. This
decrease of the entropy would, as already noted, violate the second law
of thermodynamics, if the piston remained mechanical, see further
discussions in \cite{Li,CPS} and critical remarks in \cite{G,GF}.

Therefore, the evolution of the system beyond the mechanical
equilibrium cannot be described by macroscopic thermodynamics (beyond
the statement that any evolution in an isolated macroscopic system will
not decrease the total entropy). The actual evolution is a result of
microscopic energy transfer between the gases via collisions with the
piston. This process is purely microscopic and, in a sense,
counterintuitive, as we explain next. Under the collisions with gas
atoms on both sides the piston vibrates, i.e.\ it jiggles back and
forth. When the piston moves toward the hotter side, the atoms of the
hotter gas bounce off the piston with an increased speed and so gain
energy, while the atoms of the cooler gas collide with the piston and
slow down, hence lose some energy. When the piston moves toward the
cooler side it is vice versa. Since, on the average, the hotter gas
must cool down and the cooler gas must heat up, one may conclude that
the piston's movements toward the cooler side dominate. On the other
hand, the piston has to slowly move toward the hotter side in order to
maintain the pressure balance, see above, so its displacements in the
direction of the hotter gas actually dominate. It is not quite clear
how these seemingly opposite trends manage to coexist. Some physicists
joke about a ``conspiracy'' between the microscopic vibrations of the
piston and the incoming atoms of the gases \cite{GF,GP}. In the words
of Callen \cite{Ca}, ``the movable adiabatic wall presents a unique
problem with subtleties''.

In order to understand the mechanism of the heat transfer across the
piston at mechanical equilibrium ($P_L=P_R$), one usually considers the
simplest gas of noninteracting particles, that is an {\em ideal gas}.
As early as in 1959, Lebowitz \cite{L1} studied a piston interacting
with two infinite reservoirs filled with ideal gases held at different
temperatures $T_L\neq T_R$. His piston also interacted with an external
potential, e.g.\ a spring, and it could therefore come to a stationary
nonequilibrium state under the influence of the infinite reservoirs. He
used an approximation by a Markov process and found the distribution of
the piston velocity to be Maxwellian corresponding to some intermediate
temperature $T\in (T_L,T_R)$, which led to a systematic heat transfer
between the gases. Recently, Gruber and others \cite{GF,PG,GP} used
kinetic theory to study a freely movable piston of mass $M$ interacting
with two infinite ideal gases of atoms of mass $m\ll M$ at equal
pressures but different temperatures. They use the expansion of the
Boltzmann equation in $\varepsilon=m/M$ to show that a macroscopic heat
flux across the piston does occur whenever $T_L\neq T_R$, hence the
system gradually approaches thermal equilibrium. They also found a
stationary distribution of the piston velocity, whose average value is
given by
\be
        \la V\ra =
    \frac{\sqrt{2\pi m}\, (\, \sqrt{k_BT_R}-\sqrt{k_BT_L}\, )}{4M}
    + o\left (\frac mM\right)
       \label{Vav}
\ee
(it is independent of the gas densities). We note that if $T_L<T_R$,
then $\la V\ra >0$, confirming our previous observation that the piston
moves from the cooler side to the hotter side.

Equation (\ref{Vav}) shows that the average velocity of the piston
is different from zero, albeit just of order $O(m/M)$, despite the
perfect pressure balance $P_L=P_R$. We note, however, that for a
macroscopic-size piston the ratio $m/M$ is so small that the time
it takes the piston to cover any noticeable distance is much
longer than the age of the universe \cite{GF}, so such a
phenomenon cannot be observed experimentally.

We conclude that the evolution of the system proceeds in two different
stages. The first one is the convergence to a mechanical equilibrium,
which is relatively fast and can, in principle, be computed on the
basis of macroscopic equations. The second stage is the transition of
the system from mechanical equilibrium to thermal equilibrium. This
process is very slow and much less understood.

In addition, for the ideal gas (in which the atoms do not interact) a
new problem arises. At time $t=0$, before the piston is released, the
gas atoms move independently of each other -- every atom bounces off
the walls and the clamped piston, without exchanging momentum or energy
with other atoms. Therefore, the velocity distribution does not have to
be Maxwellian. A stationary state of the ideal gas can be described by
any Poisson process with a uniform spacial density and a symmetric
velocity distribution. For example, half of the atoms may move toward
the piston with unit velocity $v=1$ and the other half -- in the
opposite direction with velocity $v=-1$, and this state will be
stationary. However, once the piston is released, the atoms start
interacting with each other, indirectly, via collisions with the
piston. This provides a way to exchange momentum and energy between the
atoms. One can expect that these interactions will lead, ultimately, to
a true thermal equilibration, when the velocity distribution becomes
Maxwellian, as we explain in Section~\ref{secBORI}. This process,
however, may take even longer than the equilibration of the mean
kinetic energies described above.

To consider this new process in its ``pure'' form, we assume that
initially the system is already in a homogeneous state -- the gas
density is constant across the entire cylinder and the velocity
distribution is the same in both compartments (but different from
Maxwellian). Then there seem to be no forces of any kind that would
drive the piston anywhere. In particular, when the piston is initially
placed in the middle of the cylinder, then by symmetry there should be
no reason for it to move either way! On the other hand, the system is
not in equilibrium until the velocity distribution becomes Maxwellian,
hence it should find ways to evolve toward equilibrium, thus changing
its macroscopic state. We discuss this further in
Section~\ref{secBORI}.

One can also consider a simpler case when that the container is
infinitely long on both sides of the piston and the ideal gases have
infinite number of atoms, as in \cite{L1,GP,PG}. In that case the
problem reduces to the classical Rayleigh gas -- a big massive particle
submerged in an ideal gas. In particular, our piston in an infinite
cylinder becomes a one-dimensional Rayleigh gas, which we describe in
some detail.

Let a heavy tagged particle (called molecule) of mass $M$ move on a
line under elastic collisions with atoms of mass $m$ of an ideal gas
with a uniform density $n$ and some velocity distribution $f(v)\, dv$.
Denote by $X(t)$ and $V(t)=\dot{X}(t)$ the position and velocity of the
molecule at time $t$. Even though $f(v)$ need not be Maxwellian, the
velocity function $V(t)$ and the coordinate function $X(t)$ can be
approximated by certain Gaussian stochastic processes:

\begin{theorem}[Holley \cite{H}]
Let the density $f(v)$ be symmetric $f(v)=f(-v)$ and have a finite
fourth moment $\int v^4f(v)\, dv<\infty$. Then for every finite
$t_0<\infty$, the function $V(t)\sqrt{M}$ on the interval $[0,t_0]$
converges, in distribution, as $M,n\to\infty$ and $M/n\to\,{\rm
const}$, to an Ornstein-Uhlenbeck velocity process ${\cal V}_t$, while
$X(t)\sqrt{M}$ converges to an Ornstein-Uhlenbeck position process
${\cal X}_t$. \label{tmH}
\end{theorem}

An Ornstein-Uhlenbeck process $({\cal X}_t,{\cal V}_t)$ is defined
by \cite{Ne}
$$
    d{\cal X}_t = {\cal V}_t\, dt,
    \ \ \ \ \ \
    d{\cal V}_t = -a{\cal V}_t\, dt +\sqrt{D}\, d{\cal W}_t
$$
where $a>0$, $D>0$ are constants and ${\cal W}_t$ a Wiener
process. The Ornstein-Uhlenbeck position process ${\cal X}_t$
converges in an appropriate limit (e.g.\ $a\to\infty$,
$a^2/D=\,$const) to a Wiener process.

We note that the typical velocity of the molecule $V(t)$ is of order
$O(1/\sqrt{M})$, which agrees with the equipartition of energy in the
system requiring that average energies of all particles be equal, i.e.\
$M\la V^2\ra = m\la v^2\ra = m\int v^2f(v)\, dv$.

D\"urr et al \cite{DGL} extended the above theorem to arbitrary
dimension and to asymmetric velocity distributions. The main technical
difficulty in the proof of this theorem comes from the so called {\em
recollisions}, which occur when an atom collides with the molecule more
than once. Recollisions result in intricate autocorrelations in the
process $(X(t),V(t))$, which otherwise would be Markovian. The proof
essentially consists in estimating the undesirable effect of
recollisions and showing that it vanishes in the limit $M\to\infty$.

When the gas is confined in a finite cylinder, though, the effect
of recollisions becomes crucial. All atoms will travel to the
walls, bounce off it and come back to the piston for more and more
collisions. The induced autocorrelations will build up. There is
no standard techniques available to estimate (let alone eliminate)
the effect of recollisions in general, but we make partial
progress in this direction, see Section~\ref{secORI}.

In summary, the piston problem raises serious mathematical questions
and even leads to confusions in the physical theories. The ``notorious
piston'', as it is known among physicists, again attracted much
attention recently due to a series of papers \cite{GF,GP,PG,LPS} where
a more extensive mathematical apparatus was developed. At the same
time, many new numerical experiments led to better theoretical
understanding of the underlying dynamics but also raised some new
questions. We emphasize that very few rigorous results are available,
even for ideal gases, apart from the Rayleigh-type stochastic
approximations in the infinite cylinder mentioned above.

We study the piston in a finite cylinder filled with ideal gases. Since
our gases are ideal, we will not need to assume that the velocity
distribution of atoms is Maxwellian. Since autocorrelations induced by
recollisions present a major difficulty, we specify the initial state
in such a way, that during a certain interval of time each gas atom
collides with the piston at most twice. The main goal of our work is to
describe rigorously the dynamics of the piston during that time
interval. We show that, in an appropriate limit, the evolution of the
piston and the gas converges to a deterministic process, which
satisfies a certain closed system of differential equations. The
assumptions that we make here simplify technical considerations but by
no means reduce the problem to a triviality. In fact, many intriguing
questions still remain open in our context, and we discuss them in the
last two sections of the paper.

\medskip\noindent {\bf Precise statement of problem and main results}.
Consider a cubical domain $\Lambda_L$ of size $L$ separated into two
parts by a movable wall (piston). Each part of $\Lambda_L$ contains a
gas of noninteracting particles of mass $m=1$. The particles collide
with the outer (fixed) walls of $\Lambda_L$ and with the moving piston
elastically. The piston has mass $M=M_L$ and moves along the $x$-axis
under the collisions with the gas particles on both sides. The size $L$
of the cube is a large parameter of our model, and we are interested in
the behavior as $L\to\infty$. We will assume that $M_L$ is proportional
to the area of the piston, i.e.\ $M_L\sim L^2$, and the number of gas
particles $N$ is proportional to the volume of the cube $\Lambda_L$,
i.e.\ $N\sim L^3$, while the particle velocities remain of order one.

The position of the piston at time $t$ is specified by a single
coordinate $X=X_L(t)$, $0 \leq X \leq L$, its velocity is then
given by $V=V_L(t)=\dot{X}_L(t)$. Since the components of the particle
velocities perpendicular to the $x$-axis play no role in the
dynamics, we may assume that each particle has only one
coordinate, $x$, and one component of velocity, $v$, directed
along the $x$-axis.

When a particle with velocity $v$ collides with the piston with velocity
$V$, their velocities after the collision, $v^\prime$ and
$V^\prime$, respectively, are given by
\be
    V^\prime = (1-\varepsilon)V + \varepsilon v
      \label{V'}
\ee
\be
    v^\prime = -(1-\varepsilon)v + (2-\varepsilon)V
      \label{v'}
\ee
where $ \varepsilon = 2m/(M+m)$. We assume that
$M+m=2mL^2/a$, where $a>0$ is a constant, so that
\be
    \varepsilon = \frac{2m}{M+m}=\frac{a}{L^2}
      \label{varepsMm}
\ee
When a particle collides with a wall at $x=0$ or $x=L$,
its velocity just changes sign.

The evolution of the system is then completely deterministic, but
one needs to specify the initial conditions. We shall assume that
the piston starts at the midpoint $X_L(0)=L/2$ with zero velocity
$V_L(0)=0$ (see also Section~\ref{secHE}). The initial
configuration of gas particles and their velocities is chosen at
random as a realization of a (two-dimensional) Poisson process on
the $(x,v)$-plane (restricted to $0\leq x\leq L$) with density
$L^2p_L(x,v)$, where $p_L(x,v)$ is a function satisfying certain
conditions, see below, and the factor of $L^2$ is the
cross-sectional area of the container. This means that for any
domain $D\subset [0,L]\times\IR^1$ the number $N_D$ of gas
particles $(x,v)\in D$ at time $t=0$ has a Poisson distribution
with parameter
$$
          \lambda_D=L^2\int\!\!\int_Dp_L(x,v)\, dx\, dv
$$
For any two nonoverlapping domains, say $D_1\cap D_2=\emptyset$,
the corresponding numbers $N_{D_1}$ and $N_{D_2}$
are statistically independent. We remark that the total
number of gas particles $N$
is a Poisson random variable, too. The total
energy and the total initial momentum are random as well.

Let $\Omega_L$ denote the space of all possible configurations of gas
particles in $\Lambda_L$ (i.e., all countable subsets of
$[0,L]\times\IR^1$). For each realization\footnote{Technically, it is
possible that two or more particles collide with the piston
simultaneously, and then the dynamics will no longer be defined, but
multiple collisions are known to occur with probability zero \cite{H},
so we will ignore such anomalies.} $\omega\in\Omega_L$ the
deterministic piston trajectory will be denoted by $X_L(t,\omega)$ and
its velocity by $V_L(t,\omega)$.

The above model is a mechanical system whose dynamical characteristics
$X_L(t,\omega)$ and $V_L(t,\omega)$ depend on the large parameter $L$
and, for each $L$, are random (depend on $\omega$).

In order to obtain a deterministic description of the dynamics of the
piston one needs to take a limit as $L\to\infty$ and simultaneously
rescale space and time. We introduce new space and time coordinates by
\be
     y=x/L\ \ \ \ {\rm and}\ \ \ \  \tau=t/L.
     \label{ytau}
\ee
which corresponds to Euler scaling for the hydrodynamical limit
transition. We call $y$ and $\tau$ the {\em macroscopic} (``slow'')
variables, as opposed to the original {\em microscopic} (``fast'') $x$
and $t$. Now let
\be
           Y_L(\tau,\omega)=X_L(\tau L,\omega)/L,
       \ \ \ \ \ \ \ \ \ \
       W_L(\tau,\omega)=V_L(\tau L,\omega)
          \label{YWL}
\ee
denote the position and velocity of the piston in the
macroscopic context. The initial conditions are then $Y_L(0)=
X_L(0)/L=0.5$ and $W(0)=V(0)=0$.

It is now very natural to assume that the initial density $p_L(x,v)$
agrees with our rescaling:
\be
      p_L(x,v) = \pi_0(x/L,v)
         \label{pLpi}
\ee
where the function $\pi_0(y,v)$ is independent of $L$. Without
loss of generality, we can assume that $\pi_0$ is normalized so
that
$$
     \int_0^1\int_{-\infty}^{\infty}\pi_0(y,v)\, dv\, dy = 1
$$
Then the mean number of particles in the entire container
$\Lambda_L$ is exactly equal to $L^3$:
$$
         E(N) = \int\!\!\int L^2p_L(x,v)\, dv\, dx = L^3
$$
where $E(\cdot)$ is the expected value.

Furthermore, we assume that the function $\pi_0(y,v)$ satisfies several
technical requirements stated below. The meaning and purpose of these
assumptions will become clear later.

\begin{itemize}
\item[(P1)] {\em Smoothness}. $\pi_0(y,v)$ is a piecewise $C^1$
function with uniformly bounded partial derivatives, i.e.\
$|\partial\pi_0/\partial y|\leq D_1$ and $|\partial\pi_0/\partial
v|\leq D_1$ for some $D_1>0$.

\item[(P2)] {\em Discontinuity lines}. $\pi_0(y,v)$ may be
discontinuous on the line $y=Y_L(0)$ (i.e., ``on the piston''). In
addition, it may have a finite number ($\leq K_1$) of other
discontinuity lines in the $(y,v)$-plane with strictly positive
slopes (each line is given by an equation $v=f(y)$ where $f(y)$ is
$C^1$ and $0<c_1<f'(y)<c_2<\infty$).

\item[(P3)] {\em Density bounds}. Let
\be
         \pi_0(y,v)>\pi_{\min}>0\ \ \ \ \ \ {\rm for}\ \ v_1<|v|<v_2
                \label{pmin0}
\ee
for some $0<v_1<v_2<\infty$, and
\be
   \sup_{y,v}\pi_0(y,v)=\pi_{\max}<\infty
      \label{pmax0}
\ee
The requirements (\ref{pmin0}) and (\ref{pmax0}) basically mean
that $\pi_0(y,v)$ takes values of order one.

\item[(P4)] {\em Velocity ``cutoff''}. Let
\be
     \pi_0(y,v) = 0, \quad {\rm if} \quad |v| \leq v_{\rm min}
     \quad {\rm or} \quad |v| \geq v_{\max}
        \label{0cutoff}
\ee
with some $0<v_{\min}<v_{\max}<\infty$. This means that the
speed of gas particles is bounded from above by $v_{\max}$ and
from below by $v_{\min}$.

\item[(P5)] {\em Approximate pressure balance}. $\pi_0(y,v)$ must
be nearly symmetric about the piston, i.e.
\be
         |\pi_0(y,v)-\pi_0(1-y,-v)|<\varepsilon_0
             \label{0symmetry}
\ee
for all $0<y<1$ and some sufficiently small $\varepsilon_0>0$.
\end{itemize}

The requirements (P4) and (P5) are crucial. We will see that they
are made to ensure that the speed of the piston $|V_L(t,\omega)|$
will be smaller than the minimum speed of the gas particles, with
probability close to one, for times $t=O(L)$. Such assumptions
were first made in \cite{LPS}.

We think of $D_1$, $K_1$, $c_1$, $c_2$, $v_1$, $v_2$, $v_{\min}$,
$v_{\max}$, $\pi_{\min}$ and $\pi_{\max}$ in (P1)--(P4) as fixed
(global) constants and $\varepsilon_0$ in (P5) as an adjustable
small parameter. We will assume throughout the paper that
$\varepsilon_0$ is small enough, meaning that
$$
   \varepsilon_0<\bar{\varepsilon}_0
   (D_1,K_1,c_1,c_2,v_1,v_2,v_{\min},v_{\max},\pi_{\min},\pi_{\max})
$$
It is important to note that the hydrodynamic limit does {\em not}
require that $\varepsilon_0\to 0$. The parameter $\varepsilon_0$
stays positive and fixed as $L\to\infty$.

Now we state our main result:

\begin{theorem}
There is an $L$-independent function $Y(\tau)$ defined for
all $\tau\geq 0$ and a positive $\tau_{\ast}\approx 2/v_{\max}$
(actually, $\tau_{\ast}\to 2/v_{\max}$ as $\varepsilon_0\to 0$),
such that
\be
   \sup_{0\leq\tau\leq\tau_{\ast}}
        |Y_L(\tau,\omega) - Y(\tau)| \to 0
        \label{YY}
\ee
and
\be
   \sup_{0\leq\tau\leq\tau_{\ast}}
        |W_L(\tau,\omega) - W(\tau)| \to 0
       \label{WW}
\ee
in probability, as $L\to\infty$. Here $W(\tau)=\dot{Y}(\tau)$.
\label{tmmain}
\end{theorem}

This theorem establishes the convergence in probability of the
random functions $Y_L(\tau,\omega),W(\tau,\omega)$ characterizing
the mechanical evolution of the piston
to the deterministic functions $Y(\tau),W(\tau)$, in
the hydrodynamical limit $L\to\infty$. \medskip

The functions $Y(\tau)$ and $W(\tau)$ satisfy certain (Euler-type)
differential equations stated in the next section. Those equations have
solutions for all $\tau\geq 0$, but we can only guarantee the
convergence (\ref{YY}) and (\ref{WW}) for $\tau<\tau_{\ast}$. What
happens for $\tau>\tau_{\ast}$, especially as $\tau\to\infty$, remains
an open problem. Some numerical results and heuristic observations in
this direction are presented in Section~\ref{secBORI}.

\medskip\noindent{\bf Remarks}.
The function $Y(\tau)$ is at least $C^1$ and,
furthermore, piecewise $C^2$.
On the interval $(0,\tau_{\ast})$, its first derivative
$W=\dot{Y}$ (velocity) and its second derivative
$A=\ddot{Y}$ (acceleration) remain $\varepsilon_0$-small:
$\sup_{\tau} |W(\tau)|\leq\,$const$\cdot\varepsilon_0$ and
$\sup_{\tau} |A(\tau)|\leq\,$const$\cdot\varepsilon_0$,
see the next section.

We will also estimate the speed of convergence in (\ref{YY})
and (\ref{WW}). Precisely, we  show that there is a $\tau_0>0$
($\tau_0\approx 1/v_{\max}$) such that
$$
     |Y_L(\tau,\omega) -Y(\tau)|=O(\ln L/L)
$$
for $0<\tau<\tau_0$ and
$$
     |Y_L(\tau,\omega)-Y(\tau)|=O(\ln L/L^{1/7})
$$
for $\tau_0<\tau<\tau_{\ast}$. The same bounds are valid for
$|W_L(\tau,\omega)-W(\tau)|$, see Sections~\ref{secZRI}
and \ref{secORI}. These estimates hold with
``overwhelming'' probability, specifically they hold for
all $\omega\in\Omega^{\ast}_L\subset\Omega_L$ such that
$P(\Omega^{\ast}_L)=1-O(L^{-\ln L})$.

\section{Hydrodynamical equations}
\label{secHE} \setcounter{equation}{0}

The equations describing the deterministic function $Y(\tau)$ involve
another deterministic function -- the scaled density of the gas
$\pi(y,v,\tau)$. Initially, $\pi(y,v,0)=\pi_0(y,v)$, and for $\tau>0$
the density $\pi(y,v,\tau)$ evolves according to the following rules.

\begin{itemize}
\item[(H1)] {\em Free motion}. Inside the container
the density satisfies the standard continuity equation for
a noninteracting particle system without external forces:
\be
     \left ( \frac{\partial}{\partial \tau}+
      v\, \frac{\partial}{\partial y} \right )
        \, \pi(y,v,\tau)=0
          \label{pdinside}
\ee
for all $y$ except $y=0$, $y=1$ and $y=Y(\tau)$.
\end{itemize}

\noindent
Equation (\ref{pdinside}) has a simple solution
\be
      \pi(y,v,\tau)=\pi(y-vs,v,\tau-s)
         \label{pinside}
\ee
for $0<s<\tau$ such that $y-vr\notin\{0,Y(\tau-r),1\}$ for all
$r\in (0,s)$. Equation (\ref{pinside}) has one advantage over
(\ref{pdinside}): it applies to all points $(y,v)$, including
those where the function $\pi$ is not differentiable.

\begin{itemize}

\item[(H2)] {\em Collisions with the walls}.
At the walls $y=0$ and $y=1$ we have
\be
       \pi(0,v,\tau) = \pi(0,-v,\tau)
          \label{pwall0}
\ee
\be
       \pi(1,v,\tau) = \pi(1,-v,\tau)
          \label{pwall1}
\ee

\item[(H3)] {\em Collisions with the piston}. At the piston
$y=Y(\tau)$ we have
\begin{eqnarray}
      \pi(Y(\tau)-0,v,\tau) &=& \pi(Y(\tau)-0,2W(\tau)-v,\tau)
      \ \ \ \ \ {\rm for}\ \ v<W(\tau)\nonumber\\
      \pi(Y(\tau)+0,v,\tau) &=& \pi(Y(\tau)+0,2W(\tau)-v,\tau)
      \ \ \ \ \ {\rm for}\ \ v>W(\tau)
        \label{ponpiston}
\end{eqnarray}
where $v$ represents the velocity after the collision and
$2W(\tau)-v$ that before the collision; here
\be
        W(\tau) = \frac{d}{d\tau}Y(\tau)
          \label{W=Y'}
\ee
is the (deterministic) velocity of the piston.

\end{itemize}

It remains to describe the evolution of $W(\tau)$.
Suppose the piston's position at time $\tau$ is $Y$ and its
velocity $W$. The piston is affected by the particles $(y,v)$ hitting
it from the right (such that $y=Y+0$ and $v<W$) and from the
left (such that $y=Y-0$ and $v>W$).

\begin{itemize}
\item[(H4)] {\em Piston's velocity}.
The velocity $W=W(\tau)$ of the piston must satisfy the equation
\be
     \int_{W}^{\infty} (v-W)^2 \pi(Y-0,v,\tau)\, dv =
     \int_{-\infty}^{W} (v-W)^2 \pi(Y+0,v,\tau)\, dv
            \label{quadraticint}
\ee
see also an  additional requirement (H4') below.
\end{itemize}

In physical terms, (\ref{quadraticint}) is a pressure balance: the
piston ``chooses'' velocity $W$ so that the pressure of the
incoming particles balances out. Equation (\ref{quadraticint}) is
instrumental for our deterministic approximation of the piston
dynamics.

One can combine the two integrals in (\ref{quadraticint})
into one by introducing the density of
the particles colliding with the piston
(``density on the piston'') by
\be
    q(v,\tau;Y,W)=\left\{\begin{array}{ll}
       \pi(Y+0,v,\tau)  &  {\rm if}\ \ v<W\\
       \pi(Y-0,v,\tau)  &  {\rm if}\ \ v>W\\
          \end{array}\right .
            \label{qp}
\ee
Then (\ref{quadraticint}) can be rewritten as
$$
       \int_{-\infty}^{\infty} (v-W(\tau))^2\,{\rm sgn}
       (v-W(\tau))\, q(v,\tau;Y(\tau),W(\tau))\, dv =0
$$

We also remark that for $\tau>0$, when (\ref{ponpiston}) holds,
$$
    W(\tau) =
    \frac{\int v \pi(Y-0,v,\tau)\, dv}{\int \pi(Y-0,v,\tau)\, dv}=
    \frac{\int v \pi(Y+0,v,\tau)\, dv}{\int \pi(Y+0,v,\tau)\, dv}
$$
i.e.\ the piston's velocity is the average of the nearby particle
velocities on each side.

The system of (hydrodynamical) equations given in (H1)--(H4) is
closed and, given appropriate initial conditions, should
completely determine the functions $Y(\tau)$, $W(\tau)$ and
$\pi(y,v,\tau)$ for $\tau>0$, as we will see shortly.

To specify the initial conditions, we set $\pi(y,v,0)=\pi_0(y,v)$
and $Y(0)=0.5$. The initial velocity $W(0)$ does not have to be
specified, it comes ``for free'' as the solution of the equation
(\ref{quadraticint}) at time $\tau=0$. It is easy to check that
the initial speed $|W(0)|$ will be smaller than $v_{\min}$, in
fact $W(0)\to 0$ as $\varepsilon_0\to 0$ in (P5).

We first determine conditions under which equation
(\ref{quadraticint}) has a solution $W$. Let
$$
   v^-_{\sup}(\tau)=\sup\{v:\, \pi(Y-0,v,\tau)>0\}
$$
(with the convention that the supremum of an empty set
is $-\infty$) and
$$
   v^+_{\inf}(\tau)=\inf\{v:\, \pi(Y+0,v,\tau)>0\}
$$
(similarly, the infimum of an empty set must be set to $+\infty$).

\begin{lemma}
We have three cases:
\begin{itemize}
\item[{\rm (a)}] If $v^-_{\sup}>v^+_{\inf}$ or
$v^-_{\sup}=v^+_{\inf}\in\IR$, then (\ref{quadraticint}) has a
unique solution $W\in [v^+_{\inf},v^-_{\sup}]$. \item[{\rm (b)}]
If $v^-_{\sup}<v^+_{\inf}$, then the solutions of
(\ref{quadraticint}) occupy the entire interval
$[v^-_{\sup},v^+_{\inf}]$. \item[{\rm (c)}] If
$v^-_{\sup}=v^+_{\inf}=\infty$ or $v^-_{\sup}=v^+_{\inf}=-\infty$,
then (\ref{quadraticint}) has no real solutions.
\end{itemize}
\end{lemma}

\noindent{\em Proof}. In the case (a), the difference between the
left hand side and  the right hand side of (\ref{quadraticint}) is
a continuous and strictly monotonically decreasing function of
$W$. For $W<v^+_{\inf}$ it is positive, and for $W>v^-_{\sup}$
negative. The rest of the proof goes by direct inspection.
$\Box$\medskip

It is easy to show (we do not elaborate) that under our
assumptions (P1)--(P4) for every $\tau>0$ the density
$\pi(y,v,\tau)$ has a compact support on the $y,v$ plane, i.e.\
$\pi(y,v,\tau)\equiv 0$ for all $|v|>v_{\max}(\tau)$. Therefore,
the ``no solution'' case (c) never occurs. The multiple solution
case (b) is very unlikely, but not impossible. If that happens,
the velocity $W(\tau)$ must be defined uniquely by an additional
requirement:

\begin{itemize}
\item[(H4')]
If $W(\tau-0)\in [v^-_{\sup},v^+_{\inf}]$, we define $W(\tau)$
by continuity, $W(\tau)=W(\tau-0)$. If  $W(\tau-0)<v^-_{\sup}$
or $W(\tau-0)>v^+_{\inf}$, we set $W(\tau)=v^-_{\sup}$ or
$W=v^+_{\inf}$, respectively.
\end{itemize}

This completes the definition of $W(\tau)$ started by (H4).

For generic piecewise smooth densities $\pi(y,v,\tau)$, the
velocity $W(\tau)$ is continuous, but in some cases the continuity
of $W(\tau)$ might be broken. The following simple lemma will be
helpful, though:

\begin{lemma}
Suppose that for every $\tau\in [a,b]$ the density $\pi(y,v,\tau)$
is piecewise $C^1$ and has a finite number of $C^1$ smooth
discontinuity lines on the $y,v$ plane with positive slopes, as we
require of $\pi_0(y,v)$ in Section~\ref{secI}. Then $W(\tau)$ will
be continuous and piecewise differentiable on the interval
$[a,b]$.
\end{lemma}

We now pause to make a few remarks. The piston mass is never used
in our equations, because its macroscopic mass is zero. Indeed,
for the mechanical system described in Section~\ref{secI}, the
piston mass is $\sim L^2$, while the total mass of the gas
particles is $\sim L^3$, hence the relative mass of the piston
vanishes as $L\to\infty$. Consider now the total (macroscopic)
mass of the gas
$$
    {\cal M}_{\rm tot}(\tau)=\int_0^1\!\int \pi(y,v,\tau)\, dv\, dy
$$
and the mass in the left and right compartments, separately,
$$
    {\cal M}_L(\tau)=\int_0^{Y(\tau)}\!\int \pi(y,v,\tau)\, dv\, dy
$$
$$
    {\cal M}_R(\tau)=\int_{Y(\tau)}^1\!\int \pi(y,v,\tau)\, dv\, dy
$$
and the total kinetic energy
$$
     2{\cal E}_{\rm tot}(\tau)=\int_0^1\!\int v^2\pi(y,v,\tau)\, dv\, dy
$$
The following lemma is left as a (simple) exercise:

\begin{lemma}
The quantities ${\cal M}_{\rm tot}$, ${\cal M}_L$, ${\cal M}_R$,
and ${\cal E}_{\rm tot}$ remain constant in $\tau$.
\end{lemma}

The main equation (\ref{quadraticint}) also preserves the total
momentum $\int\!\int v\pi(y,v,\tau)\, dv\, dy$, but this quantity
changes due to collisions with the walls.

\medskip\noindent
{\bf Remark}. Previously, Lebowitz, Piasecki and Sinai \cite{LPS}
studied the piston dynamics under essentially the same initial
conditions as our (P1)--(P5). They argued heuristically that the
piston dynamics could be approximated by certain deterministic
equations in the original (microscopic) variables $x$ and $t$. In
fact, the present work grew as a continuation of \cite{LPS}. The
deterministic equations found in \cite{LPS} correspond to our
(\ref{pinside})--(\ref{W=Y'}) with obvious transformation back to
the variables $x,t$, but our main equation (\ref{quadraticint})
has a different counterpart in the context of \cite{LPS}, which
reads
\be
    \frac{d}{dt}V(t)=a\,\left [ \int_V^{\infty} (v-V(t))^2 \pi(Y-0,v,t)\, dv
      - \int_{-\infty}^V (v-V(t))^2 \pi(Y+0,v,t)\, dv \right ]
        \label{tquadraticint}
\ee
Here $X=X(t)$ and $V=V(t)=\dot{X}(t)$ denote the deterministic
position and velocity of the piston and $\pi(x,v,t)$ the density
of the gas (the constant $a$ appeared in (\ref{varepsMm})). We
refer to \cite{LPS} for more details and a heuristic derivation of
(\ref{tquadraticint}). Since (\ref{tquadraticint}), unlike our
(\ref{quadraticint}), is a differential equation, the initial
velocity $V(0)$ has to be specified separately, and it is
customary to set $V(0)=0$. Equation (\ref{tquadraticint}) can be
reduced to (\ref{quadraticint}) in the limit $L\to\infty$ as
follows. One can show (we omit details) that (\ref{tquadraticint})
is a dissipative equation whose solution with any (small enough)
initial condition $V(0)$ converges to the solution of
(\ref{quadraticint}) during a $t$-time interval of length $\sim\ln
L$. That interval has length $\sim L^{-1}\ln L$ on the $\tau$
axis, and so it vanishes as $L\to\infty$, this is why we replace
(\ref{tquadraticint}) with (\ref{quadraticint}) and ignore the
initial condition $V(0)$ when working with the thermodynamic
variables $\tau$ and $y$. For the same reasons, it will be
convenient to reset the initial value of the piston velocity in
the mechanical model of Section~\ref{secI} to from $V(0)=0$ to
$V(0)=W(0)$, see Theorem~\ref{tmdV2} below. The equation
(\ref{tquadraticint}) will not be used anymore in this paper.
\medskip

We now describe the solution of the hydrodynamical equations
(H1)--(H4) in more detail. Assume that for some $\tau
> 0$ the gas density $\pi(y,v,\tau)$ satisfies the following
requirements, similar to (P1)--(P5) imposed on the initial
function $\pi_0(y,v)$ in Section~\ref{secI}:

\begin{itemize}
\item[(P1')] {\em Smoothness}. $\pi(y,v,\tau)$ is a piecewise
$C^1$ function with uniformly bounded partial derivatives, i.e.\
$|\partial \pi/\partial y|\leq D_1'$ and $|\partial \pi/\partial
v|\leq D_1'$ for some $D_1'>0$.

\item[(P2')] {\em Discontinuity lines}. $\pi(y,v,\tau)$ has a
finite number ($\leq K_1'$) of discontinuity lines in the
$(y,v)$-plane with strictly positive slopes (each line is given by
an equation $v=f(y)$ where $f(y)$ is $C^1$ and
$0<c_1'<f'(y)<c_2'<\infty$).

\item[(P3')] {\em Density bounds}. Let
\be
         \pi(y,v,\tau)>\pi_{\min}'>0\ \ \ \ \ \ {\rm for}\ \ v_1'<|v|<v_2'
                \label{pmin}
\ee
for some $0<v_1'<v_2'<\infty$, and
\be
   \sup_{y,v}\pi(y,v,\tau)=\pi_{\max}'<\infty
      \label{pmax}
\ee

\item[(P4')] {\em Velocity ``cutoff''}. Let
\be
     \pi(y,v,\tau) = 0, \quad {\rm if} \quad |v| \leq v_{\rm min}'
     \quad {\rm or} \quad |v| \geq v_{\max}'
        \label{cutoff}
\ee
with some $0<v_{\min}'<v_{\max}'<\infty$.
\end{itemize}

\noindent Lastly, we want to assume, similarly to (P5), that
$\pi(y,v,\tau)$ is nearly symmetric about the piston, but this
assumption requires a little extra work, since the piston does not
have to stay at the middle point $Y(0)=0.5$. For every $Y\in
(0,1)$ denote by $h_Y$ the unique homeomorphism of $[0,1]$ such
that $h_Y(0)=1$, $h_Y(1)=0$, $h_Y(Y)=Y$ and $h_Y$ is linear on the
subsegments $[0,Y]$ and $[Y,1]$. Next, we consider
$[0,Y]\times\IR$ as a manifold in which points $(0,v)$ and
$(0,-v)$ are identified for all $v>0$, and so are the points
$(Y,v)$ and $(Y,-v)$ for $v>0$. Similarly, let $[Y,1]\times\IR$ be
a manifold in which one identifies $(1,v)$ with $(1,-v)$ and
$(Y,v)$ and $(Y,-v)$ for all $v>0$. We denote by $d_Y$ the
distance on each of these two manifolds induced by the Euclidean
metric $(dy^2+dv^2)^{1/2}$. The reason why we need this special
distance will be clear later, in the proof of
Proposition~\ref{prpropagate}.

\begin{itemize}
\item[(P5')] {\em Approximate pressure balance}. We require that
\be
             |Y(\tau)-0.5|<\varepsilon_0'
               \label{Y05}
\ee
and for any point $(y,v)$ with $0\leq y\leq 1$ and $v_{\min}'\leq
|v|\leq v_{\max}'$ there is another point $(y_{\ast},v_{\ast})$
``across the piston'', i.e.\ such that $(y-Y)(y_{\ast}-Y)<0$,
where $Y=Y(\tau)$, satisfying
\be
         d_Y((y_{\ast},v_{\ast}),(h_Y(y),-v))<\varepsilon_0'
           \label{Rtau}
\ee
and
\be
         |\pi(y,v,\tau)-\pi(y_{\ast},v_{\ast},\tau)|<\varepsilon_0'
             \label{symmetry}
\ee
for some sufficiently small $\varepsilon_0'>0$. In addition,
we require that
\be
         \varepsilon_0' < C_0'\varepsilon_0
         \label{ee0}
\ee
with some constant $C_0'>0$.
\end{itemize}

\noindent Actually, the map $(y,v)\mapsto (y_{\ast},v_{\ast})$
involved in (P5'), which we will denote by $R_{\tau}$, is
one-to-one and will be explicitly constructed below, in the proof
of Proposition~\ref{prpropagate}.

Again, we think of $D_1'$, $K_1'$, $c_1'$, $c_2'$, $v_1'$, $v_2'$,
$v_{\min}'$, $v_{\max}'$, $\pi_{\min}'$, $\pi_{\max}'$, and now
also $C_0'$, as global constants. They must be bounded on the time
interval on which we consider the dynamics (and $v_{\min}'$,
$\pi_{\min}'$ must be bounded away from zero), hence we may treat
all these constants as independent of $\tau$. By (\ref{ee0}),
$\varepsilon_0'$ is, just like $\varepsilon_0$ in (P5), a small
adjustable parameter.

Now we derive rather elementary but important consequences of the
above assumptions. Since the density $\pi(y,v,\tau)$ vanishes for
$|v|<v_{\min}'$, so does the function $q(v,\tau;Y,W)$ defined by
(\ref{qp}). Moreover, for all $|W|<v_{\min}'$, the function
$q(v,\tau;Y,W)$ will be independent of $W$ and can be redefined by
\be
    q(v,\tau;Y)=\left\{\begin{array}{ll}
       \pi(Y+0,v,\tau)  &  {\rm if}\ \ v<0\\
       \pi(Y-0,v,\tau)  &  {\rm if}\ \ v>0\\
          \end{array}\right .
            \label{qp0}
\ee
Also, the equation (\ref{quadraticint}) can be simplified: the
factor sgn$(v-W)$ can be replaced by sgn$\, v$. Then, expanding
the squares in (\ref{quadraticint}) reduces it to a simple
quadratic equation for $W$:
\be
       {\cal Q}_0W^2-2{\cal Q}_1W+{\cal Q}_2 = 0
         \label{quadratic}
\ee
where
\be
     {\cal Q}_{0}=\int {\rm sgn}\, v\cdot q(v,\tau;Y)\, dv
        \label{Q0}
\ee
\be
     {\cal Q}_{1}=\int v\, {\rm sgn}\, v\cdot q(v,\tau;Y)\, dv
        \label{Q1}
\ee
\be
     {\cal Q}_{2}=\int v^2\, {\rm sgn}\, v\cdot q(v,\tau;Y)\, dv
        \label{Q2}
\ee
with $Y=Y(\tau)$. The integrals ${\cal Q}_0,{\cal Q}_1,{\cal Q}_2$
have the following physical meaning:
$$
        m {\cal Q}_0=m_L-m_R
$$
$$
        m {\cal Q}_1=p_L-p_R
$$
$$
        m {\cal Q}_2=2(e_L-e_R)
$$
where $m_L,p_L,e_L$ represent the total mass, momentum and energy
of the incoming gas particles (per unit length) on the left hand
side of the piston, and  $m_R,p_R,e_R$ -- those on the right hand
side of it. The value ${\cal Q}_2$ also represents the net
pressure exerted on the piston by the gas if the piston did not
move. Of course, if ${\cal Q}_2(\tau)=0$, then we must have
$W(\tau)=0$, which agrees with (\ref{quadratic}). The following
lemma easily follows from (P1')--(P5'). It means that the function
$q(v,\tau;Y(\tau))$ is nearly symmetric in $v$ about $v=0$.

\begin{lemma}
For any smooth function $f(v)$ defined for $v>0$ we have
$$
       \left |\int_0^{\infty} f(v)\, q(v,\tau;Y(\tau))\, dv\, -\,
       \int_{-\infty}^0 f(-v)\, q(v,\tau;Y(\tau))\, dv\right |
       \leq C_f\,\varepsilon_0
$$
where the factor $C_f>0$ depends on $f$ but not on
$\varepsilon_0$.
\label{lmqq}
\end{lemma}

\noindent{\bf Convention}. We call constants that do not depend on
our small adjustable parameter $\varepsilon_0$ involved in (P5)
and (P5') {\em global constants} (such as $C_f$ in the above
lemma). All the constants in the requirements (P1)--(P5) and
(P1')--(P5') are global, except $\varepsilon_0$ itself and the
related $\varepsilon_0'$. In many cases, we will denote various
global constants by $C_i$, $i\geq 0$, or just by $C$.
\medskip

Lemma~\ref{lmqq} implies that ${\cal Q}_0$ and ${\cal Q}_2$ are
small, more precisely
\be
       \max\{|{\cal Q}_0|,|{\cal Q}_2|\}\leq C\varepsilon_0
            \label{Q0Q2}
\ee
where $C>0$ is a global constant. At the same time, the assumption
(P3') guarantees that
\be
             {\cal Q}_1\geq {\cal Q}_{1,\min}>0
            \label{Q1min}
\ee
where ${\cal Q}_{1,\min}$ is another global constant.

If $\varepsilon_0$ is small enough, there is a unique root of the
quadratic polynomial (\ref{quadratic}) on the interval
$(-v_{\min}', v_{\min}')$, which corresponds to the only solution
of (\ref{quadraticint}). Since this root is smaller, in absolute
value, than the other root of (\ref{quadratic}), it can expressed
by
\be
         W=\frac{{\cal Q}_1-\sqrt{{\cal Q}_1^2-{\cal Q}_0{\cal Q}_2}}{{\cal Q}_0}
                   \label{Wroot}
\ee
where the sign before the radical is ``$-$'', not ``$+$''. Of
course, (\ref{Wroot}) applies whenever ${\cal Q}_0\neq 0$, while
for ${\cal Q}_0=0$ we simply have
\be
             W = \frac{{\cal Q}_2}{2{\cal Q}_1}
           \label{Wroot0}
\ee

\begin{corollary}
If $\varepsilon_0$ is small enough, then
\be
            |W(\tau)|\leq {\cal B}\varepsilon_0<v_{\min}'/3
                 \label{WBeps0}
\ee
with some global constant ${\cal B}>0$.
\end{corollary}

\noindent{\em Proof}. This immediately follows from equations
(\ref{Q0Q2})--(\ref{Wroot0}). $\Box$\medskip

We now make an important remark.

\medskip\noindent
{\bf Remark (Extension)}. Consider the density of the incoming gas
particles on the left hand side of the piston, i.e.\
$\pi(y,v,\tau)$ for $y=Y(\tau)-0$ and $v>v_{\min}'$. This function
``terminates'' on the piston, i.e.\ has a discontinuity in $y$ at
$y=Y(\tau)$. But it can be naturally extended smoothly ``across
the piston'', i.e.\ for $y>Y(\tau)$ if one ignores the interaction
of the gas coming from the left compartment with the piston at
times $s\in(\tau-\delta,\tau)$ and applies the rule (H1) instead,
as if the gas ``passed through the piston''. This defines a smooth
extension of $\pi(y,v,\tau)$ from the region $y\leq Y(\tau)$ to
the region $Y(\tau)< y< Y(\tau)+O(\delta)$ for all $v\geq
v_{\min}'$. This extension allows us to differentiate
$q(v,\tau;Y)$ defined by (\ref{qp0}) with respect to $Y$ for any
$v\geq v_{\min}'$. A similar extension can be made for the density
$\pi(y,v,\tau)$ from the region $y\geq Y(\tau)$ to the region
$Y(\tau)-O(\delta)< y< Y(\tau)$ for all $v\leq -v_{\min}'$, hence
$q(v,\tau;Y)$ becomes differentiable with respect to $Y$ for
$v\leq -v_{\min}'$. We note that our extension can be
unambiguously defined because we only need it for $|v|\geq
v_{\min}'$ while the piston's velocity remains smaller than
$v_{\min}'$.\medskip

Now the quantities ${\cal Q}_0$, ${\cal Q}_1$, and ${\cal Q}_2$
defined by (\ref{Q0})--(\ref{Q2}) become differentiable in $Y$ for
each fixed $\tau$, and the assumptions (P1')--(P4') easily imply
that
\be
             |d{\cal Q}_i/dY|\leq C_1,\ \ \ \ \ i=0,1,2
               \label{dQdY}
\ee
where $C_1>0$ is a global constant.

\begin{corollary}
The piston acceleration $A(\tau)=dW(\tau)/d\tau$ satisfies
\be
       |A(\tau)|\leq\, C \varepsilon_0
         \label{ACeps0}
\ee
with a global constant $C>0$.
\end{corollary}

\noindent{\em Proof}. We differentiate the quadratic equation
(\ref{quadratic}) with respect to $\tau$ and get
$$
     A(\tau)=\frac{(d{\cal Q}_0/d\tau)W^2-2(d{\cal Q}_1
     /d\tau)W+(d{\cal Q}_2/d\tau)}{2({\cal Q}_1-{\cal Q}_0W)}
$$
Clearly, the denominator is bounded away from zero, and the
numerator has an upper bound of order $\varepsilon_0$, because
$|d{\cal Q}_i/d\tau|=|(d{\cal
Q}_i/dY)W|\leq\,$const$\cdot\varepsilon_0$ by (\ref{dQdY}) and
(\ref{WBeps0}). $\Box$\medskip

More importantly, we can now derive the existence and uniqueness
of the solution of the hydrodynamical equations (H1)--(H4) as long
as the conditions (P1')--(P5') continue holding:

\begin{lemma}
If the hydrodynamical equations {\rm (H1)--(H4)} have a solution
on an interval $0\leq\tau\leq T$ and the conditions {\rm
(P1')--(P5')} hold on this interval, then the solution is unique
at $\tau=T$ and can be extended immediately beyond the point
$\tau=T$. \label{lmpropH}
\end{lemma}

\noindent{\em Proof}. The only differential equation in our system
(H1)--(H4) is (\ref{W=Y'}), in which $W(\tau)$ is the root of the
quadratic equation (\ref{quadratic}) given by
(\ref{Wroot})--(\ref{Wroot0}). Due to the above Extension Remark
we can think of $W$ as an implicit function of $Y$, i.e.\
effectively $W=F(Y,\tau)$. Then the differential equation
(\ref{W=Y'}) takes a canonical form
\be
      \frac{d}{d\tau} Y(\tau)=F(Y(\tau),\tau)
          \label{YW1}
\ee
For this equation to have a unique solution, it suffices that
$F(Y,\tau)$ has a bounded partial derivative with respect to $Y$.

Since $W$ is a root of the quadratic equation (\ref{quadratic}),
we can differentiate (\ref{quadratic}) with respect to $Y$ and get
$$
       \frac{\partial F(Y,\tau)}{\partial Y}=
       \frac{(d{\cal Q}_0/dY)W^2-2(d{\cal Q}_1/dY)W+(d{\cal Q}_2/dY)}{2({\cal Q}_1-{\cal Q}_0W)}
$$
We already know that the denominator is bounded away from zero. It
follows from (\ref{dQdY}) that the numerator stays bounded above,
hence
\be
       \left |\frac{\partial F(Y,\tau)}{\partial Y}\right |
       \leq \kappa
        \label{dFdY}
\ee
with a global constant $\kappa>0$. $\Box$\medskip

Next we consider the evolution of a point $(y,v)$ in the domain
$$
           {\cal G}:=\{(y,v):\ 0\leq y\leq 1\}
$$
under the rules (H1)--(H3), i.e.\ as it moves freely with constant
velocity and collides elastically with the walls and the piston.
Denote by $(y_{\tau},v_{\tau})$ its position and velocity at time
$\tau\geq 0$. Then (H1) translates into $\dot{y}_{\tau} =v_{\tau}$
and $\dot{v}_{\tau}=0$ whenever $y_{\tau} \notin\{0,1,Y(\tau)\}$,
(H2) becomes $(y_{\tau+0},v_{\tau+0})= (y_{\tau-0},-v_{\tau-0})$
whenever $y_{\tau-0} \in\{0,1\}$, and (H3) gives
\be
       (y_{\tau+0},v_{\tau+0})=(y_{\tau-0},2W(\tau)-v_{\tau-0})
          \label{tildeFpiston}
\ee
whenever $y_{\tau-0}=Y(\tau)$. Note that (\ref{tildeFpiston})
corresponds to a special case of the mechanical collision rules
(\ref{V'})--(\ref{v'}) with $\varepsilon=0$ (equivalently, $m=0$).
Hence the point $(y,v)$ moves in ${\cal G}$ as if it was a gas
particle with zero mass.

The motion of points in $(y,v)$ is described by a one-parameter
family of transformations ${\cal F}^{\tau}:\, {\cal G}\to {\cal
G}$ defined by ${\cal F}^{\tau}(y_0,v_0)=(y_{\tau},v_{\tau})$ for
$\tau>0$. We will also write ${\cal
F}^{-\tau}(y_{\tau},v_{\tau})=(y_0,v_0)$. According to (H1)--(H3),
the density $\pi(y,v,\tau)$ satisfies a simple equation
\be
       \pi(y_{\tau},v_{\tau},\tau)=
       \pi({\cal F}^{-\tau}(y_{\tau},v_{\tau}),0)
       =\pi_0(y_0,v_0)
         \label{pFp}
\ee
for all $\tau\geq 0$. Also, it is easy to see that for each
$\tau>0$ the map ${\cal F}^{\tau}$ is one-to-one and preserves
area, i.e. ${\rm det}\,|D{\cal F}^{\tau}(y,v)|=1$.

Now, because of (P4), the initial density $\pi_0(y,v)$ can only be
positive in the region
$$
     {\cal G}^+:=\{ (y,v):\  0\leq y\leq 1,\ v_{\min}\leq |v|\leq v_{\max}\}
$$
hence we will restrict ourselves to points $(y,v)\in {\cal G}^+$
only. At any time $\tau>0$, the images of those points will be
confined to the region ${\cal G}^+(\tau):={\cal F}^{\tau}({\cal
G}^+)$. In particular, $\pi(y,v,\tau)=0$ for $(y,v)\notin {\cal
G}^+(\tau)$.

We now make an important observation. If a point
$(y_{\tau},v_{\tau})$ collides with a piston whose velocity is
slow, $|W(\tau)|\ll |v_{\tau}|$, they cannot recollide too soon:
the point must travel to a wall, bounce off it, and then travel
back to the piston before it hits it again. This is quantified in
the following lemma:

\begin{lemma}
Let a point $(y_{\tau},v_{\tau})\in{\cal G}^+(\tau)$ collide with
the piston, i.e.\ $y_{\tau}=Y(\tau)$. Then during the interval
$(\tau,\tau+\Delta)$ with
$$
      \Delta = \frac{1-2{\cal B}\varepsilon_0\tau}{v_{\max}' +3{\cal B}\varepsilon_0}
$$
it cannot recollide with the piston, i.e.\ $y_s\neq Y(s)$ for
$s\in (\tau,\tau+\Delta)$, provided {\rm (P1')--(P5')} continue
holding during this interval. \label{lmtau0geq}
\end{lemma}

\noindent{\em Proof}. The point's speed after the collision is at
least $v_{\min}'-2{\cal B}\varepsilon_0$ and at most
$v_{\max}'+2{\cal B}\varepsilon_0$. The piston cannot ``catch up''
with it, since $|W(\tau )|<v_{\min}'-2{\cal B}\varepsilon_0$ by
(\ref{WBeps0}). So, the point travels to the wall, bounces off it,
and travels back to the piston, and all that will take time
$$
      \Delta \geq 2D/(v_{\max}'+3{\cal B}\varepsilon_0)
$$
where $D=\min\{Y(\tau),1-Y(\tau)\}\geq 0.5-{\cal B}\varepsilon_0
\tau$. $\Box$\medskip

Therefore, as long as (P1')--(P4') hold, the collisions of each
moving point $(y_{\tau},v_{\tau})\in {\cal G}^+(\tau)$ with the
piston occur at well separated time moments, which allows us to
effectively count them. For $(x,v)\in {\cal G}^+$
$$
     N(y,v,\tau)=\#\{s\in (0,\tau):\ y_s=Y(s),\ v_s\neq W(s)\}
$$
is the number of collisions of the point $(y,v)$ with the piston
during the interval $(0,\tau)$. For each $\tau>0$, we partition
the region ${\cal G}^+(\tau)$ into subregions
$$
     {\cal G}_{n}^+(\tau):=\{ {\cal F}^{\tau}(y,v):\,
     (y,v)\in {\cal G}^+\ \ \&\ \ N(y,v,\tau)=n\}
$$
so ${\cal G}^+_{n}(\tau)$ is occupied by the points that at time
$\tau$ have experienced exactly $n$ collisions with the piston
during the interval $(0,\tau)$.

Now, for each $n\geq 1$ we define $\tau_n>0$ to be the first time
when a point $(y_{\tau},v_{\tau})\in {\cal G}^+(\tau)$ experiences
its $(n+1)$-st collision with the piston, i.e.
$$
   \tau_n=\sup\{\tau>0:\, {\cal G}_{n+1}^+(\tau) = \emptyset\}
$$
In particular, $\tau_1>0$ is the earliest time when a point
$(y_{\tau},v_{\tau})\in{\cal G}^+(\tau)$ experiences its first
recollision with the piston. Hence, no recollisions occur on the
interval $[0,\tau_1)$, and we call it the {\em zero-recollision
interval}. Similarly, on the interval $(\tau_1,\tau_2)$ no more
than one recollision with the piston is possible for any point,
and we call it the {\em one-recollision interval}.

The time moment $\tau_{\ast}$ mentioned in Theorem~\ref{tmmain} is
the earliest time when a point $(y_{\tau},v_{\tau})\in{\cal
G}^+(\tau)$ either experiences its third collision with the piston
or has its second collision with the piston given that the first
one occurred after $\tau_1$. Hence, $\tau_{\ast}\leq \tau_2$, and
actually $\tau_{\ast}$ is very close to $\tau_2$, see the next
lemma.

\begin{lemma}
Let {\rm (P1')--(P5')} hold on the interval
$(0,n/v_{\max}+\delta)$ for some $n\geq 1$ and $\delta>0$. Then,
for all sufficiently small $\varepsilon_0$
$$
     |\tau_k - k/v_{\max}| \leq C\varepsilon_0
$$
for all $1\leq k\leq n$, where $C>0$ is a global constant that may
depend on $n$. Also,
$$
     |\tau_{\ast} - 2/v_{\max}| \leq C\varepsilon_0
$$
\label{lm29}
\end{lemma}

\noindent{\em Proof}. The necessary lower bounds on $\tau_k$
follow from Lemma~\ref{lmtau0geq}. The necessary upper bounds are
just as easy to obtain, we omit details. $\Box$\medskip

It is clear at this point that the hydrodynamical equations
(H1)--(H4) will have a unique and ``well behaved'' solution as
long as the conditions (P1')--(P5') continue holding with some
small $\varepsilon_0$. Our next goal is to show that this is
indeed the case.

\begin{proposition}
Let $T>0$. If the initial density $\pi_0(y,v)$ satisfies {\rm
(P1)--(P5)} and $\varepsilon_0$ in {\rm (P5)} is small enough (for
the given $T$), then the conditions {\rm (P1')--(P5')} will hold
on the interval $0<\tau<T$. \label{prpropagate}
\end{proposition}

Note: The corresponding global constants in (P1')--(P5') will
depend on $T$ as specified below.\medskip

\noindent{\em Proof}. The main idea is to show that the
restrictions on $\pi(y,v,\tau)$ imposed by (P1)--(P5) at $\tau=0$
``deteriorate'' very slowly, as time goes on, so that (P1')--(P5')
will continue holding (``propagate'') with the respective global
constants slowly changing in time.

We first note that as long as (P1')--(P5') hold, the number of
collisions grows at most linearly in $\tau$, i.e.\ on any interval
$(0,\tau)$ on which (P1')--(P5') hold, every moving point
$(y_s,v_s)\in {\cal G}^+(s)$ experiences at most $\tau
v_{\max}'+1$ collisions with the piston, if $\varepsilon_0$ is
small enough, see Lemmas~\ref{lmtau0geq}--\ref{lm29}. Next, we
examine the conditions (P1')--(P5') individually and show that
each of them should hold up to time $T$, provided that the others
do.

We start with (P1). Due to (H1) we have
$$
             \frac{\partial \pi(y,v,\tau+s)}{\partial y}=
             \frac{\partial \pi(y-sv,v,\tau)}{\partial y}
$$
and
$$
             \frac{\partial \pi(y,v,\tau+s)}{\partial v}=
             \frac{\partial \pi(y-sv,v,t)}{\partial v}
             -s \frac{\partial \pi(y-sv,v,\tau)}{\partial y}
$$
for all $s>0$ such that the moving point located at $(y,v)$ at
time $\tau+s$ did not experience collisions with the piston during
the interval $(\tau,\tau+s)$. Thus, between collisions with the
walls and the piston, the partial derivatives of $\pi(y,v,\tau)$
can grow at most linearly with $\tau$. Collisions with the walls
could only change the sign of the derivatives of $p$, but not
their absolute values.

Now consider the effect of interactions with the piston. We
evaluate the partial derivatives of $\pi(y,v,\tau)$ at a point
$(y,v)$ after a collision with the piston at some earlier time
$s\in (0,\tau)$. For simplicity, assume that there are no other
collisions of the moving point $(y,v)$ with the piston or the
walls on the interval $(s,\tau)$. Then $s$ satisfies the equation
\be
              Y(s)=y-(\tau-s)v
             \label{Ys}
\ee
Due to (H3) and (H1) we have
\begin{eqnarray*}
            \pi(y,v,\tau) &=& \pi(y-(\tau-s)v,v,s+0)\\
              &=& \pi(y-(\tau-s)v,2W-v,s-0)\\
              &=& \pi(y-(\tau-s)v-(s-s_0)(2W-v),2W-v,s_0)
\end{eqnarray*}
where $s_0<s$ is any earlier time (that we consider fixed) and
$W=W(s)$ is the piston velocity at the time of collision. Let
$y_0=y-(\tau-s)v-(s-s_0)(2W-v)$ and $v_0=2W-v$. Then
\begin{eqnarray*}
      \frac{\partial \pi(y,v,\tau)}{\partial y} &=&
      \frac{\partial \pi(y_0,v_0,s_0)}{\partial y}
              \left [ 1+v\frac{ds}{dy}-2(s-s_0)\frac{dW}{dy}-
           (2W-v)\frac{ds}{dy}\right ]\\
              & &\ +  \frac{\partial \pi(y_0,v_0,s_0)}{\partial v}\cdot 2\,\frac{dW}{dy}
\end{eqnarray*}
Differentiating (\ref{Ys}) with respect to $y$ gives
$$
                \frac{dY}{ds}\cdot\frac{ds}{dy}=1+v\frac{ds}{dy}
$$
hence
$$
                \frac{ds}{dy}=\frac{1}{W-v}
$$
Also,
$$
                \frac{dW}{dy}=\frac{dW}{ds}\cdot\frac{ds}{dy}=\frac{A}{W-v}
$$
where $A=A(s)$ is the piston acceleration at the time of
collision. Now, as long as (P1')--(P5') hold, we have
$W=O(\varepsilon_0)$ and $A=O(\varepsilon_0)$, hence
$ds/dy=-v^{-1}+O(\varepsilon_0)$ and so
$$
          \frac{\partial \pi(y,v,\tau)}{\partial y}=
          -\frac{\partial \pi(y_0,v_0,s_0)}{\partial y} +
          O(\varepsilon_0)
$$
In other words, the piston (due to its low speed and acceleration)
acts almost as a wall, which only changes the sign of $\partial
\pi/\partial y$. A similar calculation (we omit it) holds for the
partial derivative with respect to $v$.

Thus, as long as (P1')--(P5') hold, the density $\pi(y,v,\tau)$
remains piecewise $C^1$ and its partial derivatives can grow at
most linearly with $\tau$.

Next, we check the condition (P2'). We begin with three special
discontinuity lines that do not explicitly appear in (P2). They
are created immediately by the reflections at the walls and the
piston at time $\tau=0$, since the initial density $\pi(y,v,0)$
does not have to satisfy (H2)--(H3). Those discontinuity lines are
$y=0$, $y=0.5$ and $y=1$ at $\tau=0$, and their images at $\tau>0$
will be slanted lines
\be
       y=v\tau,\ \ \ \ y=0.5+v\tau,\ \ \ \ y=1+v\tau
          \label{slanted}
\ee
respectively, see Fig.~2. So, their slope at any time $\tau$ is
positive and constant: $dy/dv=\tau$. It is not bounded away from
zero as $\tau\to 0$, so we have a technical violation of (P2') for
small $\tau$, but it will be clear immediately why this does not
bother us.

\begin{figure}[h]
\centering
\epsfig{figure=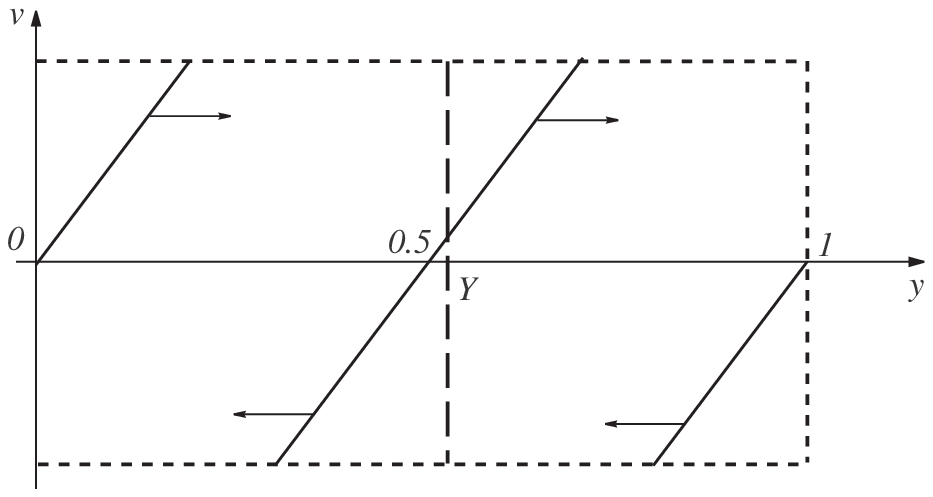}\caption{Slanted discontinuity lines.}
{The dashed vertical line shows the piston position.}
\end{figure}

The singularity lines (\ref{slanted}) only exist in the region
$v_{\min}\leq |v|\leq v_{\max}$ (elsewhere $p\equiv 0$), hence
they cannot intersect the piston $y=Y(\tau)$ for small $\tau$. It
will take some time, at least
$$
    \tau^{\ast}=\frac{0.5}{v_{\max}+{\cal B}\varepsilon_0} >0
$$
before any of these singularity lines ``reaches'' the piston and
its effect has to be reckoned with. At that time the slopes of
those lines will be bounded away from zero: $dy/dv\geq
\tau^{\ast}>0$, hence (P2') will hold.

We now consider the evolution of all discontinuity curves of the
function $\pi(y,v,\tau)$ as $\tau$ increases.  Let a discontinuity
curve of the function $\pi(y,v,s)$ at time $s$ be given by
equation $y=g_s(v)$, and its slope is then $h_s(v)=dg_s(v)/dv$.
Since the curve and its slope change in time, the function $g_s$
and its derivative $h_s$ depend on $s$. According to
(\ref{pinside}), we have $g_{s+r}(v)=g_s(v)+vr$ between collisions
with the piston and the walls, hence
\be
        \frac{dg_s(v)}{ds}=v
        \ \ \ \ \ {\rm and}\ \ \ \ \
        \frac{dh_s(v)}{ds}=1
       \label{dgds}
\ee
Hence, between collisions with the piston, the slope of
discontinuity curves grows linearly with $\tau$ (note that, in
particular, it remains positive).

Now, let the curve $y=g_s(v)$ cross the piston at some point
\be
           g_s(v)=Y(s)
             \label{gY}
\ee
(this equation makes $v$ a function of $s$). After the collision
with the piston, this point transforms to $(Y,2W-v)$, according to
the rule (H3), here $W=W(s)$ is the piston velocity. If $\tau>s$
is some fixed time, then the image of our point at time $\tau$ is
$(Y+(\tau-s)(2W-v),2W-v)$. Such points make a curve on the $y,v$
plane, parameterized by $s$ (the collision time). This will be the
discontinuity curve for the density $\pi(y,v,\tau)$ at time
$\tau$. Let $y_s=Y+(\tau-s)(2W-v)$ and $v_s=2W-v$ be the
coordinates of a point on that curve. To compute the slope
$dy_s/dv_s$ of that curve, we first differentiate $y_s$ and $v_s$
with respect to the parameter $s$:
\begin{eqnarray*}
       \frac{dy_s}{ds} &=& W+(\tau-s)\left [
       2\frac{dW}{ds}-\frac{dv}{ds}\right ]-(2W-v)\\
        &=& v-W+(\tau-s)\left [2A-\frac{dv}{ds}\right ]
\end{eqnarray*}
and
$$
       \frac{dv_s}{ds}=2\frac{dW}{ds}-\frac{dv}{ds}=2A-\frac{dv}{ds}
$$
where $A=A(s)$ is the piston acceleration  (at the collision time
$s$). Also, differentiating (\ref{gY}) with respect to $s$ and
using (\ref{dgds}) gives
$$
           \frac{dg_s(v)}{dv}\cdot\frac{dv}{ds}+v=W
$$
hence
$$
                \frac{dv}{ds}=\frac{W-v}{h_s(v)}
$$
Therefore, the slope of our singularity curve at time $\tau$ is
\be
         \frac{dy}{dv}(\tau)=
         \frac{(v-W)\, [h_s(v)+\tau-s]+2Ah_s(v)(\tau-s)}{v-W+2Ah_s(v)}
                 \label{slope}
\ee
As long as (P1')--(P5') hold, we have $W=O(\varepsilon_0)$ and
$A=O(\varepsilon_0)$, hence
\be
           \frac{dy}{dv}(\tau) = h_s(v)+\tau-s +
           O(\varepsilon_0)
              \label{slopea}
\ee
Hence, every collision with the piston only adds a
$O(\varepsilon_0)$ correction to the linear growth of the slopes
of discontinuity curves.

Next we check the conditions (P3')--(P5') based on the following
lemma:

\begin{lemma}
Let {\rm (P1')--(P5')} hold on an interval $(0,\tau)$. Then for
every point $(y,v)\in {\cal G}^+(\tau)$ there is another point
$(y_0,v_0)\in {\cal G}^+$ such that $\pi(y,v,\tau)=\pi(y_0,v_0,0)$
and
$$
       |\, |v|-|v_0|\, |=2(v_{\max}'\tau+1){\cal B}\varepsilon_0
$$
\label{lmvv}
\end{lemma}

\noindent{\em Proof}. We set $(y_0,v_0)={\cal F}^{-\tau}(y,v)$ and
use (\ref{pFp}). At each collision of the point $(y_0,v_0)$ with
the piston, its speed $|v|$ changes by $2|W|\leq 2{\cal
B}\varepsilon_0$ according to (\ref{tildeFpiston}) and
(\ref{WBeps0}), and the number of collisions is bounded by
$v_{\max}'\tau+1$. $\Box$\medskip

Lemma~\ref{lmvv} immediately implies that (P3') and (P4') continue
holding with global constants $v_1'$, $v_2'$, $v_{\min}'$ and
$v_{\max}'$ slowly changing with time -- they change at most by
$CT\varepsilon_0$ on the interval $(0,T)$, with a global constant
$C>0$. In particular, $v_1'$ and $v_{\min}'$ remain positive,
provided $\varepsilon_0$ is small enough. The constants
$\pi_{\min}'$ and $\pi_{\max}'$ do not change at all.

To check (P5'), we explicitly construct the map
$R_{\tau}:(y,v)\mapsto (y_{\ast},v_{\ast})$ involved in
(\ref{Rtau}) and (\ref{symmetry}), it is defined here by
$R_{\tau}={\cal F}^{\tau}\circ R_0\circ {\cal F}^{-\tau}$, where
$R_0(y,v)=(1-y,-v)$ is a simple reflection ``across the piston''
at time $\tau=0$. Now, (\ref{Y05}) follows from (\ref{WBeps0}),
and (\ref{symmetry}) follows from (P5).

Lastly, we derive (\ref{Rtau}) from the Lemma~\ref{lmvv}. Let
$(y,v)$ be a moving point at time $\tau$ and $(y_0,v_0)= {\cal
F}^{-\tau}(y,v)\in {\cal G}^+$ its preimage to time zero. Compare
the evolution of the point $(y_0,v_0)$ and its mirror image
$R_0(y_0,v_0)= (1-y_0,-v_0)\in {\cal G}^+$ ``across the piston''
during the interval $(0,\tau)$. Due to (\ref{Y05}) and
(\ref{WBeps0}), these two points will experience collisions with
the walls and the piston at time moments that differ at most by
$O(\varepsilon_0)$. And their velocities will also differ at most
by $O(\varepsilon_0)$, hence their positions at time $T$ will be
almost symmetric about the piston, up to $O(\varepsilon_0)$. This
implies (\ref{Rtau}).

Note that by the given time $T$ the above two moving points may
have experienced a different number of collisions, as one point
may have just collided with the piston or a wall, while the other
may be about to collide with it. To take care of this case, we
introduced the special distance $d_Y$ in (P5'). $\Box$\medskip

We summarize our main results in the following theorem:

\begin{theorem}
Let $T>0$ be given. If the initial density $\pi_0(y,v)$ satisfies
{\rm (P1)--(P5)} with a sufficiently small $\varepsilon_0$, then
\begin{itemize}
\item[{\rm (a)}] the solution of our hydrodynamical equations {\rm
(H1)--(H4)} exists and is unique on the interval $(0,T)$;
\item[{\rm (b)}] the density $\pi(y,v,\tau)$ satisfies {\rm
(P1')--(P5')} for all $0<\tau<T$; \item[{\rm (c)}] The piston
velocity and acceleration remain small,
$|W(\tau)|=O(\varepsilon_0)$ and $|A(\tau)|=O(\varepsilon_0)$, and
its position remains close to the midpoint $0.5$ in the sense
$|Y(\tau)-0.5|=O(\varepsilon_0)$, for all $0<\tau<T$; \item[{\rm
(d)}] we have $|\tau_k-k/v_{\max}|=O(\varepsilon_0)$ for all
$1\leq k< Tv_{\max}$, and if $Tv_{\max}>2$, then also
$|\tau_{\ast}-\tau_2|=O(\varepsilon_0)$.
\end{itemize}
\label{tmprop}
\end{theorem}

\begin{corollary}
If $\varepsilon_0=0$, so that the initial density $\pi_0(y,v)$ is
completely symmetric about the piston, the solution is trivial:
$Y(\tau)\equiv 0.5$ and $W(\tau)\equiv 0$ for all $\tau>0$.
\label{crprop}
\end{corollary}

Lastly, we demonstrate the reason for  our assumption that all the
discontinuity curves of the initial density $\pi(y,v)$ must have
positive slopes. It would be quite tempting to let $\pi(y,v)$ have
more general discontinuity lines, e.g. allow it be smooth for
$v_{\min}<|v|<v_{\max}$ and abruptly drop to 0 at $v=v_{\min}$ and
$v=v_{\max}$. The following example shows why this is not
acceptable.

\begin{figure}[h]
\centering
\epsfig{figure=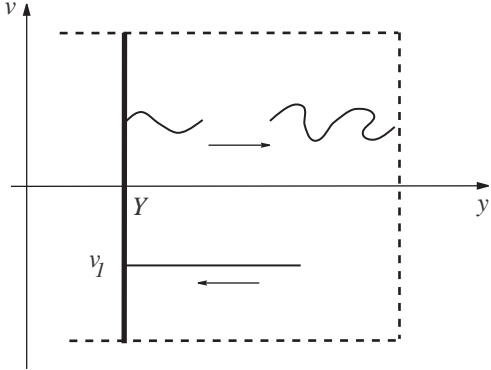}\caption{A horizontal discontinuity line
(bottom) comes off the piston as an oscillating curve (top).}
\end{figure}

\medskip\noindent{\bf Example}. Suppose the initial density
$\pi_0(y,v)$ has a horizontal discontinuity line $v=v_0$ (say,
$v_0=v_{\min}$ or $v_0=v_{\max}$). After one interaction with the
piston the image of this discontinuity line can oscillate up and
down, due to the fluctuations of the piston acceleration (Fig.~3).
As time goes on, this oscillating curve will ``travel'' to the
wall and come back to the piston, experiencing some distortions on
its way, caused by the differences in velocities of its points
(Fig.~3). When this curve comes back to the piston again, it may
well have ``turning points'' where its tangent line is vertical,
or even contain vertical segments of positive length. This
produces unwanted singularities or even discontinuities of the
piston velocity and acceleration. The same phenomena can also
occur when a discontinuity line of the initial density
$\pi_0(y,v)$ has a negative slope.

\section{Dynamics before the first recollision}
\label{secZRI} \setcounter{equation}{0}

In this section we begin to study the mechanical model of the
piston in the ideal gas described in Section~\ref{secI}. We will
show that the random trajectory of the piston described by the
functions $Y_L(\tau,\omega)=X_L(\tau L,\omega)/L$ and
$W_L(\tau,\omega)=V_L(\tau L,\omega)$, cf.\ (\ref{YWL}), converges
in probability, as $L\to\infty$, to the solution of the
hydrodynamical equations $Y(\tau)$ and $W(\tau)$ found in the
previous section, on the zero-recollision interval $(0,\tau_1)$.

\medskip
\noindent{\bf Convention}. For brevity of notation, we will
suppress the dependence of $L$ and $\omega$ in our expressions,
when it does not cause confusion. For example, we will write
$X(t)$ and $V(t)$ instead of $X_L(t,\omega)$ and $V_L(t,\omega)$,
respectively, etc.
\medskip

We will work here with the microscopic time $t$. First, we define
the ``microscopic'' gas density, which we will denote by
$p(x,v,t)$, for all $t\geq 0$. For $t=0$ it is initialized by
$p(x,v,0)=\pi_0(x/L,v)$, see (\ref{pLpi}). For $t>0$, its
evolution is defined by the rules similar to (H1)--(H3): the free
motion between collisions
\be
      p(x,v,t)=p(x-vs,v,t-s)
         \label{pinsideL}
\ee
for $s>0$ such that $x-vr\notin\{0,X(t-r),L\}$ for all $r\in
(0,s)$; reflections at the walls
\be
       p(0,v,t) = p(0,-v,t)
       \ \ \ \ \ {\rm and}\ \ \ \ \
       p(L,v,t) = p(L,-v,t)
          \label{pwallL}
\ee
and elastic collisions with the piston
\begin{eqnarray}
      p(X(t)\pm 0,v,t) &=& p(X(t)\pm 0,2V(t)-v,t)
        \label{ponpistonL}
\end{eqnarray}
Since the last equation involves the random functions $X(t)$ and
$V(t)$, the density $p(x,v,t)$ will depend on $\omega$, i.e.\ it
is now a random function.

The evolution of the density $p(x,v,t)$ can be conveniently
described with the help of a one-parameter family of
transformations $F^t$ similar to ${\cal F}^{\tau}$ defined in
Section~\ref{secHE}. Let $(x,v)$ be a point in the domain
$$
           G:=\{(x,v):\ 0\leq x\leq L\}
$$
Its trajectory $(x_t,v_t)$ for $t>0$ is defined by the free motion
inside the container, $\dot{x}_t =v_t$ and $\dot{v}_t=0$ whenever
$x_t \notin\{0,X(t),L\}$, reflections at the walls
$(x_{t+0},v_{t+0})= (x_{t-0},-v_{t-0})$ whenever $x_{t-0}
\in\{0,L\}$, and collisions with the piston
\be
       (x_{t+0},v_{t+0})=(x_{t-0},2V(t)-v_{t-0})
          \label{Fpiston}
\ee
whenever $x_{t-0}=X(t)$. Now the family of transformations $F^t$
is defined by $F^t(x_0,v_0)=(x_t,v_t)$ for $t>0$. We will also
write $F^{-t}(x_t,v_t)=(x_0,v_0)$. Now we simply have
\be
       p(x_t,v_t,t)=
       p(F^{-t}(x_t,v_t),0)=p(x_0,v_0,0)
         \label{pFpL}
\ee
Note that for each $t>0$ the map $F^t$ is a bijection of $G$ and
preserves area, i.e.
\be
          {\rm det} \,|DF^t(x,v)|=1
            \label{det1}
\ee
We emphasize that the transformation $F^t$, just as the density
$p(x,v,t)$, is random, i.e. depends on $\omega$.

\medskip\noindent
{\bf Remark}. The piston velocity $V(t)$ is a piecewise constant
function updated at the moments of collision with gas atoms by the
rules (\ref{V'})--(\ref{v'}). If $t$ is such a collision moment,
then $V(t)$ in equation (\ref{ponpistonL}) must be replaced by the
average of its one-sided limit values $(V(t-0)+V(t+0))/2$. This
modification is important, since it makes the rule
(\ref{ponpistonL}) equivalent to (\ref{V'})--(\ref{v'}) when
$(x_t,v_t)$ represents an actual gas particle of mass $m$.
Otherwise, it will correspond to the motion of a particle of zero
mass, and we may call it a {\em virtual particle}.
\medskip

Because of (P4), the initial density $p(x,v,0)$ can only be
positive in the region
$$
    G^+:=\{ (x,v):\  0\leq x\leq L,\ v_{\min}\leq |v|\leq v_{\max}\}
$$
which therefore contains all the gas particles at time $t=0$. For
any $t>0$, the region $G^+(t):=F^t(G^+)$ contains all the actual
gas particles at time $t$, and $p(x,v,t)=0$ for all $(x,v)\notin
G^+(t)$.

For each point $(x,v)\in G$ and $t>0$ we define the number of
collisions with the piston during the interval $(0,t)$
$$
     N(x,v,t)=\#\{s\in (0,t):\ x_s=X(s),\ v_s\neq V(s)\}
$$
Then we partition the region $G$ into subregions
$$
   G_n(t):=\{F^t(x,v):\ (x,v)\in G\ \ \&\ \ N(x,v,t)=n\}
$$
and put $G_n^+(t):=G^+(t)\cap G_n(t)$. The region $G_n^+(t)$ is
occupied by the points that have experienced exactly $n$
collisions with the piston during the time interval $(0,t)$.

We emphasize that our transformations $F^t$ and the regions
$G_n(t)$ and $G_n^+(t)$ depend on $\omega$, i.e.\ are random. We
note, however, that they are completely determined by the
trajectory of the piston, i.e. by the function $X(s)$, $0<s<t$.

The family of transformations ${\cal F}^{\tau}:\, {\cal G}\to
{\cal G}$ introduced in Section~\ref{secHE} induces a
(deterministic) family $\tilde{F}^t:\, G\to G$ defined as follows:
if ${\cal F}^{\tau}(y,v)= (y_{\tau},v_{\tau})$, then we put
$\tilde{F}^{\tau L}(yL,v):= (y_{\tau}L,v)$. The transformations
$\tilde{F}^t$ define a deterministic density function on $G$ by
$\tilde{p}(x,v,t):= p(\tilde{F}^t(x,v),0)$, which is related to
the density $\pi(y,v,\tau)$ studied in Section~\ref{secHE} by
$\tilde{p} (x,v,t)=\pi(x/L,v,t/L)$. We also put
$\tilde{G}^+(t):={\cal F}^t(G^+)$.

The gas particles in $G_0(t)$ make a Poisson process, as the
following lemma shows. Let $\omega\in\Omega$ and $t>0$. Fix the
trajectory of the piston $X(s)$, $0<s<t$. That completely
specifies the region $G_0(t)$ and the density $p(x,v,t)$.

\begin{lemma}
The conditional distribution of the gas particles in $G_0(t)$
(given the trajectory $X(s)$, $0<s<t$, of the piston) is Poisson
with density function $L^2p(x,v,t)$. \label{lmpot1}
\end{lemma}

\noindent{\em Proof}. Let $D\subset G_0(t)$ be any domain. Then
its preimage $F^{-(t-s)}(D)$ stays positive distance away from the
piston $X(s)$ for all $s\in (0,t)$. Hence, the particles starting
out in the region $F^{-t}(D)$ and ending up in the region $D$
could not affect the piston during the time interval $(0,t)$.
Therefore, the number of particles in $D$ at time $t$, being equal
to the number of particles in $F^{-t}(D)$ at time 0, is
independent of the piston trajectory, so it is a Poisson random
variable with parameter
$$
    \lambda_D(t)=L^2\int\!\!\int_{F^{-t}(D)}p(x,v,0)\, dx\, dv=
    L^2\int\!\!\int_{D} p(x,v,t)\, dx\, dv
$$
The identity of the above integrals follows from (\ref{pFpL}) and
(\ref{det1}). $\Box$\medskip

\noindent{\bf Remark}. For any domain $D\subset G_0(t)$ its
preimage $F^{-t}(D)$ is actually independent of $\omega$. Indeed,
let $F^t_0$ be another family of transformations on $G$ defined by
the free motion on the entire interval $0<x<L$ and elastic
reflections at the walls $x=0$ and $x=L$ only (as if the piston
did not exist). Then we have $F^{-t}(D)=F^{-t}_0(D)$ for any
domain $D\subset G_0(t)$.
\medskip

For $n\geq 1$, we define $T_n$ to be the earliest time the piston
interacts with points from $G_{n}^+(t)$ (thus creating the region
$G_{n+1}^+(t)$), or, equivalently,
\be
   T_n=\sup_{t>0}\{G_{n+1}^+(t) = \emptyset\}
     \label{Tn}
\ee
The time moments $T_n=T_n(\omega)$ are random analogues of
$\tau_n$ introduced in Section~\ref{secHE}. In particular, $T_1$
is the time of the first recollision in the system (by an actual
or a virtual particle).

\begin{lemma}
For all $\omega\in\Omega$
\be
         T_1 \leq T_{1,\max}:=L/v_{\max}
            \label{T0max}
\ee
\label{lmT0max}
\end{lemma}

\noindent{\em Proof}. The fastest particles $(x,v)\in G^+$ that
collide with the piston at time 0 will move with the speed
$v_{\max}$ and recollide with the piston at time $t$ that
satisfies $v_{\max}t+|X(t)-L/2|=L$. For all such $t$ we have
$$
          T_1\leq t = \frac{L-|X(t)-L/2|}{v_{\max}}
$$
This proves the lemma. $\Box$ \medskip

During the time interval $(0,T_1)$ the piston interacts with
particles in $G_0^+(t)$. Denote by
\begin{eqnarray*}
     {\cal X}_0(t)=\{(x,v):\, x=X(t)+0,\, -v_{\max}<v<-v_{\min}\}
       \\
       \cup \{(x,v):\, x=X(t)-0,\, v_{\min}<v<v_{\max}\}
\end{eqnarray*}
two immediate one-sided vicinities of the piston which contain all
``incoming'' particles, which are about to collide with the
piston.

In order to study the piston dynamics on the zero-recollision
interval, it is convenient to assume that the piston is slow
enough and only interacts with the original particles that started
in $G^+$ at time $0$. A subinterval $(0,S_1) \subset (0,T_1)$
where the piston satisfies these requirements will be called a
``slow'' interval:

\medskip
\noindent{\bf Definition} We define $(0,S_1)\subset (0,T_1)$ to
be the maximal time interval on which \\ (a) $|V(t)|<v_{\min}$;\\
(b) ${\cal X}_0(t) \subset G_0(t)$.
\medskip

Note that by the condition (b) all the incoming particles that
are about to hit the piston at time $t$ have started out in $G^+$
at time zero and have never interacted with the piston during the
time interval $(0,t)$.

Almost all of our considerations in this section are restricted to
the ``slow'' interval $(0,S_1)$. But in the end of the section we
will see that for typical $\omega\in\Omega$ the ``slow'' interval
coincides with the entire interval $(0,T_1)$, i.e.\ $S_1=T_1$.

Now, for every $t\in (0,S_1)$ we define the density of colliding
particles ``on the piston'' $q(v,t)$ by
\be
    q(v,t)=\left\{\begin{array}{ll}
       p(X(t)-0,v,t)  &  {\rm if}\ \ v>0\\
       p(X(t)+0,v,t)  &  {\rm if}\ \ v<0\\
          \end{array}\right .
            \label{qp00}
\ee
cf.\ (\ref{qp0}). Next, we define
\be
     Q_{0}(t)=\int ({\rm sgn}\, v)\, q(v,t)\, dv
        \label{Q00}
\ee
\be
     Q_{1}(t)=\int v\, ({\rm sgn}\, v)\, q(v,t)\, dv
        \label{Q10}
\ee
\be
     Q_{2}(t)=\int v^2\, ({\rm sgn}\, v)\, q(v,t)\, dv
        \label{Q20}
\ee
in a way similar to (\ref{Q0})--(\ref{Q2}) in Section~\ref{secHE}.

Since $p(x,v,t)$, restricted to the domain $G_0(t)$, coincides
with $\tilde{p}(x,v,t)=\pi(x/L,v,t/L)$, the conditions
(P1')--(P2') imply

\begin{lemma}
The density $p(x,v,t)$ restricted to the region $G_0(t)$ is
piecewise $C^1$ smooth on the $x,v$ plane. We also have $\left
|\partial p(x,v,t)/\partial x \right | \leq D_1'/L$. The
discontinuity lines of  $p(x,v,t)$ within the region $G_0(t)$ have
slope of order $O(1/L)$ (for large $L$, they are almost parallel
to the $x$ axis). \label{lmpx}
\end{lemma}

Due to the above lemma the quantities $Q_i$, $i=0,1,2$, are, as
functions of the piston position $X$, smooth and have derivatives
\be
    \left |\frac{\partial Q_i}{\partial X}\right |
      \leq \frac{\rm const}{L}
        \label{Qx}
\ee
where const is a global constant.

The following theorem gives a key technical estimate of this
section.

\begin{theorem}
For sufficiently large $L$ there is a set
$\Omega_{0}^{\ast}\subset\Omega$ of initial configurations of
gas particles such that \\
{\rm (i)} for some constant $c>0$
$$
         P(\Omega_{0}^{\ast})>1-L^{-c\,\ln\ln L}
$$
{\rm (ii)} for each configuration $\omega\in\Omega^{\ast}_{0}$,
for each time interval
$$
       (t,t+\Delta t) \subset (0,S_1)
$$
such that
\be
             \frac{1}{L^{2}}<\Delta t <\frac{1}{L^{2/3}\ln L}
                 \label{Deltat}
\ee
the change of the velocity of the piston during $(t,t+\Delta t)$
satisfies
\be
        V(t+\Delta t)-V(t)=
        {\cal D}(t)\, \Delta t + \chi
           \label{VV}
\ee
where
\be
       {\cal D}(t)/a =
       Q_0(t)V^2(t)-2Q_1(t)V(t)+Q_2(t)
          \label{calD}
\ee
and
\be
        |\chi| \leq C\,\frac{\ln L\,\sqrt{\Delta t}}{L}
       \label{chibound}
\ee
with some constant $C>0$. \label{tmdV1}
\end{theorem}

\noindent {\em Remark}. The function ${\cal D}(t)/a$ in
(\ref{calD}) is the random analogue of the quadratic polynomial
(\ref{quadratic}). The term ${\cal D}(t)\, \Delta t$ in (\ref{VV})
is the main (``deterministic'') component of the dynamics of the
piston velocity. The term $\chi$ represents random fluctuations.
\medskip

\noindent{\em Proof}. The set $\Omega^{\ast}_{0}$ will consist of
all configurations that satisfy certain requirements. We start
with {\em preliminary requirements}.

Consider a discrete set of time moments $t_i=i/L^2$, where
$i=0,1,\ldots,I$ and $I=[T_{1,\max}L^2]$. Partition the domain $G$
into the strips $S_j:=\{ (x,v):\ j/L^2\leq x<(j+1)/L^2\}$, where
$j=0,1,\ldots,L^3-1$. For each $i$ and $j$ denote by $N_{i,j}$ the
number of gas particles in the region $S_j\cap G_0(t_i)$ at time
$t_i$. Our preliminary requirements are
\be
             N_{i,j}\leq \ln L
               \label{Nij}
\ee
for all $0\leq i\leq I$ and $0\leq j<L^3$. We observe that
$N_{i,j}$ equals the number of gas particles in the region
$F^{-t_i}(S_j\cap G_0(t_i)) $ at time $0$, and
$$
    F^{-t_i}(S_j\cap G_0(t_i)) \subset F^{-t_i}_0(S_j)
$$
So, $N_{i,j}$ does not exceed the number of gas particles in
$F^{-t_i}_0(S_j)$ at time $0$, denote the latter by
$\tilde{N}_{i,j}$. Now, $\tilde{N}_{i,j}$ is a Poisson random
variable whose parameter $\lambda_{i,j}$ is bounded by
$$
      \lambda_{i,j}\leq L^2\pi_{\max}|S_j|\leq
      2\pi_{\max}(v_{\max}-v_{\min})
$$
According to Corollary~\ref{crPoi1}, for each $i,j$ our
requirement (\ref{Nij}) will fail with probability $<L^{-d\,\ln\ln
L}$ with some $d>0$. The total number of pairs $i,j$ equals
$L^3I=L^5T_{1,\max}\leq L^{6}/2v_{\max}$. Hence, all our
preliminary requirements hold with probability $>1-L^{-c'\,\ln\ln
L}$ with some global constant $c'>0$.

We now turn to the proof of (\ref{Deltat})--(\ref{calD}). Let
$i=[L^2t]$ and $t_1=(i+1)/L^2$. Note that $t_1-t\leq L^{-2}$. One
can easily derive from our preliminary requirements that the
number of gas particles colliding with the piston on the time
interval $(t,t_1)$ is less than const$\cdot \ln L$. Hence, the
piston velocity $V$ does not change by more than const$\cdot\ln
L/L^2$ during this interval. This amount is less than the bound on
$\chi$ in (\ref{chibound}) for all $\Delta t$ satisfying
(\ref{Deltat}). Therefore, we can ignore the interval $(t,t_1)$
and assume that $t=t_1$. Note that the quantities $Q_0,Q_1,Q_2$ in
(\ref{calD}) will change, as the result of the substitution
$t=t_1$, also, but only by the amount $<\,{\rm const}/L^3$ due to
Lemma~\ref{lmpx}. This change can be obviously ignored, too. So,
we suppose that $t=i/L^2$ for some $i=0,1,\ldots,I$

Now we state our {\em main requirements}. We again partition the $x$
axis into intervals $S_j:=\{j/L^2\leq x<(j+1)/L^2\}$, where
$j=0,1,\ldots,L^3-1$. For each $j$ we put $x_j=j/L^2$. For each integer
$p$, $|p|\leq v_{\min}L^2$, we put $v_p=p/L^2$ and for each integer
$1\leq q\leq L^2$ we put $d_q=q/L^2$. For each triple $(j,p,q)$ we now
define two trapezoid-like domains on the $x,v$ plane (see Fig.~4
below):
\be
           D^-_{j,p,q}:=
           \left \{(x,v):\  \frac{v-v_p}{x-x_j}
           < -\frac{1}{d_q},\ \ v_{\min}<v<v_{\max}\right \}
              \label{Djpq-}
\ee
and
\be
           D^+_{j,p,q}:=
           \left \{(x,v):\  \frac{v-v_p}{x-x_j}
           < -\frac{1}{d_q},\ \ -v_{\max}<v<-v_{\min}\right \}
              \label{Djpq+}
\ee
(here $-1/d_q$ is the slope of the oblique side of these
trapezoids), and two strips
$$
           U^-_{j,p,q}:= \{(x,v):\   |x-x_j +d_q(v-v_p)|<10v_{\max}/L^{2},
           \ \ v_{\min}<v<v_{\max}\}
$$
and
$$
           U^+_{j,p,q}:= \{(x,v):\   |x-x_j +d_q(v-v_p)|<10v_{\max}/L^{2},
           \ \ -v_{\max}<v<-v_{\min}\}
$$
Note that $U^{\pm}_{j,p,q}$ are the neighborhoods of the oblique
sides of the trapezoids $D^{\pm}_{j,p,q}$.

Consider all time moments $t_i=i/L^2$ for $i=0,1,\ldots,I$. Denote
by $N^{\pm}_{i,j,p,q}$ the number of particles in the region
$F_0^{-t_i}(D^{\pm}_{j,p,q})$ at time $0$. And denote by
$M^{\pm}_{i,j,p,q}$ the number of particles in the region
$F_0^{-t_i}(U^{\pm}_{j,p,q})$ at time $0$. These are Poisson
random variables. The parameter of the variable
$N^{\pm}_{i,j,p,q}$ is
$$
     \lambda^{\pm}_{i,j,p,q}=E(N^{\pm}_{i,j,p,q})=
     L^2\int_{F_0^{-t_i}(D^{\pm}_{j,p,q})}p(x,v,0)\, dx\, dv
$$
One can verify directly that $\lambda^{\pm}_{i,j,p,q}$ are
uniformly bounded below by a positive constant (even for the
smallest $d_p$, i.e. for $d_q=1/L^2$), due to the assumption
(\ref{pmin0}) on the initial density. Also note that the
parameters of $M^{\pm}_{i,j,p,q}$ are uniformly bounded above, by
$$
   E(M^{\pm}_{i,j,p,q})<\pi_{\max}|U^{\pm}_{j,p,q}|
   <20v_{\max}(v_{\max}-v_{\min})\pi_{\max}
$$

Our main requirements are
\be
       |N^{\pm}_{i,j,p,q}-\lambda^{\pm}_{i,j,p,q}|\leq
       \ln L\,\sqrt{\lambda^{\pm}_{i,j,p,q}}
           \label{Nijpq}
\ee
and
\be
             M^{\pm}_{i,j,p,q}\leq \ln L
            \label{Mijpq}
\ee

By Lemma~\ref{lmPoi3} and Corollary~\ref{crPoi1} the probability
that any of these requirements fails will be less than
$L^{-d\,\ln\ln L}$ with some constant $d>0$. The total number of
quadruples $(i,j,p,q)$ does not exceed $L^9T_{1,\max}\leq L^{A''}$
with some fixed $A''>0$. Therefore, all our main requirements hold
with probability $>1-L^{-c''\,\ln\ln L}$ with some constant
$c''>0$.

In addition, let
$$
         Z^{\pm}_{i,j,p,q}=\sum_{(x,v)\in F_0^{-t_i}(D^{\pm}_{j,p,q})}v
$$
taken at time $0$. This is an ``integrated'' Poisson random
variable, as defined in Appendix. (Technically, we require there
that the domain must be on one side of the $x$ axis, and now it
may happen here that the region $F_0^{-t_i}(D^{\pm}_{j,p,q})$
crosses the wall $x=0$ or $x=L$ and then lies on both sides on the
$x$-axis; in that case we need to replace $t_i$ by a nearby time
moment $t_{i'}<t_i$ so that $F_0^{-t_{i'}}(D^{\pm}_{j,p,q})$ lies
entirely on one side of the $x$ axis and define
$Z^{\pm}_{i,j,p,q}$ at time $t_i-t_{i'}$ rather than $0$; Some
obvious modifications need to be made then, we omit details.) The
estimates obtained in Appendix yield
\be
       E(Z^{\pm}_{i,j,p,q})=L^2 \int_{F_0^{-t_i}(D^{\pm}_{j,p,q})}
       v\, p(x,v,0)\, dx \, dv
          \label{EZijpq}
\ee
\be
       {\rm Var}(Z^{\pm}_{i,j,p,q})=L^2\int_{F_0^{-t_i}(D^{\pm}_{j,p,q})}
       v^2p(x,v,0)\, dx\, dv
           \label{VarZijpq}
\ee

Our last main requirement is
\be
          |Z^{\pm}_{i,j,p,q}-E(Z^{\pm}_{i,j,p,q})|\leq
          \ln L\,\sqrt{{\rm Var}(Z^{\pm}_{i,j,p,q})}
           \label{Wijpq}
\ee
for all $i,j,p,q$. The probability of failure for these
requirements is estimated exactly as above, by using
Lemma~\ref{lmZ2}.

We now turn to the estimation of the piston velocity on the time
interval $(t,t+\Delta t)$.  Recall that $t=t_i$ for some
$i=0,1,\ldots,I$.

\medskip
\noindent {\bf Velocity decomposition scheme}. Here we obtain a
general formula for the piston velocity, which we will use in the
proof of several theorems. The laws of elastic collisions imply
\cite{LPS}
\be
    V(t+\Delta t)=(1-\varepsilon)^kV(t) +
    \varepsilon\sum_{j=1}^k(1-\varepsilon)^{k-j}\cdot v_j
       \label{main}
\ee
Here $k$ is the number of particles colliding with the piston
during the time interval $(t,t+\Delta t)$, and $v_j$ are their
velocities numbered in the order in which the particles collide.
Equation (\ref{main}) can be easily verified by induction on $k$.

We modify the formula (\ref{main}) as follows:
\be
    V(t+\Delta t)=(1-\varepsilon k)V(t)
    + \varepsilon\sum_{j=1}^k v_j+\chi^{(1)}+\chi^{(2)}
          \label{main1}
\ee
where
$$
     \chi^{(1)}=V(t)[(1-\varepsilon)^k-1+\varepsilon k]
$$
and
$$
     \chi^{(2)}=\varepsilon\sum_{j=1}^kv_j[(1-\varepsilon)^{k-j}-1]
$$
Let us assume that the fluctuations of the velocity $V(s)$ on the
interval $(t,t+\Delta t)$, are bounded: \be
       \sup_{s\in (t,t+\Delta t)} |V(s) - V(t)| \leq \delta V
        \label{deltaV}
\ee
Consider two regions on the $x,v$ plane:
\be
      D_1=\left \{(x,v):\  \frac{v-V(t)-({\rm sgn}\, v)\, \delta V}{x-X(t)}
         < -\frac{1}{\Delta t},\ \ v_{\min}<|v|<v_{\max} \right \}
               \label{D1}
\ee
and
\be
      D_2=\left \{(x,v):\  \frac{v-V(t)+({\rm sgn}\, v)\, \delta V}{x-X(t)}
         < -\frac{1}{\Delta t},\ \ v_{\min}<|v|<v_{\max} \right \}
               \label{D2}
\ee
Each of them is a union of two trapezoids $D_i=D_i^+\cup D_i^-$,
$i=1,2$, where $D_i^-$ denotes the upper and $D_i^+$ the lower
trapezoid, see Fig.~4.

\begin{figure}[h]
\centering
\epsfig{figure=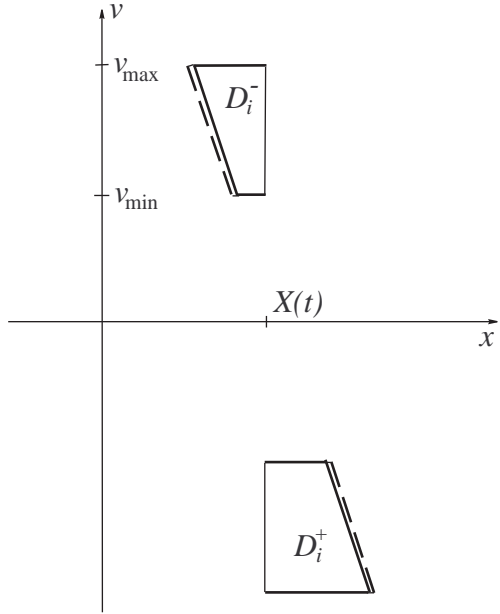} \caption{Region $D_1$ is bounded by
solid lines. Region $D_2$ is bounded by a dashed line.}
\end{figure}

Note that $D_1\subset D_2$. The bound (\ref{deltaV}) implies that
all the particles in the region $D_1$ necessarily collide with the
piston during the time interval $(t,t+\Delta t)$. Moreover, the
trajectory of every point $(x,v)\in D_1$ hits the piston within
time $\Delta t$, hence $D_1\subset G_0(t)$ by the condition (b) in
the definition of the ``slow'' interval $(0,S_1)$. The bound
(\ref{deltaV}) also implies that all the particles actually
colliding with the piston during the interval $(t,t+\Delta t)$ are
contained in $D_2$ (though it is not necessarily true that
$D_2\subset G_0(t)$).

We now obtain an upper bound on $k$. Since the velocities of the
gas particles and the piston are bounded by $v_{\max}$ all the
particles colliding with the piston during the interval
$(t,t+\Delta t)$ are contained in the region $S'\cap G_0(t)$,
where
$$
           S'=\{(x,y):\ |x-X(t)| < 2v_{\max}\Delta t\}
$$
Hence, $k$ does not exceed the number of particles in the region
$F^{-t}_0(S')$ at time zero. Our preliminary requirements imply
\be
         k\leq 2v_{\max}\Delta t \, L^2\ln L
           \label{kbound}
\ee
This allows us to bound the quantities $\chi^{(1)}$ and
$\chi^{(2)}$, for large $L$, by
\be
         |\chi^{(u)}|\leq v_{\max}\varepsilon^2 k^2\leq 4a^2v_{\max}^3(\ln L)^2\,(\Delta t)^2
            \label{chi12}
\ee
for $u=1,2$. It also follows from (\ref{main1}) that
\be
        \delta V \leq 5av_{\max}\Delta t \, \ln L
        \label{deltaV1}
\ee
We denote by $k^{\pm}_r$ the number of particles in the regions
$D^{\pm}_r\cap G_0(t)$ for $r=1,2$ at time $t=t_i$. We also denote
by $k^-$ the number of particles colliding with the piston ``on
the left'', and by $k^+$ that number ``on the right'' (of course,
$k^-+k^+=k$). Clearly,
\be
              k_1^{\pm}\leq k^{\pm}\leq k_2^{\pm}
                \label{kkk}
\ee

Since $D_1\subset G_0(t)$, then  $k^{\pm}_1$ equals the number of
particles in the region $F^{-t}_0(D^{\pm}_1)$ at time 0.
Similarly,  $k^{\pm}_2$ equals the number of particles in the
region $F^{-t}_0(D^{\pm}_2\cap G_0(t))$ at time 0.

The trapezoids $D^{\pm}_r$, $r=1,2$, can be well approximated by
some trapezoids $D^{\pm}_{j,p,q}$ defined earlier in our main
requirements. In fact, the horizontal sides $v=\pm v_{\min}$ and
$v=\pm v_{\max}$ are common for all trapezoids, only the vertical
side $x=X(t)$ and the oblique side need approximation. The
symmetric difference between $D^{\pm}_r$ and the approximating
$D^{\pm}_{j,p,q}$ will lie inside some strips $S_j$ and
$U^{\pm}_{j,p,q}$ also defined above. Then our main requirements
will guarantee that
$$
      k_1^{\pm} \geq \lambda_1^{\pm} - \ln L\, \sqrt{\lambda_1^{\pm}}
$$
and
$$
      k_2^{\pm} \leq \lambda_2^{\pm} + \ln L\, \sqrt{\lambda_2^{\pm}}
$$
where
$$
         \lambda^{\pm}_1=L^2 \int_{F_0^{-t}(D^{\pm}_1)}p(x,v,0)\, dx\, dv
          =L^2 \int_{D^{\pm}_1}p(x,v,t)\, dx\, dv
$$
and
\begin{eqnarray*}
         \lambda^{\pm}_2 & = & L^2 \int_{F_0^{-t}(D^{\pm}_2\cap G_0(t))}p(x,v,0)\, dx\, dv \\
          & \leq &  \lambda^{\pm}_1 + L^2\pi_{\max}|D^{\pm}_2\setminus D^{\pm}_1|
\end{eqnarray*}
where we used the boundedness of the density (\ref{pmax0}). The
area of the domain $D^{\pm}_2\setminus D^{\pm}_1$ is bounded by
$$
         |D^{\pm}_2\setminus D^{\pm}_1| < 4v_{\max}\, \delta V\,\Delta t
$$
Also, (\ref{pmax0}) implies that
$$
      \lambda^{\pm}_1\leq L^2\pi_{\max}|D^{\pm}_1|\leq
      2\pi_{\max}v_{\max}^2 L^2\, \Delta t
$$
Combining the above estimates gives
\be
      |k^{\pm}-\lambda_1^{\pm}|\leq
      {\rm const}\cdot (L \ln L\, \sqrt{\Delta t}+L^2\, \delta V\, \Delta t)
          \label{klambda}
\ee

We now turn to the quantity
$$
         Z=\sum_{j=1}^k v_j
$$
also involved in the main equation (\ref{main1}). Again we can
decompose $Z=Z^-+Z^+$, where $Z^-$ and $Z^+$ denote the sum of
velocities of the particles colliding with the piston ``on the
left'' and ``on the right'', respectively. We put
$$
        Z^{\pm}_1=\sum_{(x,v)\in D^{\pm}_1}v
$$
(taken at time $t$). Analysis similar to the previous one and the
requirement (\ref{Wijpq}) with formulas
(\ref{EZijpq})--(\ref{VarZijpq}) give
\be
        |Z^{\pm}-E(Z_1^{\pm})|\leq {\rm const}\cdot
        (L \ln L\, \sqrt{\Delta t}+L^2\, \delta V\, \Delta t)
          \label{Wlambda}
\ee

We now combine (\ref{main1}) with all the subsequent estimates and
obtain
\be
    V(t+\Delta t)-V(t)=-\varepsilon \lambda_1V(t) + \varepsilon E(Z_1)+\chi^{(3)}
          \label{main2}
\ee
where
$$
       \lambda_1=\lambda_1^++\lambda_1^-=L^2\int_{D_1}p(x,v,t)\,dx\,dv
$$
$$
       E(Z_1)=E(Z_1^+)+E(Z_1^-)=L^2\int_{D_1}v\, p(x,v,t)\,dx\,dv
$$
and
$$
    |\chi^{(3)}|\leq {\rm const}\cdot \Big [ L^{-1} \ln L\, \sqrt{\Delta t}
    +\delta V\, \Delta t+(\ln L)^2\,(\Delta t)^2 \Big ]
$$

Using Lemma~\ref{lmpx} (as we already did in deriving (\ref{Qx}))
to bound possible fluctuations of the density $p(x,v,t)$ within
$D_1$ gives
\be
       \int_{D_1}p(x,v,t)\,dx\,dv=
       (Q_1(t)-Q_0(t))\, V(t)\,\Delta t + \chi^{(4)}
          \label{intQ10}
\ee
and
\be
       \int_{D_1}v\, p(x,v,t)\,dx\,dv=
       (Q_2(t)-Q_1(t))\, V(t)\,\Delta t + \chi^{(5)}
          \label{intQ21}
\ee
with
\be
        |\chi^{(u)}| \leq {\rm const}\cdot \Big [ (\Delta t)^2/L
        +\delta V\, \Delta t\Big ]
          \label{chiuQ}
\ee
for $u=4,5$.

Therefore, we get
\be
        V(t+\Delta t)-V(t)={\cal D}(t)\, \Delta t + \chi
           \label{VVlast}
\ee
where
\be
     |\chi|\leq {\rm const}\cdot \left [
     L^{-1} \ln L\, \sqrt{\Delta t}+\delta V\, \Delta t+
     (\ln L)^2\,(\Delta t)^2\right ]
       \label{chilast}
\ee
By using (\ref{Deltat}) and (\ref{deltaV1}) it is easy to
incorporate the second and the third terms in (\ref{chilast}) into
the first one. Theorem~\ref{tmdV1} is proved. $\Box$ \medskip

For the next theorem, we rewrite (P5) in terms of the microscopic
coordinates:
\be
     |p(x,v,0)-p(L-x,-v,0)|<\varepsilon_0
        \label{eps0}
\ee
We also assume also that the initial velocity of the piston is set
to $V(0)=W(0)$ rather than to $0$, see a remark
Section~\ref{secHE} and another remark below.

\begin{theorem}
Assume that $\varepsilon_0>0$ is small enough. For all
sufficiently large $L$, for each configuration
$\omega\in\Omega^{\ast}_{0}$ and for all $t\in (0,S_1)$ we have \\
{\rm (i)} there is a constant $B>0$ such that
\be
                 |V(t)|<B\varepsilon_0
            \label{VtB0}
\ee
{\rm (ii)} there is a constant $C_0>0$ such that
\be
     |V(t)-V_0(t)|<\frac{C_0(\ln L)^{3/2}}{L^{2/3}}
        \label{VV0}
\ee
where $V_0(t)$ is defined by
\be
    V_0(t)=\frac{Q_1(t)-\sqrt{Q_1^2(t)-Q_0(t)Q_2(t)}}{Q_0(t)}
        \label{V0}
\ee
whenever $Q_0(t)\neq 0$ and by
\be
    V_0(t)=\frac{Q_2(t)}{2Q_1(t)}
        \label{V00}
\ee
otherwise. \label{tmdV2}
\end{theorem}

\noindent{\em Proof}. We will start proving (i) and obtain (ii) as
a ``side result''.

Assume that (i) is false and let $t_{\ast}<S_1$ be the first time
(\ref{VtB0}) fails, i.e. let
\be
          |V(t_{\ast}+0)| \geq B\varepsilon_0
            \label{tast}
\ee
Since (\ref{VtB0}) holds for $t<t_{\ast}$, the piston's position
satisfies
\be
    |X(t)-L/2|\leq B\varepsilon_0 t
         \label{XL2eps0}
\ee
for all $t<t_{\ast}$. Assume for the moment that the piston did
not move at all, i.e. $X(s)=L/2$ for all $0<s<t$. Then, by
(\ref{eps0}), the density $q(v,t)$ of the gas ``on the piston''
would be almost symmetric, i.e.
$$
         |q(v,t)-q(-v,t)|<\varepsilon_0
$$
Hence, we would have
\be
     |Q_i(t)|<C_1\varepsilon_0,\ \ \ \ \ \ \ i=0,2
       \label{Q0Q2ini}
\ee
with $C_1=v_{\max}^3$. We emphasize that $C_1$ does not depend on
the choice of $B$ in (\ref{VtB0}). Below we will introduce some
more constants $C_i$, $i\geq 2$, so that none of them will depend
on $B$.

When the piston actually moves and covers the distance $X(t)-L/2$,
then (\ref{Qx}) and (\ref{Q0Q2ini}) imply
\be
          |Q_i(t)| < C_1\varepsilon_0+C_2L^{-1}|X(t)-L/2|,
             \ \ \ \ \ i=0,2
               \label{Q0Q2aaa}
\ee
with some constant $C_2>0$. At the same time, $Q_1(t)$ stays
bounded above and below by positive constants for all
$t<t_{\ast}$:
\be
        0<Q_{1,\min}\leq Q_1(t)\leq Q_{1,\max}<\infty
          \label{Q1Qmin}
\ee
where $Q_{1,\min}$ and $Q_{1,\max}$ are constants determined by
$\pi_{\min}$ and $\pi_{\max}$ in (P3).

By (\ref{XL2eps0}), (\ref{Q0Q2aaa}) and (\ref{T0max}) we have
\be
          |Q_i(t)|< (C_1+Bv_{\max}^{-1})\,\varepsilon_0,
          \ \ \ \ \ \ i=0,2
             \label{Qi0}
\ee
for all $t<t_{\ast}$. This and (\ref{Q1Qmin}) imply that the
quadratic polynomial (\ref{calD}) has two real roots, and the
smaller one (the one closer to zero) is given by
(\ref{V0})--(\ref{V00}), cf. (\ref{Wroot}) and (\ref{Wroot0}) in
Section~\ref{secHE}. Due to (\ref{Q0Q2}) and (\ref{Q1Qmin}) we
have
\be
    |V_0(t)| < C_3\varepsilon_0 + C_4L^{-1}|X(t)-L/2|
      \label{C3C4}
\ee
for all $t<t_{\ast}$ and some constants $C_3,C_4>0$ (independent
of the choice of $B$).

We note that (\ref{C3C4}) and (\ref{T0max}) imply the boundedness
of $V_0(t)$
\be
      |V_0(t)|\leq \,{\rm const}\, =
      C_3\varepsilon_0+C_4v_{\max}^{-1}
        \label{V0tmax}
\ee
for all $t<t_{\ast}$.

Next, we need to estimate the derivative $dV_0(t)/dt$. (Note: the
function $V_0(t)$ defined by (\ref{V0})--(\ref{V00}) is continuous
and piecewise differentiable, it should not be confused with the
piston velocity $V(t)$, which is piecewise constant and hence not
even continuous). Due to (\ref{Qx})
\be
     \left |\frac{dQ_i(t)}{dt}\right |
     =\left |\frac{dQ_i(t)}{dX}\cdot V(t)\right |\leq
     \frac{{\rm const}\cdot \varepsilon_0}{L}
        \label{Qt}
\ee
for $i=0,1,2$. Differentiating the quadratic equation
$$
   Q_0(t)V_0^2(t)-2Q_1(t)V_0(t)+Q_2(t)=0
$$
with respect to $t$ gives
\be
     \frac{dV_0(t)}{dt}=
     \frac{(dQ_0/dt)V_0^2-2(dQ_1/dt)V_0+(dQ_2/dt)}
     {2(Q_1-Q_0V_0)}
\ee
Due to (\ref{Q0Q2}), (\ref{Q1Qmin}), (\ref{V0tmax}) and (\ref{Qt})
we have
\be
     \left |\frac{dV_0(t)}{dt}\right |\leq
     \frac{E_0\,\varepsilon_0}{L}
       \label{dV0t}
\ee
where $E_0>0$ is a constant.

Now consider the quantity $\cal D$ defined by (\ref{calD}) as a
function of $V$ (with fixed $Q_i$, $i=0,1,2$). Its derivative is
$$
     \frac{\partial{\cal D}}{\partial V}=
     2a[Q_0V-Q_1]
$$
Due to (\ref{Q1Qmin}) and (\ref{Qi0}) there are positive constants
$0<E_1<E_2$ such that for all $t<t_{\ast}$ we have
$$
     -E_2<\frac{\partial{\cal D}}{\partial V}<-E_1
$$
therefore, by the mean value theorem, for all $t<t_{\ast}$
\be
       -E_2<\frac{{\cal D}(t)}{V(t)-V_0(t)}<-E_1
          \label{D1D2}
\ee

We now prove (\ref{VV0}) for all $t<t_{\ast}$ with some constant
$C_0>0$ independent of the choice of $B$ in (\ref{VtB0}). Recall
that we have set the initial velocity of the piston to $V(0)=W(0)$
and that $W(0)=V_0(0)$, see Section~\ref{secHE}. Now, by way of
contradiction, let $t\in (0,t_{\ast})$ be the first time
(\ref{VV0}) fails. Denote by
$$
     \Delta_0=\frac{1}{L^{2/3}\ln L}
$$
the maximal allowed time increment in Theorem~\ref{tmdV1}. Let
$s=t-\Delta_0$. Due to Theorem~\ref{tmdV1}
\be
     V(t)=V(s)+{\cal D}(s)\Delta_0+\chi
       \label{VsD0}
\ee
with
$$
     |\chi|\leq \frac{C\sqrt{\Delta_0}\ln L}{L}
     =\frac{C\sqrt{\ln L}}{L^{4/3}}
$$
Due to (\ref{dV0t}) we have
\be
      V_0(t)=V_0(s)+\chi_0
        \label{V0sD0}
\ee
with
$$
    |\chi_0|\leq\frac{E_0\,\varepsilon_0\,\Delta_0}{L}
    =\frac{E_0\,\varepsilon_0}{L^{5/3}\ln L}
$$
For brevity, put $U(s)=V(s)-V_0(s)$ for all $s$. Subtracting
(\ref{V0sD0}) from (\ref{VsD0}) then gives
\be
    U(t)=U(s)+{\cal D}(s)\Delta_0+\chi'
       \label{UtUs}
\ee
with $\chi'=\chi-\chi_0$, so that for large $L$
\be
     |\chi'|\leq \frac{2C\sqrt{\ln L}}{L^{4/3}}
         \label{chiprime}
\ee
Now assume, without loss of generality, that $U(t)>0$. Since
(\ref{VV0}) fails at time $t$, we have
\be
     U(t)\geq \frac{C_0(\ln L)^{3/2}}{L^{2/3}}
        \label{Ut}
\ee
Now consider two cases. If $U(s)\leq 0$, then by (\ref{D1D2})
$$
    U(t)\leq |{\cal D}(s)|\Delta_0+|\chi'|\leq E_2\,
    |U(s)|\Delta_0+|\chi'|\ll \frac{(\ln L)^{3/2}}{L^{2/3}}
$$
for large $L$, which contradicts to (\ref{Ut}). If $U(s)>0$, then,
again due to (\ref{UtUs}) and (\ref{D1D2}),
$$
    U(t)<U(s)[1-E_1\Delta_0]+\chi',
$$
hence
\begin{eqnarray}
      U(s) & > & \frac{U(t)-\chi'}{1-E_1\Delta_0}
      > (U(t)-\chi')(1+E_1\Delta_0) \nonumber\\
      & > &
      U(t)+U(t)E_1\Delta_0-2\chi'
         \label{Ut>Us}
\end{eqnarray}
Now, if $C_0$ in (\ref{VV0}) is large enough, say $C_0=5C/E_1$,
then $U(t)E_1\Delta_0>2\chi'$ by (\ref{Ut}) and (\ref{chiprime}).
This fact and (\ref{Ut>Us}) imply $U(s)>U(t)$, so (\ref{VV0})
fails at an earlier time $s<t$, a contradiction. Hence,
(\ref{VV0}) is proved for all $t<t_{\ast}$ and $C_0=5C/E_1$.

Now, combining (\ref{C3C4}) and (\ref{VV0}) gives, for large $L$
and all $t<t_{\ast}$
$$
    | dX(t)/dt | < 2C_3\varepsilon_0+C_4L^{-1}|X(t)-L/2|
$$
Using the standard Gronwall inequality in differential equations,
see, e.g., Lemma~2.1 in \cite{TVS}, gives
$$
     |X(t)-L/2| < 2\varepsilon_0C_3C_4^{-1}L(e^{C_4L^{-1}t}-1)
$$
and
$$
      |V(t)| < 2\varepsilon_0C_3e^{C_4L^{-1}t}
$$
for all $t<t_{\ast}$. By (\ref{T0max}) we have
\be
      |V(t)| < 2\varepsilon_0C_3e^{C_4v_{\max}^{-1}}
        \label{Vtless}
\ee

Now we choose $B=3C_3e^{C_4v_{\max}^{-1}}$. Clearly,
(\ref{Vtless}) then contradicts (\ref{tast}). This completes the
proof of (\ref{VtB0}) for all $t<S_1$. Theorem~\ref{tmdV2} is now
proved. $\Box$\medskip

\begin{corollary}
Assume that $\varepsilon_0>0$ in (\ref{0symmetry}) is small
enough. Then, for all large $L$ and all $\omega
\in\Omega_0^{\ast}$, we have $S_1=T_1$, i.e. the previous theorems
hold on the entire zero-recollision interval $(0,T_1)$.
\label{crS0T0}
\end{corollary}

\noindent{\em Proof}. Recall that the ``slow'' interval
$(0,S_1)\subset (0,T_1)$ is defined by two conditions, (a) and
(b). If $S_1<T_1$, then either (a) or (b) fails at $S_1$. Clearly,
(a) cannot fail ``abruptly'' since (\ref{VtB0}) holds for all
$t<S_1$.

Suppose (b) fails at some $s<T_1$, while (a) still holds. The
failure of (b) means that at time $s$ the piston ``collides'' with
a point $(x,v)$ such that $v_{\min}<|v|<v_{\max}$ and $(x,v)\notin
G_0(s)$. Therefore, the backward trajectory $F^{-(s-t)}(x,v)$,
$0<t<s$, of the point $(x,y)$ hits the piston at some time $t>0$.
Now, during the time interval $(0,t)$ the piston covers the
distance $|X(t)-L/2|\leq B\varepsilon_0t$, and during the time
interval $(t,s)$ the trajectory of our point covers the distance
$|v|(s-t)<v_{\max}(s-t)$. Hence we have
\begin{eqnarray*}
     L & \leq & |X(t)-L/2|+|v|(s-t)+|X(s)-L/2| \\
     & \leq & |X(s)-L/2|+
     v_{\max}s-(v_{\max}-B\varepsilon_0)t
\end{eqnarray*} On the other hand, since $s<T_1$, we have
$$
         L>|X(s)-L/2|+v_{\max}s
$$
see the proof of Lemma~\ref{lmT0max}. This contradiction shows
that (b) cannot fail either. The proof of Corollary~\ref{crS0T0}
is completed. $\Box$ \medskip

\noindent{\bf Remark}. We have reset the initial velocity of the
piston to $V(0)=W(0)$ here, while in Section~\ref{secI} it was set
to zero. If $V(0)=0$, then Theorem~\ref{tmdV1} would imply that
$V(t)$ converges to $V_0(t)$ exponentially fast in $t$, until it
gets $\delta$-close to $V_0(t)$ with $\delta=C_0(\ln
L)^{3/2}L^{-2/3}$. After that all our results will apply without
change. The initial interval on which the convergence takes place
will be of order $\ln L$, and in the hydrodynamical time it is
$L^{-1}\ln L$, which vanishes as $L\to\infty$. This is why we
simply opted for the most convenient setting $V(0)=W(0)$ here.
\medskip

The following theorem improves the results of Theorems~\ref{tmdV1}
and \ref{tmdV2}.

\begin{theorem}
Assume that $\varepsilon_0>0$ in (\ref{0symmetry}) is small
enough. Then there is a constant $C>0$ such that for all large
$L$, all $\omega\in\Omega_0^{\ast}$ and all $t<T_1$
\be
       |V(t)-V_0(t)| < \frac{C\ln L}{L}
         \label{VtV0L}
\ee
and for any time interval $(t,t+\Delta t)\subset (0,T_1)$ such
that $L^{-2}<\Delta t \leq 1$ we have
\be
     |V(t+\Delta t)-V(t)| < C\, \frac{\ln L\, \sqrt{\Delta t}}{L}
        \label{VDVL}
\ee
\label{tmdV3}
\end{theorem}

\noindent{\em Proof}. Due to our choice of the initial velocity,
$V(0)=V_0(0)$, hence (\ref{VtV0L}) holds for at least small $t$.
Assume that it fails at some $t_{\ast}<T_1$, and $t_{\ast}$ is the
earliest time of failure. Without loss of generality, assume
\be
      V(t_{\ast})-V_0(t_{\ast}) \geq \frac{C\ln L}{L}
         \label{VVtast}
\ee
Let $0<t_0<t_{\ast}$ be the latest time when
\be
       V(t_0)-V_0(t_0) \leq \frac{C\ln L}{2L}
          \label{VVt0}
\ee
Then we have
\be
       V(t_0)-V_0(t_0) \leq V(t)-V_0(t) \leq V(t_{\ast})-V_0(t_{\ast})
         \label{3VV}
\ee
for all $t\in (t_0,t_{\ast})$. Let
\be
         \Delta t=\min\{1,t_{\ast}-t_0\}
        \label{Deltat1}
\ee
We will analyze the dynamics of the piston during the time
interval $(t_0,t_0+\Delta t)$. Due to (\ref{dV0t}) we have
\be
     |V_0(t)-V_0(t_0)| <  \delta V:=E_0\,\varepsilon_0L^{-1}\Delta t
        \label{deltaVE0}
\ee
for all $t\in(t_0,t_0+\Delta t)$, hence (\ref{3VV}) implies
\be
     V(t)>V(t_0)-\delta V
       \label{3VVa}
\ee
We note that $\Delta t$ is not too small, it is at least $\Delta t
> (L^{2/3}\ln L)^{-1}$. Indeed, otherwise we would have
$t_{\ast}=t_0+\Delta t$ and then (\ref{VVtast}), (\ref{VVt0}) and
(\ref{deltaVE0}) would imply
$$
    V(t_{\ast})-V(t_0) \geq 2^{-1}CL^{-1}\ln L-E_0\,\varepsilon_0L^{-5/3}
$$
which would contradict Theorem~\ref{tmdV1}, since ${\cal
D}(t_0)<0$ (because $V(t_0)>V_0(t_0)$).

Next, we develop a generalized version of the velocity
decomposition (\ref{main1}) in the proof of Theorem~\ref{tmdV1}.
We partition the interval $(t_0,t_0+\Delta t)$ into subintervals
of length $\delta$ (to be chosen shortly) with endpoints
$t_i=t_0+i\delta$, $i=0,1,\ldots, I$, where $I=\Delta t/\delta$.
We select $\delta$ so that
\be
         \frac{0.5}{L\ln L}<\delta<\frac{1}{L\ln L}
        \label{deltachoose}
\ee
and $\Delta t/\delta$ is an integer (for convenience). Preliminary
requirements in the proof of  Theorem~\ref{tmdV1} allow us to
adjust time so that $L^2t_0$ and $L^2\delta$ are integers, hence
$L^2t_i$ will be an integer for every $i$. The velocity
decomposition in the proof of Theorem~\ref{tmdV1} now applies to
each subinterval $(t_i,t_{i+1})$ of length $\delta$. In
particular, (\ref{main1}) implies
\be
    V(t_{i+1})-V(t_i)=-\varepsilon k_iV(t_i)
    +\varepsilon\sum_{j=1}^{k_i}v_j+\chi_i^{(1)}
      \label{VVi1}
\ee
where $k_i$ is the number of particles colliding with the piston
during the time interval $(t_i,t_{i+1})$ and $v_j$, $1\leq j\leq
k_i$, are their velocities. The fluctuation term $\chi_i^{(1)}$
can be bounded by (\ref{chi12}):
\be
      |\chi_i^{(1)}| \leq 8a^2v_{\max}^3(\ln L)^2\delta^2
         \label{chii1}
\ee

Due to (\ref{3VVa}), the expansion (\ref{VVi1}) can be rewritten
as
\be
    V(t_{i+1})-V(t_i) \leq -\varepsilon k_iV(t_0)
    +\varepsilon\sum_{j=1}^{k_i}v_j+\chi_i^{(1)}+\chi_i^{(2)}
      \label{VVi2}
\ee
with
$$
      |\chi_i^{(2)}| \leq
      \varepsilon k_i\, \delta V \leq
      2av_{\max}E_0\,\varepsilon_0L^{-1}\ln L\, \Delta t\,\delta
$$
where in the last step we used (\ref{kbound}) and (\ref{deltaVE0}).
Summing (\ref{VVi2}) up over $i$
yields
\be
    V(t_0+\Delta t)-V(t_0) \leq -\varepsilon k V(t_0)
    +\varepsilon\sum_{j=1}^{k}v_j+\chi^{(3)}
      \label{VVtot1}
\ee
where $k$ is the number of particles colliding with the piston
during the time interval $(t_0,t_0+\Delta t)$ and $v_j$, $1\leq
j\leq k$, are their velocities, and we have
\begin{eqnarray}
      |\chi^{(3)}| & \leq &
      8a^2v_{\max}^3(\ln L)^2\delta \Delta t +
      2av_{\max}E_0\,\varepsilon_0L^{-1}\ln L\, (\Delta t)^2\nonumber\\
       & \leq &
     (8a^2v_{\max}^3 + 2av_{\max}E_0\,\varepsilon_0)\, L^{-1}\ln L\, \Delta t
        \label{chi3a}
\end{eqnarray}
(where we used (\ref{deltaVE0}) and the assumption $\Delta t \leq 1$).

The expansion (\ref{VVtot1}) can be analyzed similarly to (\ref{main1}) in the
proof of Theorem~\ref{tmdV1}. Define a region on the $x,v$
plane:
\be
      D_1=\left \{(x,v):\  \frac{v-V(t_0)+\delta V}{x-X(t_0)}
         < -\frac{1}{\Delta t},\ \ v_{\min}<|v|<v_{\max} \right \}
               \label{D1a}
\ee
It is the union of two trapezoids $D_1=D_1^+\cup D_1^-$, where $D_1^-$
denotes the upper and $D_1^+$ the lower one, see Fig.~4. The bound
(\ref{3VVa}) implies that all the particles in the region $D_1^+$
necessarily collide with the piston during the time interval
$(t_0,t_{0}+\Delta t)$ and all the particles actually colliding with
the piston on its left hand side during this interval of time are
contained in $D_1^-$.

Since $v>V(t_0)$ for all particles $(x,v)\in D_1^-$ and $v<V(t_0)$
for all $(x,v)\in D_1^+$, we can remove from (\ref{VVtot1}) the
particles that do not belong in $D_1^+$ and simultaneously
add to (\ref{VVtot1})
the particles that belong in $D_1^-$ but do not collide with the
piston. This modification only makes the right hand side of
(\ref{VVtot1}) larger, hence
\be
    V(t_0+\Delta t)-V(t_0) \leq -\varepsilon k_1 V(t_0)
    +\varepsilon\sum_{j=1}^{k_1}v_j+\chi^{(3)}
      \label{VVtot2}
\ee
where $k_1$ is the number of particles in $D_1$ at time
$t_0$, and the summation runs over all those
particles. Let $Z_1=\sum_{(x,v)\in D_1}v$. Just like in
the proof of Theorem~\ref{tmdV1}, our main requirements stated there
guarantee that
$$
             |k_1-E(k_1)| \leq c_3 L\ln L\, \sqrt{\Delta t}
$$
and
$$
             |Z_1-E(Z_1)| \leq c_4 L\ln L\, \sqrt{\Delta t}
$$
where the constant $c_3,c_4>0$ do not depend on the choice of $C$
in (\ref{VtV0L}), which we have not made yet. Now, computing the
mean values of $k_1$ and $Z_1$ as in the proof of
Theorem~\ref{tmdV1} we arrive at
\be
    V(t_0+\Delta t)-V(t_0) \leq {\cal D}(t_0)\,\Delta t
    +\chi^{(3)}+\chi^{(4)}
      \label{VVtot3}
\ee
with
\be
    |\chi^{(4)}|\leq a(c_3+c_4)L^{-1}\ln L\, \sqrt{\Delta t} +
    c_5\,\delta V\, \Delta t
      \label{chi4a}
\ee
where $c_5>0$ is a constant independent of the choice of $C$ in
(\ref{VtV0L}). The last term in (\ref{chi4a}) comes from the
adjustment $\delta V$ to the velocity $V(t_0)$ in (\ref{D1a}).
This last term is bounded by $c_5E_0\,\varepsilon_0L^{-1}(\Delta
t)^2$, and since $\Delta t\leq 1$, it can be incorporated into the
first term in (\ref{chi4a}). Recall that $V(t_0)-V_0(t_0)\approx
2^{-1}CL^{-1}\ln L>0$ (here we have an approximation up to a
quantity of order $1/L^2$, since the piston velocity changes by
$O(1/L^2)$ at each collision). Then due to (\ref{D1D2}) we have
$$
     {\cal D}(t_0)\leq -E_1(V(t_0)-V_0(t_0))\approx -2^{-1}CE_1L^{-1}\ln L
$$
Therefore, combining the above estimates gives
\be
    V(t_0+\Delta t)-V(t_0) \leq -2^{-1}CE_1L^{-1}\ln L\, \Delta t+\chi^{(5)}
      \label{VVtot4}
\ee
with $\chi^{(5)}=\chi^{(3)}+\chi^{(4)}$ bounded by (\ref{chi3a}) and
(\ref{chi4a}):
$$
    |\chi^{(5)}|\leq c_6 (L^{-1}\ln L\, \Delta t+L^{-1}\ln L\, \sqrt{\Delta t})
$$
where $c_6>0$ is a constant independent of the choice of $C$ in
(\ref{VtV0L}). Now we chose the constant $C$ there as
\be
         C=\max\{c_6,6E_1^{-1}c_6\}
        \label{Cchoose}
\ee
Then (\ref{VVtot4}) implies
$$
    V(t_0+\Delta t)-V(t_0) \leq c_6 L^{-1}\ln L \,
     (-2\Delta t+\sqrt{\Delta t})
$$
and hence, due to (\ref{deltaVE0}),
\begin{eqnarray}
    V(t_0+\Delta t)-V_0(t_0+\Delta t) & \leq & V(t_0) - V_0(t_0) \nonumber\\
    & & + c_6 L^{-1}\ln L \, (-2\Delta t+\sqrt{\Delta t})+E_0\,\varepsilon_0L^{-1}\Delta t
      \label{VVtot5}
\end{eqnarray}
We now have two cases. First, let $t_{\ast}-t_0\leq 1$, hence
$\Delta t \leq 1$. The expression $-2\Delta t+\sqrt{\Delta t}$ has
a maximum, equal to $1/8$, at the point $\Delta t=1/16$.
Therefore, (\ref{VVtot5}) implies
$$
        V(t_{\ast})-V_0(t_{\ast})<
    V(t_0)-V_0(t_0)+8^{-1}c_6 L^{-1}\ln L+E_0\,\varepsilon_0L^{-1}
$$
This contradicts (\ref{VVtast}) and (\ref{VVt0}) when $L$ is large
enough, recall our choice of $C$ in (\ref{Cchoose}). Consider the
second case: $t_{\ast}-t_0>1$. Then $\Delta t=1$ and
(\ref{VVtot5}) implies, for large $L$,
$$
    V(t_0+1)-V_0(t_0+1)<V(t_0)-V_0(t_0)-2^{-1}c_6L^{-1}\ln L
$$
which contradicts (\ref{3VV}). This completes the proof of
(\ref{VtV0L}).

We now prove (\ref{VDVL}). If $\Delta t<(L^{2/3}\ln L)^{-1}$, then
we can use Theorem~\ref{tmdV1}:
$$
     |V(t+\Delta t)-V(t)| \leq |{\cal D}(t)|\, \Delta t+CL^{-1}\ln L\, \sqrt{\Delta t}
$$
The early estimates (\ref{D1D2}) and (\ref{VtV0L}) imply
\be
         |{\cal D}(t)| \leq E_2|V(t)-V_0(t)| \leq CE_2L^{-1}\ln L
        \label{calDabs}
\ee
so that (\ref{VDVL}) follows (with some larger value of $C$ than
above).

Now let  $(L^{2/3}\ln L)^{-1} \leq \Delta t\leq 1$. Without loss of generality,
assume that $V(t+\Delta t)>V(t)$. Moreover, we can assume that
\be
      V(s)>V(t)\ \ \ \ \ \ {\rm for}\ {\rm all}\ s\in (t,t+\Delta t)
              \label{VsVt}
\ee
Indeed, if this is not the case, we can replace $t$
by $t'=\max\{s<t+\Delta t:\, V(s)\leq V(t)\}$ and prove (\ref{VDVL})
for the smaller interval $(t',t+\Delta t)$.

Next, our plan is to apply some estimates from the proof of
(\ref{VtV0L}) and then argue along the lines of the proof of
Theorem~\ref{tmdV1}. Denote $t_0=t$ and partition the interval
$(t_0,t_0+\Delta t)$ into subintervals of length $\delta$
satisfying (\ref{deltachoose}). Then we again have decomposition
(\ref{VVi1})--(\ref{chii1}). Due to (\ref{VsVt}) we have
$V(t_i)>V(t_0)$ for all $i$, hence (\ref{VVi1}) implies
$$
    V(t_{i+1})-V(t_i)<-\varepsilon k_iV(t_0)
    +\varepsilon\sum_{j=1}^{k_i}v_j+\chi_i^{(1)}
$$
Summing this up over $i$ gives (\ref{VVtot1}) with the bound
(\ref{chi3a}), in which the second term can be simply removed,
since we do not have $\chi_i^{(2)}$ anymore. Next, possible
fluctuations of the piston velocity $V(s)$ during the time
interval $(t_0,t_0+\Delta t)$ can be estimated with the help of
(\ref{VtV0L}) and (\ref{deltaVE0}):
$$
      |V(s)-V(t_0)| \leq \delta V :=
      2CL^{-1}\ln L + E_0\,\varepsilon_0L^{-1}\Delta t
$$
for all $s\in (t_0,t_0+\Delta t)$. Then we estimate the random
variables $k$ and $Z=\sum_j v_j$ along the lines of the proof of
Theorem~\ref{tmdV1}, starting with construction of two domains
$D_1$ and $D_2$ by (\ref{D1})--(\ref{D2}), etc. Repeating the
argument almost word by word we arrive at an analogue of
(\ref{VVlast}):
$$
        V(t+\Delta t)-V(t) < {\cal D}(t)\, \Delta t + \chi'
$$
where
$$
     |\chi'|\leq {\rm const}\cdot \left [
     L^{-1} \ln L\, \sqrt{\Delta t}+\delta V\, \Delta t+
     L^{-1}\ln L\,\Delta t \right ]
$$
where the last term comes from (\ref{chi3a}), which we have now,
instead of (\ref{chi12}) (note: (\ref{chi12}) would not be nearly
enough anymore, since $\Delta t$ is large; this is why we needed
to partition the interval $(t,t+\Delta t)$ into smaller
subintervals). We now combine the above estimates with
(\ref{calDabs}) and complete the proof of (\ref{VDVL}) and
Theorem~\ref{tmdV3}. $\Box$ \medskip

We finally prove the convergence, as $L\to\infty$, of the random
trajectory of the piston to the solution $Y(\tau),W(\tau)$ of the
hydrodynamical equations found in Section~\ref{secHE}.

\begin{theorem}
Assume that $\varepsilon_0>0$ in (\ref{0symmetry}) is small
enough. Then, for all large $L$ and all
$\omega\in\Omega_0^{\ast}$, there is a constant $C>0$ such that
\be
          |Y_L(\tau,\omega)-Y(\tau)|\leq \frac{C\,\ln L}{L}
          \label{YLY0}
\ee
and
\be
          |W_L(\tau,\omega)-W(\tau)|\leq \frac{C\,\ln L}{L}
          \label{WLW0}
\ee
for all $0<\tau < \min\{\tau_1,T_1/L\}$ and
\be
           |\tau_1-T_1/L|\leq \frac{C\,\ln L}{L}
          \label{tau0T0L}
\ee
\label{tmzerorec}
\end{theorem}

\noindent{\em Proof}. In Section~\ref{secHE} we defined the
function $F(Y,\tau)$ so that the hydrodynamical solution $Y(\tau)$
satisfies
\be
          dY(\tau)/d\tau=F(Y,\tau),\ \ \ \ \ \ \ \ Y(0)=1/2
          \label{Y'tau}
\ee
see (\ref{YW1}). Now Theorem~\ref{tmdV3} implies that for all
$\omega\in\Omega_0^{\ast}$ the random trajectory satisfies
\be
         \partial Y_L(\tau,\omega)/\partial \tau =
         F(Y,\tau)+\chi(\tau,\omega),
         \ \ \ \ \ \ Y_L(0,\omega)=1/2
          \label{Y'tauomega}
\ee
with some
$$
        |\chi(\tau,\omega)|\leq\frac{C\,\ln L}{L}
$$
Recall that $|\partial F(Y,\tau)/\partial Y|\leq \kappa$, see
(\ref{dFdY}). Therefore, the difference
$Z_L(\tau,\omega):=Y_L(\tau,\omega)-Y(\tau)$ satisfies
$$
          |Z_L'(\tau,\omega)|\leq \kappa|Z_L(\tau,\omega)|+
          \frac{C\,\ln L}{L}
$$
and $Z_L(0,\omega)=0$. By the standard Gronwall inequality in
differential equations, see, e.g., Lemma~2.1 in \cite{TVS}, we
have
$$
          |Z_L(\tau,\omega)|\leq
          \frac{C\,\ln L}{\kappa L}\,\Big (e^{\kappa\tau}-1\Big )
$$
and
$$
          |Z_L'(\tau,\omega)|\leq
          \frac{C\,\ln L}{L}\, e^{\kappa\tau}
$$
for all $\tau < \min\{\tau_1,T_1/L\}$, which imply (\ref{YLY0})
and (\ref{WLW0}).

Lastly, we verify (\ref{tau0T0L}). By (\ref{WLW0}), random
fluctuations of the piston velocity are bounded by $CL^{-1}\ln L$.
Hence, random fluctuations of the velocities of particles that
have had one collision with the piston are bounded by $2CL^{-1}\ln
L$. The random fluctuations of the positions of both the piston
and particles at every moment of time $t<\min\{\tau_1L,T_1\}$ are
bounded by the same quantities (with, possibly, a different value
of $C$) in the coordinate $y=x/L$. On the other hand, the relative
velocity of the piston and the particles stays bounded away from
zero (by, say, $v_{\min}-4B\varepsilon_0>0$). Hence the time of
the first recollision $T_1/L$ can differ from $\tau_1$ by at most
const$\cdot L^{-1}\ln L$. Theorem~\ref{tmzerorec} is proved.
$\Box$

\section{Dynamics between the first and second recollisions}
\label{secORI} \setcounter{equation}{0}

In this section we study the one-recollision interval
$(\tau_1,\tau_2)$, on which gas particles experience the second
collision (i.e., the first {\em re}collision) with the piston.

The particles that have collided with the piston no longer make a
Poisson process, hence their distribution is much harder to
control. This is our main trouble. On the other hand, we will be
satisfied with much weaker estimates than those in the previous
section. Also, many arguments and constructions in this section
are similar to those in Section~\ref{secZRI}, and we omit some
details. We will focus on new ideas.

Here our analysis is always restricted to the configurations
$\omega\in\Omega_0^{\ast}$. Later on we will put additional
requirements on $\omega$.

Recall that for $\omega\in\Omega_0^{\ast}$ the piston velocity is
small, $|V(t)|<B\varepsilon_0$, see (\ref{VtB0}), on the zero
recollision interval $(0,T_1)$. Hence, the velocities of gas
particles that experience one collision with the piston on the
interval $(0,T_1)$ are bounded
\be
       v_{1,\min}<|v|<v_{1,\max}
          \label{v1minmax}
\ee
with
\be
       v_{1,\min}:=v_{\min}-2B\varepsilon_0
       \ \ \ \ \ {\rm and}\ \ \ \ \
       v_{1,\max}:=v_{\max}+2B\varepsilon_0
\ee
The first time of the second recollision $T_2=T_2(\omega)$ is
defined by $T_2=\sup_{t>0}\{G_3^+=\emptyset\}$, see (\ref{Tn}).
Due to (\ref{VtB0}) and (\ref{v1minmax}) the following bound can
be easily obtained as in the proof of Lemma~\ref{lmT0max}:
\be
      T_2\leq \frac{L}{v_{\max}}
      +\frac{L}{v_{\max}-2B\varepsilon_0}\leq \frac{3L}{v_{\max}}
         \label{T1max}
\ee
Now let $(x,v)\in G^+$ and $(x_t,v_t)=F^t(x,v)$ for $t>0$. Denote
by
$$
       s_1(x,v)=\min\{t:\, x_t=X(t)\}
$$
the time of the first collision with the piston. The region
$$
    G_{\ast}^+(t):=\{(x_t,y_t)\in G^+_1(t):\ s_1(x,v)<T_1\}
$$
is occupied by points that by the time $t$ have experienced one
collision with the piston, which occurred before time $T_1$. By
removing the superscript $+$ in the above formula we define
$G_{\ast}(t)$. Let $T_{\ast}\leq T_2$ be the earliest time the
piston interacts with the particles
$$
    (x,v)\in [G^+_1(t)\setminus G^+_{\ast}(t)]\cup G_2^+(t)
$$
The time $T_{\ast}$ is a random analogue of $\tau_{\ast}$
introduced in Section~\ref{secHE}. During the interval
$(T_1,T_{\ast})$ the piston only interacts with the particles from
$G^+_0(t)\cup G^+_{\ast}(t)$, hence their velocities must be
bounded by (\ref{v1minmax}). Denote by
\begin{eqnarray}
     {\cal X}_1(t)=\{(x,v):\, x=X(t)+0,\, -v_{1,\max}<v<-v_{1,\min}\}
       \nonumber\\
       \cup \{(x,v):\, x=X(t)-0,\, v_{1,\min}<v<v_{1,\max}\}
        \label{calX1}
\end{eqnarray}
two immediate one-sided vicinities of the piston which contain all
``incoming'' particles for every $t\in (T_1,T_{\ast})$.

Again, as in the previous section, we define a subinterval
$(T_1,S_2)\subset (T_1,T_{\ast})$ on which the piston is slow
enough:

\medskip
\noindent{\bf Definition} Let $(T_1,S_2)\subset (T_1,T_{\ast})$
be the maximal time interval during which \\
(a) $|V(t)|<v_{1,\min}$;\\
(b) ${\cal X}_1(t)\subset G_0(t)\cup G_{\ast}(t)$.
\medskip

The condition (b) means that the particles with velocities
$v_{1,\min}<|v|<v_{1,\max}$ that are about to interact with the
piston at time $t$ have interacted with the piston during the
interval $(0,t)$ at most once, and if they did, the interaction
occurred before $T_1$.

Next we estimate how large the interval $(T_1,S_2)$ is. Suppose
\be
     |V(t)-W(t/L)|\leq \Delta
       \label{VWDelta}
\ee
for all $t<S_2$ and some small $\Delta$ (we will later estimate
$\Delta$ and show that $\Delta\to 0$ as $L\to\infty$). This
immediately implies $|V(t)|\leq \Delta+{\cal B}\varepsilon_0$,
according to (\ref{WBeps0}). Integrating (\ref{VWDelta}) with
respect to $t$ gives
$$
    |X(t)-LY(t/L)|\leq t\Delta
$$
and hence
\be
    |X(t)-L/2|\leq (\Delta+{\cal B}\varepsilon_0)t
      \label{XtL2D}
\ee

\begin{proposition}
If (\ref{VWDelta}) holds for $t<S_2$ with some small $\Delta>0$,
then $T_2-S_2\leq CL(\Delta+\varepsilon_0)$, where $C>0$ is a
constant. \label{prS1}
\end{proposition}

\noindent{\em Proof}. If $S_2=T_2$, then the statement is trivial.
If $S_2<T_2$, then either $S_2<T_{\ast}$, and so the condition (a)
or (b) in the previous definition fails at time $S_2$, or
$S_2=T_{\ast}<T_2$. Note that the condition (a) cannot fail
abruptly, since we assume $|V(t)|\leq \Delta+{\cal
B}\varepsilon_0$, on $(0,S_2)$, i.e. $V(t)$ remains small. If (b)
fails, then at time $S_2$ the piston collides with a point $(x,v)$
such that $v_{1,\min}\leq |v|\leq v_{1,\max}$ and the past
trajectory $(x_t,v_t):=F^{t-S_2}(x,v)$ of that point for $t\in
(0,S_2)$ hits the piston at some time $t_1\geq T_1$. If
$S_2=T_{\ast}<T_2$, then at time $S_2$ the piston recollides with
a gas particle $(x,v)$ whose past trajectory
$(x_t,v_t):=F^{t-S_2}(x,v)$ experiences the first collision with
the piston at some time $t_1\geq T_1$. In the last case, by
(\ref{VWDelta})
$$
   |v|\leq v_{\max}+2|V(t_1)|\leq
   v_{\max}+2(\Delta+{\cal B}\varepsilon_0)
$$
In either of the above two cases, the trajectory $(x_t,v_t)$
collides with the piston twice - once at time $t_1\geq T_1$ and
the second time at $t_2=S_2$. Denote by $X_1,X_2$ the positions of
the piston and by $V_1,V_2$ its velocities at times $t_1,t_2$,
respectively. Without loss of generality, assume that our
trajectory $(x_t,v_t)$ lies to the right of the piston. Note that
the speed $|v|=|v_t|$ for $t_1<t<t_2$ satisfies
\begin{eqnarray}
    |v| &\leq &
    \max\{v_{1,\max},v_{\max}+2
    (\Delta+{\cal B}\varepsilon_0)\}\nonumber\\
    & \leq & v_{\max}+2(\Delta+B_1\varepsilon_0)
    \label{ava}
\end{eqnarray}
with $B_1=\max\{B,{\cal B}\}$. Then we write an obvious identity
$$
   |v|(t_2-t_1)=(L-X_1)+(L-X_2)
$$
hence, by (\ref{XtL2D}) and (\ref{ava})
$$
    t_2-t_1 \geq \frac{L-(\Delta + {\cal B}\varepsilon_0)
    (t_1+t_2)}{v_{\max}+2(\Delta+B_1\varepsilon_0)}
$$
On the other hand, consider the particle that experiences the {\em
very first} recollision with the piston (this happens at time
$T_1$). After the collision, that particle acquires velocity
$|v(T_1+0)|\geq v_{\max}-2|V(T_1)|$. The next collision of this
particle with the piston occurs {\em after} $T_2$. Therefore,
\begin{eqnarray*}
    T_2-T_1 & \leq & \frac{L/2+|X(T_1)-L/2| + L/2+|X(T_2)-L/2|}
    {v_{\max}-2|V(T_1)|}\\
    & \leq & \frac{L+(\Delta + {\cal B}\varepsilon_0)
    (T_1+T_2)}{v_{\max}-2(\Delta + {\cal B}\varepsilon_0)}
\end{eqnarray*}
Combining the above estimates gives
$$
   T_2-S_2\leq (T_2-T_1)-(t_2-t_1)\leq
   CL(\Delta+\varepsilon_0)
$$
with some $C>0$ determined by $B$ and ${\cal B}$. $\Box$ \medskip

Next, we study the dynamics during the time interval $(T_1,S_2)$.
We again define the density of colliding particles ``on the
piston'' $q(v,t)$ by the equation (\ref{qp00}) and the functions
$Q_i(t)$, $i=0,1,2$, by (\ref{Q00})--(\ref{Q20}). We emphasize
that now, unlike what we had in the previous section, the density
$p(x,v,t)$ {\em essentially} depends on $\omega$ (at least for
$(x,v)\in G^+_{\ast}(t)$), hence $q(v,t)$ and $Q_i(t)$ will depend
on $\omega$ not only through the piston position $X(t)$ but also
through the surrounding density $p(x,v,t)$.

In the previous sections we also introduced the deterministic
density $\tilde{p}(x,v,t)$. So, now we can define the
corresponding deterministic density ``on the piston''
\be
    \tilde{q}(v,t)=\left\{\begin{array}{ll}
       \tilde{p}(X(t)+0,v,t)  &  {\rm if}\ \ v<0\\
       \tilde{p}(X(t)-0,v,t)  &  {\rm if}\ \ v>0\\
          \end{array}\right .
            \label{qpt00}
\ee
cf.\  (\ref{qp0}) and the deterministic functions
$\tilde{Q}_i(t)$, $i=0,1,2$ by the equations similar to
(\ref{Q00})--(\ref{Q20}) but with tildes over the corresponding
functions. We use tildes to distinguish these deterministic
functions from the random ones.

We now compare the random functions $p,q,Q_i$ with their
deterministic counterparts on the interval $(T_1,S_2)$. Since the
transformations $F^t:G^+\to G^+(t)$ and
$\tilde{F}^t:\tilde{G}^+\to \tilde{G}^+(t)$ (recall that
$\tilde{G}^+(t)=\tilde{F}^t(G^+)$) are invertible area-preserving
maps, then so is the map
$$
        \Psi^t=\tilde{F}^t\circ F^{-t}
$$
which takes $G^+(t)$ onto $\tilde{G}^+(t)$. The next lemma easily
follows from Theorem~\ref{tmzerorec}:

\begin{lemma}
Let $t<S_2$ and $\omega\in\Omega_0^{\ast}$. For every $(x,v)\in
G_{\ast}^+(t)$ put $(\tilde{x},\tilde{v}):=\Psi^t(x,v)$. Then
$$
   |x-\tilde{x}|\leq C\ln L
   \ \ \ \ \ \ {\rm and}\ \ \ \ \ \
   |v-\tilde{v}|\leq CL^{-1}\ln L
$$
with a constant $C>0$. We also have
$$
    p(x,v,t)=\tilde{p}(\tilde{x},\tilde{v},t)
$$
\label{lmPsi}
\end{lemma}

Next, the properties of the density $p(x,v,t)$ stated in
Lemma~\ref{lmpx} obviously hold for the deterministic density
$\tilde{p}(x,v,t)$ on the entire region $\tilde{G}^+$. That is,
the density $\tilde{p}(x,v,t)$ is piecewise $C^1$ smooth with
$\left |\partial \tilde{p}(x,v,t)/\partial x\right | \leq D_1' /L$
and the discontinuity lines of $\tilde{p}(x,v,t)$ have slope of
order $O(1/L)$ (so they are almost parallel to the $x$ axis).
Therefore, as in (\ref{Qx}), we have \be
    \left |\frac{\partial \tilde{Q}_i}{\partial X}\right |
      \leq \frac{\rm const}{L}
        \label{tildeQx}
\ee
Lemma~\ref{lmPsi} allows us to compare $Q_i$ and $\tilde{Q}_i$
considered as functions of $X$ in the following way:

\begin{lemma}
Let $t<S_2$ and $\omega\in\Omega_0^{\ast}$. Then for $i=0,1,2$
$$
     |Q_i(t)-\tilde{Q}_i(t)|\leq CL^{-1}\ln L
$$
\end{lemma}

Note that $\tilde{Q}_i(t)$ is defined through $\tilde{q}(v,t)$ which
uses the (random) position of the piston $X(t)$ for the given $\omega$,
see (\ref{qpt00}). The above lemma shows that the dependence of
$Q_i(t)$ on $\omega$ through the density $p(x,v,t)$ (which itself
depends on $\omega$) is very weak, because $L^{-1}\ln L$ is small. In
other words, the density $p(x,v,t)$ of the gas surrounding the piston
fluctuates with $\omega$ very little.

The following theorem is an analogue of Theorem~\ref{tmdV1}.

\begin{theorem}
For all sufficiently large $L$ there is a set
$\Omega_{1}^{\ast}\subset\Omega_0^{\ast}$ of initial
configurations of particles such that \\ {\rm (i)} there is a
constant $c>0$ such that
\be
         P(\Omega_{1}^{\ast})>1-L^{-c\,\ln\ln L}
            \label{POmega1ast}
\ee
{\rm (ii)} for each configuration $\omega\in\Omega^{\ast}_{1}$,
for each time interval
$$
       (t,t+\Delta t) \subset (T_1,S_2)
$$
such that
\be
             \frac{(\ln L)^2}{L^{1/3}}
             \leq \Delta t \leq
             \frac{1}{L^{1/7}}
                 \label{Deltat17}
\ee
the change of the velocity of  the piston satisfies
\be
        V(t+\Delta t)-V(t)=
        \tilde{\cal D}(t)\, \Delta t + \chi
           \label{VV17}
\ee
where
\be
       \tilde{\cal D}(t)=a
       [\tilde{Q}_0(t)V^2(t)-2\tilde{Q}_1(t)V(t)+\tilde{Q}_2(t)]
          \label{calD17}
\ee
and
\be
        |\chi| \leq C\,\frac{\ln L\, (\Delta t)^{1/4}}{L^{1/4}}
       \label{chibound17}
\ee with some global constant $C>0$. \label{tmdV17}
\end{theorem}

\noindent{\em Remark}. Note that our bound (\ref{chibound17}) on
random fluctuations represented by $\chi$ is much weaker than
(\ref{chibound}) in Theorem~\ref{tmdV1}. This is due to the lack
of a good control over large deviations for the distribution of
gas particles, as it will be clear from the proof.
\medskip

\noindent {\em Proof}. Our argument basically goes along the lines
of the proof of Theorem~\ref{tmdV1}. But it involves a good deal
of new constructions, which we describe in detail. The first step
is the velocity decomposition scheme, see (\ref{main1}),
\be
    V(t+\Delta t)=(1-\varepsilon k)V(t)
    + \varepsilon\sum_{j=1}^k v_j+\chi^{(1)}+\chi^{(2)}
          \label{main17}
\ee
The error terms $\chi^{(1)}$ and $\chi^{(2)}$ are defined in
Section~\ref{secZRI} after (\ref{main1}), and they are bounded by
$$
        |\chi^{(u)}|\leq v_{\max}\varepsilon^2k^2
$$
for $u=1,2$, see (\ref{chi12}). We will see later that
\be
         k\leq\,{\rm const}\cdot L^2\,\Delta t
           \label{kbound17}
\ee
hence
\be
    |\chi^{(u)}|\leq\, {\rm const}\cdot (\Delta t)^2
\ee
for $u=1,2$.

Next, we need a crude upper bound $\delta V$ on possible
fluctuations of the piston velocity $V(s)$ during the interval
$(t,t+\Delta t)$, as defined by (\ref{deltaV}). One can be easily
derived from (\ref{main17}):
\be
      \delta V \leq 2v_{\max}\varepsilon k\leq
      \,{\rm const}\cdot \Delta t
        \label{deltaV17}
\ee We now define two regions $D_1=D_1^+\cup D_1^-$ and
$D_2=D_2^+\cup D_2^-$ on the $x,v$ plane by equations
(\ref{D1})--(\ref{D2}), where $v_{\min}$ and $v_{\max}$ must be
replaced by $v_{1,\min}$ and $v_{1,\max}$, respectively. As it is
explained in Section~\ref{secZRI}, all the particles in $D_1$ (at
time $t$) will necessarily collide with the piston during the
interval $(t,t+\Delta t)$. And all the particle that actually
collide with the piston during that interval are contained in
$D_2$. Therefore, we have
\be
              k_1^{\pm}\leq k^{\pm}\leq k_2^{\pm}
                 \label{kkk1}
\ee
where $k^{\pm},k_1^{\pm},k_2^{\pm}$ are defined around
(\ref{kkk}). This gives upper and lower bounds on the number of
colliding particles.

Our next step is to estimate the numbers $k^{\pm}_i$, $i=1,2$.
Since $k^{\pm}_i$ no longer have Poisson distribution, we estimate
them by using a new approach. Let $D$ be one of the four
trapezoids $D_i^{\pm}$, $i=1,2$, and let $k_{D,\omega}$ be the
(random) number of particles in $D$ at time $t$. Obviously,
$k_{D,\omega}$ is equal to the number of particle in $F^{-t}(D)$
at time zero.

We note that the trapezoid $D$ has height
$v_{1,\max}-v_{1,\min}=\,$const and width $O(\Delta t)$, hence its
area is bounded by
\be
     d_1\, \Delta t \leq |D| \leq d_2\, \Delta t
        \label{areaD}
\ee
with some constants $0<d_1<d_2<\infty$.

Let us examine the region $F^{-t}(D)$ more closely. Since
$(t,t+\Delta t)\subset (T_1,S_2)$, it follows from the condition
(b) in the definition of $S_2$ that $D\subset G_0(t)\cup
G_{\ast}(t)$. Put $D_0=D\cap G_0(t)$ and $D_1=D\cap G_{\ast}(t)$,
then $F^{-t}(D)=F^{-t}(D_0)\cup F^{-t}(D_1)$. To emphasize the
dependence of the flow $F^t$ on $\omega$ we will write
$F^t_{\omega}$ for $F^t$. Now the part $F_{\omega}^{-t}(D_0)$ will
be obtained by a simple linear transformation of $D_0$ without
collisions with the piston, hence it will be actually independent
of $\omega$. The part $F_{\omega}^{-t}(D_1)$ is also obtained by
pulling the domain $D_1$ back in time, but one collision with the
piston will occur along the way. Since the position and velocity
of the piston are random, then the domain  $F^{-t}_{\omega}(D_1)$
will depend on $\omega$. Hence, the domain
$D_{\omega}:=F_{\omega}^{-t}(D)$ will depend on $\omega$. The
initial number of particles, $k_{D,\omega}$, in a randomly
selected domain $D_{\omega}$ certainly need not be a Poisson
random variable.

To estimate $k_{D,\omega}$ we fix some $\Delta t$ satisfying
(\ref{Deltat17}) and construct finitely many domains $D_n\subset
G$,  $1\leq n\leq N_{\ast}$, which have the following property.
For every $\omega\in\Omega_0^{\ast}$, every $t\in (T_1,S_2-\Delta
t)$, and every trapezoid $D$ defined above, there are two domains
$D_{n'}, D_{n''}$ such that \be
          D_{n'}\subset F_{\omega}^{-t}(D) \subset D_{n''}
             \label{DDD}
\ee
and the area of the difference is relatively small:
\be
        |D_{n''}\setminus D_{n'}| \leq \chi^{(3)}|D_{\omega}|
          \label{DDDarea}
\ee
with some $\chi^{(3)}\to 0$ as $L\to\infty$. We say that $D_{n'}$
and $D_{n''}$ approximate $D_{\omega}$ ``from inside'' and ``from
outside'', respectively. We denote the collection of the domains
$D_n$, for the given $\Delta t$, by
$$
         {\cal C}={\cal C}_{\Delta t}=\{D_n\}_{n=1}^{N_{\ast}}
$$
We postpone the construction of $D_n$'s for the moment and derive
immediate benefits of the above approximation. The inclusion
(\ref{DDD}) implies
\be
       k_{n',\omega}\leq k_{D,\omega}\leq k_{n'',\omega}
         \label{kkkD}
\ee
where $k_{n,\omega}$ is the (random) number of particles in the
domain $D_n$ at time zero. Since $D_n$, for each $n$, is fixed
(independent of $\omega$), the random variable $k_{n,\omega}$
does have a Poisson distribution with mean value
$$
    \lambda_n:=L^2\int_{D_n}p(x,v,0)\, dx\, dv
$$
According to Lemma~\ref{lmPoi2}, we have
\be
   P\left (\omega:\,
   |k_{n,\omega}-\lambda_{n}|>
   B\sqrt{\lambda_{n}}\right )
   \leq 2e^{-cB^2}
      \label{Pkon}
\ee
for any $B<b\sqrt{\lambda_{n}}$, where $b>0$ is a constant and
$c=c(b)>0$ another constant. We will specify the value of
$B=B_{\Delta t}$ (one for all $D_n$'s in ${\cal C}_{\Delta t}$)
later. Due to (\ref{areaD}), it is enough to require
\be
         B_{\Delta t}<L\sqrt{\Delta t}
             \label{Blambdan}
\ee
Then we define $\Omega_1^{\ast}(\Delta t)$ as the set of
configurations $\omega\in\Omega_0^{\ast}$ satisfying
\be
   |k_{n,\omega}-\lambda_n| \leq B_{\Delta t}\sqrt{\lambda_n}
   \ \ \ \ \ \ {\rm for}\ \ 1\leq n\leq N_{\ast}
      \label{kon}
\ee
Then by (\ref{Pkon}) we have
\be
     P(\Omega_0^{\ast}\setminus\Omega_1^{\ast}(\Delta t))
           \leq 2N_{\ast}e^{-cB_{\Delta t}^2}
             \label{POmega1}
\ee
Next, for all $\omega\in\Omega_1^{\ast}(\Delta t)$ the bounds
(\ref{kkkD}) and (\ref{kon}) imply
\be
      \lambda_{n'}-B_{\Delta t}\sqrt{\lambda_{n'}}
        \leq k_{D,\omega} \leq
      \lambda_{n''}+B_{\Delta t}\sqrt{\lambda_{n''}}
        \label{kln}
\ee
Furthermore, consider the quantity
$$
   \lambda_{D,\omega}=L^2\int_{F^{-t}_{\omega}(D)}p(x,v,0)\, dx\, dv
        =L^2\int_{D}p(x,v,t)\, dx\, dv
$$
Due to the inclusion (\ref{DDD}) we have
$$
    \lambda_{n'}\leq\lambda_{D,\omega}\leq \lambda_{n''}
$$
It follows from (\ref{DDDarea}) that
\be
         (1-c\chi^{(3)})\lambda_{D,\omega}
         \leq \lambda_{n'} \leq \lambda_{n''}\leq
         (1+c\chi^{(3)})\lambda_{D,\omega}
           \label{lamlamlam}
\ee for some constant $c>0$ (determined by $\pi_{\max}$,
$\pi_{\min}$, $v_1$, and $v_2$ in (P3)).

Consider the deterministic quantity
$$
      \tilde{\lambda}_D=L^2\int_{D}\tilde{p}(x,v,t)\, dx\, dv
$$
It easily follows from Lemma~\ref{lmPsi} and the properties of the
function $\tilde{p}(x,v,t)$ that
\be
      |\lambda_{D,\omega}-\tilde{\lambda}_D|
      \leq CL^{-1}\ln L\, \tilde{\lambda}_D
        \label{lamlamt}
\ee
with some constant $C>0$. Combining (\ref{lamlamlam}) and
(\ref{lamlamt}) we arrive at
\be
         (1-\chi^{(4)})\tilde{\lambda}_{D}
         \leq \lambda_{n'} \leq \lambda_{n''}\leq
         (1+\chi^{(4)})\tilde{\lambda}_{D}
           \label{lllt}
\ee
with
\be
        \chi^{(4)}=\, {\rm const}\cdot (\chi^{(3)}+L^{-1}\ln L)
          \label{chi4L}
\ee
The bounds (\ref{kln}) and (\ref{lllt}) will give the desired estimate
on the number $k_{D,\omega}$.

We now construct the domains $D_n$ that approximate the domains
$D_{\omega}=F_{\omega}^{-t}(D)$ for all $\omega\in\Omega_0^{\ast}$
and all trapezoids $D$ defined above. We first fix
$\omega\in\Omega_0^{\ast}$ and a trapezoid $D$ and will construct
two special domains $D',D''$ that approximate
$D_{\omega}=F^{-t}_{\omega}(D)$ from inside and from outside, i.e.
such that $D'\subset D_{\omega}\subset D''$.

The domain $D_{\omega}$ is obtained by pulling $D$ back in time. We
consider its ``trajectory'' $D^-_s=F_{\omega}^{s-t}(D)$ for $0<s<t$, so
that $D^-_0=D_{\omega}$ and $D^-_t=D$. We examine the shape of the
domain $D_s^-$ and how it changes as $s$ runs from $t$ down to $0$.
Recall that the trapezoid $D$ is adjacent to the piston at time $t$ (it
is about to run into the piston at that time). As $s$ goes from $t$
downward, the domain $D^-_s$ comes off the piston and travels to a wall
(as in a movie running backward). During that period, the map
$F_{\omega}^{-(t-s)}$ restricted to $D$ is linear, hence $D_s^-$ is
still a trapezoid. But since the velocities of points $(x,v)\in D$ vary
(from $v_{1,\min}$ to $v_{1,\max}$), the trapezoid $D_s^-$ will be
``skewed'' -- its ``outer'' edge $|v|=v_{1,\max}$ will move toward the
wall faster than the other edge $|v|=v_{1,\min}$. By the time it
reaches the wall, $D_s^-$ will be a long slanted trapezoid stretched
the distance $O(L)$ along the $x$ axis. Every vertical line (parallel
to the $v$ axis) will intersect $D_s^-$ in a segment of length
$O(\Delta t/L)$. As $s$ runs farther down, a collision with the wall
occurs, and a new part of $D_s^-$ appears directly across the $x$ axis,
moving now toward the piston. Its shape will be also that of a long
narrow trapezoid, whose vertical ``thickness'' is $O(\Delta t/L)$.
Eventually it will move all the way (the distance $O(L)$) from the wall
to the piston and contact the piston at some time
$s_{\ast}=s_{\ast}(D)$, see Fig.~5. Note that so far $D_s^-$ is
completely independent of $\omega$.

\begin{figure}[h]
\centering
\epsfig{figure=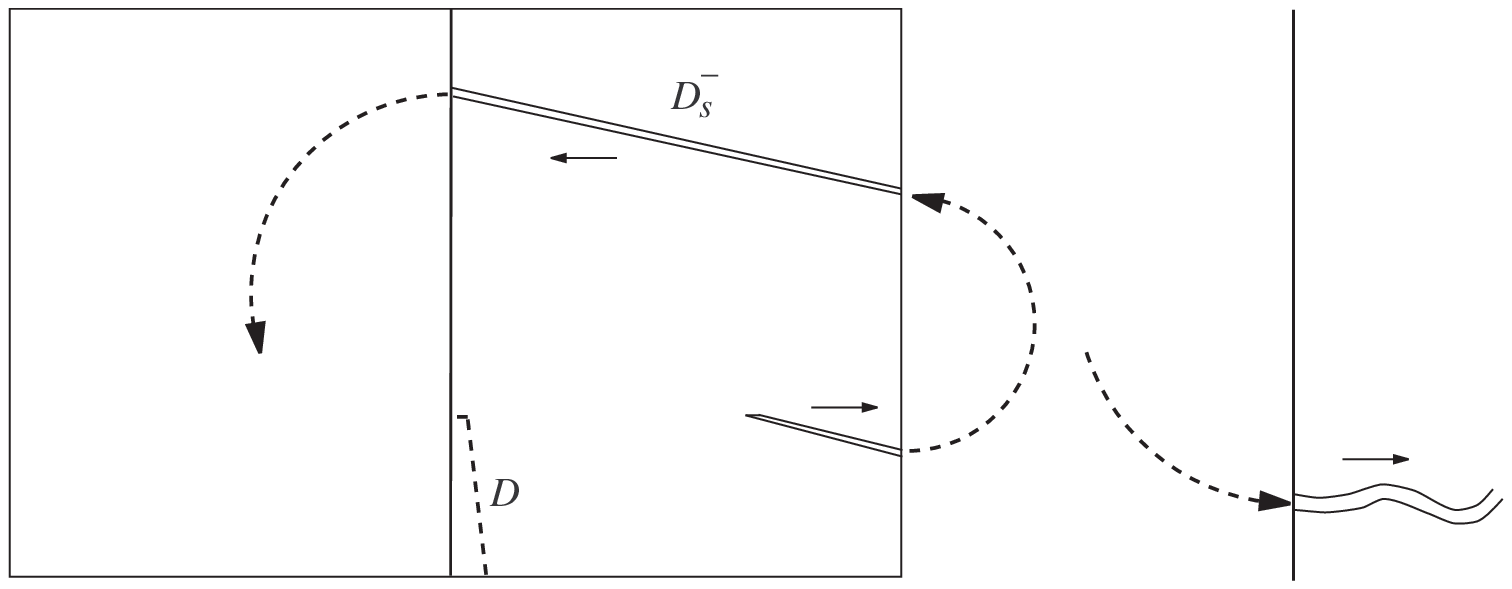}\caption{The domain $D_s^-$. The
arrows show its motion as $s$ decreases.} {The ``outgoing'' part
of $D_s^-$, as it comes off the piston, is shown on the right.}
\end{figure}

After the collision with the piston, a new part of $D_s^-$ appears
across the $x$ axis, coming off the piston and moving back to the wall.
This is the most interesting part, it will be determined by the piston
position and velocity, hence it will actually depend on $\omega$.
Recall that the part of $D_s^-$ running into the piston before the
collision (the ``incoming'' part) is a narrow trapezoid bounded by two
almost horizontal lines with slope $O(1/L)$ lying a distance $O(\Delta
t/L)$ apart. We call them the upper and lower boundaries of $D_s^-$.
The part of $D_s^-$ coming off the piston after the collision (the
``outgoing'' part) also has upper and lower boundaries (due to the
reversal of velocities at collision, though, the upper boundary after
the collision is the image of the lower boundary before the collision,
and vice versa), see Fig.~5. The boundaries of the outgoing part of
$D_s^-$ are quite irregular and depend on $\omega$. It is them who need
careful approximation.

The shape of the outgoing boundary of $D_s^-$ will be determined
by the piston velocity $V(u)$ during the interval $s<u<s_{\ast}$.
Since the collision between $D_s^-$ and the piston occurs during
the zero-recollision interval ($s_{\ast}<T_1$), we can apply the
results of Section~\ref{secZRI}. By Theorem~\ref{tmdV3} the piston
velocity behaves as a H\"older continuous\footnote{For any $d>0$,
$d$-dense sets in the space of H\"older continuous functions were
constructed by Kolmogorov and Tihomirov \cite{KT}, our
constructions here are in the same spirit.} function:
\be
     |V(u+h)-V(u)|\leq CL^{-1}\ln L\, \sqrt{h}=:d_h
        \label{VVh}
\ee
for all $h\in (L^{-2},1)$. We will only consider $h$ satisfying
$L^{-a_1}< h < L^{-a_2}$ with some $0<a_2<a_1<2$. Note that in
this case
\be
        Ld_h/h\to \infty \ \ \ \ \ \
    {\rm and} \ \ \ \ \ \ \ d_h / h\to 0
       \label{hdh}
\ee
as $L\to\infty$.

Let us consider the upper boundary of the outgoing domain $D_s^-$
after the collision. It is a continuous curve that can be
parameterized by the collision time (or contact time between
$D_s^-$ and the piston), call it $u$, and then it becomes a
parametric curve $(x_s(u),v_s(u))$ with parameter $u\in
(s,s_{\ast})$. This means, in particular, that
$$
     F^{w}(x_s(u),v_s(u))=(x_{s+w}(u),v_{s+w}(u))
$$
for $s<u<s_{\ast}$ and $w\leq u-s$, and  the point
$$
   F^{u-s}(x_s(u),v_s(u))=(x_u(u),v_u(u))
$$
is the endpoint of the upper boundary of $D_u^-$,
i.e. $x_u(u)=X(u)$ is the piston coordinate at time $u$.

\medskip\noindent
{\bf Claim 1}. For each $u\in (0,s_{\ast}-h)$ and all
sufficiently large $L$ we have
\be
           |x_{u+h}(u+h)-x_u(u)| \leq B\varepsilon_0 h,
               \label{xxh}
\ee
and
\be
     |v_{u+h}(u+h)-v_u(u)| \leq 3d_h
       \label{vvdh}
\ee

\noindent{\em Proof}. The first bound means that $|X(u+h)-X(u)|
\leq B\varepsilon_0h$ and follows from (\ref{VtB0}). To prove the
second, we use the collision rule
$$
      v_u(u)=-v_u^-(u)+2V(u)
$$
where $v_u^-(u)$ is the $v$ coordinate of the lower boundary of
the ``incoming'' part of $D_u$ where it contacts the piston (at
time $u$). Similarly,
$$
      v_{u+h}(u+h)=-v_{u+h}^-(u+h)+2V(u+h)
$$
Since the lower boundary of the ``incoming'' part of $D_s^-$ is a straight
line with slope $O(1/L)$, and the piston velocity is bounded, then
$$
      |v_{u+h}^-(u+h)-v_u^-(u)|\leq {\rm const}\cdot h/L<d_h
$$
where the last bound follows from(\ref{hdh}). This and (\ref{VVh})
imply (\ref{vvdh}). $\Box$ \medskip

We now fix some $h$ satisfying
\be
        (\Delta t/L)^{1/2}\leq h\leq 2(\Delta t/L)^{1/2}
       \label{h2}
\ee
With our restrictions (\ref{Deltat17}) on $\Delta t$, this implies
$L^{-2/3}\leq h\leq L^{-4/7}$, hence (\ref{hdh}) will hold. Also,
we note that
\be
        d_h/(\Delta t/L) \to 0 \ \ \ \ \ {\rm as}\ \ L\to\infty
       \label{dhDtL}
\ee
hence $d_h$ will be much smaller than the ``thickness'' of the domain
$D_s^-$.

Next we put a lattice on the $x,v$
plane with the $x$-spacing $rh$ and the $v$-spacing $rd_h$, where
$r>0$ is a sufficiently small constant (for example, $r=B\varepsilon_0/10$).
The lattice sites are
\be
     (x_i,v_i)=(rhi,rd_hj),\ \ \ \ \ \ \ \   i,j\in\ZZ
        \label{grid}
\ee

Now we are ready to define the boundary of $D'\subset
D_0^-=F^{-t}_{\omega}(D)$ approximating $D_0^-$ from inside. We
start with the upper boundary of the ``outgoing'' part of $D_0^-$.
It is parameterized as $(x_0(u),v_0(u))$, see above, with
$0<u<s_{\ast}$. We consider discrete parameter values $u_k=kh$, $k
=0,1,\ldots, [s_{\ast}/h]$. For each such $k$, consider the point
$$
     F^{kh}(x_0(kh),v_0(kh)) = (x_{kh}(kh),v_{kh}(kh))
$$
This is the endpoint of the upper boundary of $D_{kh}^-$, so that
$x_{kh}(kh)=X(kh)$ is the piston coordinate at time $kh$. Now we
pick a site of the lattice (\ref{grid}) closest to the point
$(x_{kh}(kh),v_{kh}(kh))$ and lying on the same side of the piston
as the domain $D_{kh}^-$. Call this site $(x_k^s,y_k^s)$. Next, we
adjust this site by moving it down (along the $v$ axis) the fixed
distance $10d_h$ and obtain the adjusted point $(x_k^a,v_k^a)$:
\be
     x_k^a=x_k^s,\ \ \ \ \ \ v_k^a=v_k^s-10d_h
        \label{adjust}
\ee
We note that Claim~1 (along with the smallness of $r$) implies
\be
   |x_k^a-x_{k+1}^a|=|x_k^s-x_{k+1}^s|\leq 2B\varepsilon_0h
      \label{xxxx}
\ee
and
\be
   |v_k^a-v_{k+1}^a|=|v_k^s-v_{k+1}^s|\leq 4d_h
      \label{vvvv}
\ee
Next, let
$$
    (x_k^0,v_k^0):=F^{-kh}(x_k^a,v_k^a)
$$
Now join the points $(x_k^0,v_k^0)$ and $(x_{k+1}^0,v_{k+1}^0)$
with a straight line segment, call it $L_k$. Then the upper
boundary of the domain $D'$ is made by the segments $L_k$, i.e. it
is
$$
     \cup_{k=0}^{[s_{\ast}/h]-1} L_k
$$

\noindent
{\bf Claim 2}. The upper boundary of $D'$ lies completely inside the
domain $D_0^-=F_{\omega}^{-t}(D)$.
\medskip

\noindent{\em Proof}.
It is enough to show that each link $L_k$ lies inside $D_0^-$.
It is easy to see that the map $F^{kh}$ restricted to the link
$L_k$ is a linear map, and the image $L_k^a:=F^{kh}(L_k)$ is a
straight line segment joining the points $(x_k^a,v_k^a)$ and
$$
      F^{-h}(x_{k+1}^a,v_{k+1}^a)=(x_{k+1}^a-v_{k+1}^ah,v_{k+1}^a)
$$
We denote by
$$
  |L_k^a|_x=|x_{k+1}^a-x_k^a-v_{k+1}^ah|,
  \ \ \ \ \ \
  |L_k^a|_v=|v_{k+1}^a-v_k^a|
$$
the lengths of the projections of $L_k^a$ onto the $x$ and $v$
axes, respectively. Note that $v_{1,\min}-B\varepsilon_0\leq
|v_{k+1}^a| \leq v_{1,\max}+B\varepsilon_0$. Hence (\ref{xxh})
implies that
$$
        (v_{1,\min}-2B\varepsilon_0)h
            \leq |L_k^a|_x \leq
        (v_{1,\max}+2B\varepsilon_0)h
$$
i.e. $|L_k^a|_x$ is of order $h$. By (\ref{vvdh}),
$|L_k^a|_v<3d_h$. Hence, the slope of $L_k^a$ is $O(d_h/h)\to 0$
by (\ref{hdh}). Therefore, $L_k^a$ is almost a horizontal segment.
The inequality (\ref{vvdh}) shows that the upper boundary
$(x_{kh}(u),v_{kh}(u))$ of $D_{kh}^-$ for $kh\leq u\leq kh+h$,
which lies directly above $L_k^a$, oscillates in the $v$ direction
by less than $3d_h$. The adjustment (\ref{adjust}) then ensures
that the segment $L_k^a$ lies entirely below the upper boundary of
the domain $D_{kh}^-$. Also, (\ref{dhDtL}) implies that the domain
$D_{kh}^-$ is much ``thicker'' than the distance between its upper
boundary and $L_k^a$, hence $L_k^a$ lies entirely inside that
domain. Hence, $L_k=F^{-kh}(L_k^a)$ lies entirely inside
$D_0^-=F^{-kh}(D_{kh}^-)$, proving our claim. $\Box$
\medskip

It is also clear from the above argument that the segment $L_k^a$
lies the distance $<20d_h$ below the actual upper boundary of the
domain $D_{kh}^-$ so that the area between them is bounded by
const$\cdot hd_h$.  Hence, the total area between the upper
boundaries of $D_0^-$ and $D'$ is bounded by
$$
        {\rm const}\cdot hd_h\cdot (s_{\ast}/h) \leq\,
        {\rm const}\cdot Ld_h
$$

In the same way we construct the lower boundary of the new domain
$D'$. The only difference is that we adjust the selected sites of
the lattice (\ref{grid}) by moving them up, so that the joining
segments will be again inside $D_0^-$. The upper and lower parts
of the boundary will give us a new domain approximating the
``outgoing'' part of $D_0^-$ after the collision with the piston,
i.e. the part $F^{-t}(D\cap G_{\ast}(t))$. Note that this part may
itself experience one collision with the wall (during the interval
$(0,s_{\ast})$), then it will consist of two connected components
lying across the $x$ axis.

It remains to approximate the ``good'' part of
$F_{\omega}^{-t}(D)=D_0^-$, which has not interacted with the
piston, i.e. the part $F_{\omega}^{-t}(D\cap G_0(t))$. That one,
as described above, consists of one or two trapezoids bounded by
two almost horizontal straight lines a distance $O(\Delta t/L)$
apart. This is an easy task. We simply replace the upper (lower)
boundary of the domain $F^{-t}(D\cap G_0(t))$ with a polygonal
line joining some sites of the lattice (\ref{grid}), picked one on
each vertical line $x=x_i$ crossing the domain $D_s^-$ and lying
inside (outside) of this domain and no farther than $d_h$ from the
original boundary (recall that $r$ in (\ref{grid}) is very small,
so the above choice is possible). That gives the boundary of $D'$.
Note that if we number the selected sites consecutively (say, from
left to right), and call them $(x_k^s,y_k^s)$, then the
neighboring sites will again satisfy (\ref{xxxx})--(\ref{vvvv}).

The complicated construction described above produces a domain
$D'\subset D_{\omega}=F_{\omega}^{-t}(D)$ such that
\be
    |D_{\omega}\setminus D'|\leq\,{\rm const}\cdot d_hL
     =\, {\rm const}\cdot \ln L\, \sqrt{h}
       \label{DDdh1}
\ee
In a completely similar way we construct another domain $D''$ that
approximates $D_{\omega}$ from outside and satisfies
\be
    |D''\setminus D_{\omega}|\leq\,{\rm const}\cdot d_hL
     =\, {\rm const}\cdot \ln L\, \sqrt{h}
       \label{DDdh2}
\ee
(we just need to adjust the sites along the upper boundary by
moving them up and the sites along the lower boundary by moving
them down, so that the new boundaries will be completely outside
of $D_{\omega}$). Since $|D|=O(\Delta t)$, the inequalities
(\ref{DDdh1})--(\ref{DDdh2}) imply
$$
      |D''\setminus D'|\leq \, {\rm const}\cdot \ln L\, \sqrt{h}
      \, (\Delta t)^{-1}\, |D|
$$
hence we get (\ref{DDDarea}) with
\be
  \chi^{(3)}= {\rm const}\cdot \ln L\, \sqrt{h} \, (\Delta t)^{-1}
     \label{chi3L}
\ee
Note that $\chi^{(3)}\to 0$ as $L\to\infty$ due to our choice
of $h$ in (\ref{h2}) and restrictions on $\Delta t$ in (\ref{Deltat17}).

Our domains $D',D''$ are constructed around $D_{\omega}$, hence
they depend on $\omega$. However, the boundaries of $D',D''$ are
polygonal lines whose vertices are derived from the sites of a
fixed lattice (\ref{grid}). Therefore, the total number of {\em
distinct} domains $D',D''$ is finite. Now we estimate their number
$N_{\ast}$.

First, the total number of the sites of the lattice (\ref{grid}) in the
relevant area $0<x<L$, $|v|<v_{1,\max}$ is
\be
   K_1=\,{\rm const}\cdot (L/h)\cdot (2v_{\max}/d_h)
   \leq \,{\rm const}\cdot L\,(hd_h)^{-1}
      \label{K1}
\ee
Recall that the domain $D_0=F^{-t}(D)$ consists of at most four
connected components. The part $F^{-t}(D\cap G_{\ast}(t))$, or the
``outgoing'' part of $D_0$ created after the collision with the
piston, consists of at most two components. And the part
$F^{-t}(D\cap G_{0}(t))$ consists of at most two components, each
is a regular, trapezoidal region.  The upper (lower) part of the
boundary of $D'$ ($D''$) will then consists of at most four
disjoint polygonal lines constructed from the lattice sites
(\ref{grid}). In each polygonal line, the first point can be
constructed from any of the $K_1$ sites of the lattice, see
(\ref{K1}). But other points can be constructed consecutively, and
each point must be constructed from a site selected from
\be
        K_0:=(4B\varepsilon_0/r)\cdot (8/r)=\,{\rm const}
\ee
nearest neighbors of the previously selected site, according to
(\ref{xxxx})--(\ref{vvvv}). It is also clear that each polygonal
line has at most
$$
          M:=\,{\rm const}\cdot L/h
$$
links (vertices).  Therefore, the total number of ways to
construct one polygonal line does not exceed $K_1K_0^M$.
The total number of ways to construct the entire boundary of
$D'$ ($D''$) is then less than $K_1^8K_0^{8M}$.
This gives an upper bound on the number of distinct domains
$D'$ ($D''$):
\be
       N_{\ast}\leq K_1^8K_0^{8M}
       \leq \, {\rm const}\cdot L^{8}(hd_h)^{-8}\cdot e^{{\rm const}\cdot L/h}
          \label{N8}
\ee
Of course, the exponential factor is dominant here and can absorb
all the others.

We now set the value of $B_{\Delta t}$ in
(\ref{Blambdan})--(\ref{kln}):
\be
    B_{\Delta t}=c_7\, \sqrt{L/h}
      \label{Bchoice}
\ee
with a sufficiently small constant $c_7>0$. Then (\ref{POmega1})
and (\ref{N8}) imply
\be
     P(\Omega_0^{\ast}\setminus\Omega_1^{\ast}(\Delta t))
           \leq {\rm const}\cdot e^{-c_8 L/h}
             \label{POmega11}
\ee
with some constant $c_8>0$. Recall that $h$ satisfies (\ref{h2}),
hence
$$
     P(\Omega_0^{\ast}\setminus\Omega_1^{\ast}(\Delta t))
           \leq {\rm const}\cdot e^{-{\rm const}\cdot L^{3/2}(\Delta t)^{-1/2}}
$$
Next, since $\Delta t$ satisfies (\ref{Deltat17}), we have
\be
     P(\Omega_0^{\ast}\setminus\Omega_1^{\ast}(\Delta t))
           \leq {\rm const}\cdot e^{-{\rm const}\cdot L^{11/7}}
             \label{POmega12}
\ee
Also, by (\ref{Bchoice}) and (\ref{h2}) $$
    B_{\Delta t}\leq c_7 L^{3/4}(\Delta t)^{-1/4}
$$
which implies (\ref{Blambdan}) since $\Delta t\gg L^{-1/3}$.

Next, for each $\omega\in\Omega_1^{\ast}(\Delta t)$, each $t\in
(T_1,S_2-\Delta t)$, and any trapezoid $D$ defined above, the
number of particles $k_{D,\omega}$ in $D$ satisfies
(\ref{kln})--(\ref{chi4L}) with $\chi^{(3)}$ given by
(\ref{chi3L}) and $h$ given by (\ref{h2}), hence
$$
     k_{D,\omega}=\tilde{\lambda}_D+\chi_{D,\omega}
$$
with
\begin{eqnarray*}
    |\chi_{D,\omega}| & \leq & \,{\rm const}\cdot
    \left (B_{\Delta t}\, L\sqrt{\Delta t}
    +L^2\ln L\,\sqrt{h}+L\ln L \right ) \\
    & \leq & \,{\rm const}\cdot L^{7/4}\ln L\, (\Delta t)^{1/4}
\end{eqnarray*}
We note that $\tilde{\lambda}_D=O(L^2\, \Delta t)$, hence
$|\chi_{D,\omega}|\ll \tilde{\lambda}_D$ for all $\Delta t$
satisfying (\ref{Deltat17}). By the way, this easily implies a
rough bound $k_{D,\omega}\leq 2\tilde{\lambda}_D\leq\,$const$\cdot
L^2\, \Delta t$, which justifies (\ref{kbound17}) and hence
(\ref{deltaV17}).

Next, using (\ref{kkk1}) and applying the above estimates to each
of $k^{\pm}_i$, $i=1,2$, gives
$$
   k=L^2\int_{D_1} \tilde{p}(x,v,t)\, dx\, dv + \chi'
$$
with
$$
    |\chi'|\leq \,{\rm const}\cdot [L^{7/4}\ln L\, (\Delta t)^{1/4}
    +L^2\,\delta V\,\Delta t]
$$
In a similar way we can estimate the other random factor in the
main decomposition formula (\ref{main17}), which is
$$
         Z=\sum_{j=1}^k v_j
$$
and get
$$
          Z=L^2\int_{D_1} v\,\tilde{p}(x,v,t)\, dx\, dv +
          \chi''
$$
with the same upper bound on $\chi''$ as that on $\chi'$ above.

Using the smoothness properties of the function $\tilde{p}(x,v,t)$
described around (\ref{tildeQx}) gives, cf.\
(\ref{intQ10})--(\ref{chiuQ}),
$$
       \int_{D_1}\tilde{p}(x,v,t)\,dx\,dv=
       (\tilde{Q}_1(t)-\tilde{Q}_0(t))\,\Delta t + \chi^{(4)}
$$
and
$$
       \int_{D_1}v\, \tilde{p}(x,v,t)\,dx\,dv=
       (\tilde{Q}_2(t)-\tilde{Q}_1(t))\,\Delta t + \chi^{(5)}
$$
with
$$
        |\chi^{(u)}| \leq {\rm const}\cdot
        \Big [ (\Delta t)^2/L+\delta V\, \Delta t\Big ]
$$
for $u=4,5$.

Therefore, we get
\be
        V(t+\Delta t)-V(t)=\tilde{\cal D}(t)\, \Delta t + \chi
\ee
where
\be
     |\chi|\leq {\rm const}\cdot \left [
     L^{-1/4} \ln L\, (\Delta t)^{1/4}+\delta V\, \Delta t+
     (\Delta t)^2\right ]
       \label{chilast17}
\ee Recall that $\delta V=O(\Delta t)$ by (\ref{deltaV17}). It is
now easy to check that the second and the third terms are much
smaller than the first one. This proves
(\ref{VV17})--(\ref{chibound17}). It does not prove
Theorem~\ref{tmdV17} yet, because we have fixed one (arbitrary)
value of $\Delta t$, and our set $\Omega_1^{\ast}$ depended on
$\Delta t$.

Actually, for our main purpose it is enough to prove
Theorem~\ref{tmdV17} for just one value of $\Delta t$, namely for
$\Delta t=1/L^{1/7}$, as we will see later. But at a little extra
effort we can prove our theorem for all $\Delta t$ satisfying
(\ref{Deltat17}). We do that next. Divide the interval
(\ref{Deltat17}) into subintervals of length $e^{-L}$. That is,
fix a finite collection of points $(\Delta t)_n=ne^{-L}$ for
$n=n_1,\ldots,n_2$ with $n_1=L^{-1/3}(\ln L)^2e^L$ and
$n_2=L^{-1/7}e^L$. Then we define
$$
    \Omega_1^{\ast}=\cap_{n=n_1}^{n_2}\Omega_1^{\ast}((\Delta t)_n)
$$
The bound (\ref{POmega12}) then implies
$$
     P(\Omega_0^{\ast}\setminus\Omega_1^{\ast})
           \leq {\rm const}\cdot e^{L-{\rm const}\cdot L^{11/7}}
$$
which is obviously sufficient to maintain the bound
(\ref{POmega1ast}).

Now for any $\Delta t$ satisfying (\ref{Deltat17}) we find
$(\Delta t)_n$ such that $|\Delta t-(\Delta t)_n|\leq e^{-L}$. For
any $\omega\in\Omega_1^{\ast}\subset \Omega_1^{\ast}((\Delta
t)_n)$ we have all the above estimates with $\Delta t$ replaced by
$(\Delta t)_n$. This replacement only causes an exponentially
small error, $e^{-L}$, in our estimates. It will not spoil our
bounds, which are all polynomial in $L$. This completes the proof
of Theorem~\ref{tmdV17}. $\Box$
\medskip

The next theorem is an analogue of Theorem~\ref{tmdV2}.

\begin{theorem}
Assume that $\varepsilon_0>0$ is small enough. For all
sufficiently large $L$, for each configuration
$\omega\in\Omega^{\ast}_{1}$ and for all $t\in (T_1,S_2)$ we have
\\ {\rm (i)} there is a constant $B>0$ such that
\be
                 |V(t)|<B\varepsilon_0
            \label{VtB017}
\ee
{\rm (ii)} there is a constant $C_0>0$ such that
\be
     |V(t)-V_0(t)|<C_0\, L^{-1/7}\ln L
        \label{VV017}
\ee
where $V_0(t)$ is defined by
\be
    V_0(t)=\frac{\tilde{Q}_1(t)-
    \sqrt{\tilde{Q}_1^2(t)-\tilde{Q}_0(t)\tilde{Q}_2(t)}}
    {\tilde{Q}_0(t)}
        \label{V017}
\ee
whenever $\tilde{Q}_0(t)\neq 0$ and by
\be
    V_0(t)=\frac{\tilde{Q}_2(t)}{2\tilde{Q}_1(t)}
        \label{V0017}
\ee
otherwise. \label{tmdV217}
\end{theorem}

\noindent{\em Proof} of this theorem very much repeats that of
Theorem~\ref{tmdV2}. The first half of it, up to the formula
(\ref{D1D2}) can be copied almost verbatim, with only replacement
of $S_1$ by $S_2$, $Q_i$ by $\tilde{Q}_i$, and ${\cal D}$ by
$\tilde{\cal D}$. We omit that part. The rest of the proof
requires more substantial modifications, and we give it in detail.

First, recall that Theorem~\ref{tmdV2} deals with $t\in (0,S_1)$.
On the interval $(0,S_1)$ the functions $Q_i(t)$, $i=0,1,2$, are
independent of $\omega$, they are defined by equations
(\ref{qp00})--(\ref{Q20}) where $p(x,v,t)$ was in fact the
deterministic density now denoted by $\tilde{p}(x,v,t)$.
Therefore, our functions $\tilde{Q}_i(t)$ for $t>T_1$, are natural
continuations of $Q_i(t)$ beyond the interval $(0,S_1)=(0,T_1)$
and they have the same properties, cf. (\ref{Qx}) and
(\ref{tildeQx}). Hence, the function $V_0(t)$ defined by
(\ref{V017})--(\ref{V0017}) for $t>T_1$ is the continuation of
$V_0(t)$ defined by (\ref{V0})--(\ref{V00}) on the interval
$(0,T_1)$.

Next, in the first half of the proof (which we omitted since it
almost coincided with that of Theorem~\ref{tmdV2}), we must
introduce $t_{\ast}<S_2$ as the first time when (\ref{VtB017})
fails. Now we prove (\ref{VV017}) for all $t<t_{\ast}$ with some
constant $C_0>0$ (independent of the choice of $B$ in
(\ref{VtB017}), which is to be made yet).

The bound (\ref{VV0}) proved for all $t\leq T_1$ implies that
(\ref{VV017}) holds for at least some $t>T_1$. Next, if
(\ref{VV017}) fails for any $t<t_{\ast}$, then let $t\in
(T_1,t_{\ast})$ be the first time (\ref{VV017}) fails. Denote by
$$
     \Delta_0=L^{-1/7}
$$
the maximal allowed time increment in Theorem~\ref{tmdV17}. Let
$s=t-\Delta_0$. Due to Theorem~\ref{tmdV17}
\be
     V(t)=V(s)+\tilde{\cal D}(s)\Delta_0+\chi
       \label{VsD017}
\ee
with
$$
     |\chi|\leq CL^{-1/4}\ln L\,(\Delta_0)^{1/4}
     =C L^{-2/7}\ln L
$$
Due to the analogue of (\ref{dV0t}) obtained in the first half of
the proof of the theorem,
\be
      V_0(t)=V_0(s)+\chi_0
        \label{V0sD017}
\ee
with
$$
    |\chi_0|\leq\frac{E_0\,\varepsilon_0\,\Delta_0}{L}
    =\frac{E_0\,\varepsilon_0}{L^{8/7}}
$$
For brevity, put $U(s)=V(s)-V_0(s)$ for all $s$. Subtracting
(\ref{V0sD017}) from (\ref{VsD017}) then gives
\be
    U(t)=U(s)+\tilde{\cal D}(s)\Delta_0+\chi'
       \label{UtUs17}
\ee
with $\chi'=\chi-\chi_0$, so that for large $L$
\be
     |\chi'|\leq 2C L^{-2/7}\ln L
         \label{chiprime17}
\ee
Now assume, without loss of generality, that $U(t)>0$. Since
(\ref{VV017}) fails at time $t$, we have
\be
     U(t)\geq C_0L^{-1/7} \ln L
        \label{Ut17}
\ee
Now consider two cases. If $U(s)\leq 0$, then by the analogue of
(\ref{D1D2})
$$
    U(t)\leq |\tilde{\cal D}(s)|\Delta_0+|\chi'|\leq E_2\,
    |U(s)|\Delta_0+|\chi'|\ll L^{-1/7}\ln L
$$
for large $L$, which contradicts to (\ref{Ut17}). If $U(s)>0$,
then, again due to (\ref{UtUs17}) and the analogue of
(\ref{D1D2}),
$$
    U(t)<U(s)[1-E_1\Delta_0]+\chi',
$$
hence
\begin{eqnarray}
      U(s) & > & \frac{U(t)-\chi'}{1-E_1\Delta_0}
      > (U(t)-\chi')(1+E_1\Delta_0) \nonumber\\
      & > &
      U(t)+U(t)E_1\Delta_0-2\chi'
         \label{Ut>Us17}
\end{eqnarray}
Now, if $C_0$ in (\ref{VV017}) is large enough, say $C_0=3C/E_1$,
then $U(t)E_1\Delta_0>2\chi'$ by (\ref{Ut17}) and
(\ref{chiprime17}). This fact and (\ref{Ut>Us17}) imply
$U(s)>U(t)$, so (\ref{VV017}) fails at an earlier time $s<t$, a
contradiction. Hence, (\ref{VV017}) is proved for all $t<t_{\ast}$
and $C_0=3C/E_1$.

Lastly, the remaining part of the proof of Theorem~\ref{tmdV2} can
be repeated verbatim, concluding the proof of
Theorem~\ref{tmdV217}.  $\Box$\medskip

We finally prove the convergence, as $L\to\infty$, of the random
trajectory of the piston to the solution $Y(t/L)$, $W(t/L)$ of the
hydrodynamical equations described in Section~\ref{secHE} for all
$t\in (T_1,S_2)$.

\begin{theorem}
Assume that $\varepsilon_0>0$ in {\rm (P5)} is small enough. Then,
for all large $L$ and all $\omega\in\Omega_1^{\ast}$, there is a
constant $C>0$ such that \be
          |Y_L(\tau,\omega)-Y(\tau)|\leq \frac{C\,\ln L}{L^{1/7}}
          \label{YLY017}
\ee
and
\be
          |W_L(\tau,\omega)-W(\tau)|\leq \frac{C\,\ln L}{L^{1/7}}
          \label{WLW017}
\ee
for all $\min\{\tau_1,T_1/L\}< \tau < \min\{\tau_2,S_2/L\}$. We
also have
\be
           T_2-S_2\leq C\varepsilon_0 L
            \label{T1S10}
\ee
and
\be
           \tau_2-S_2/L \leq C\varepsilon_0
            \label{tau1S1}
\ee
The fluctuations of the function $S_2=S_2(\omega)$ for
$\omega\in\Omega_1^{\ast}$ are bounded by
\be
           \sup_{\omega,\omega'\in\Omega_1^{\ast}}
           |S_2(\omega)-S_2(\omega')|\leq \frac{C\,\ln L}{L^{1/7}}
               \label{S2fluct}
\ee
 \label{tmonerec}
\end{theorem}

\noindent{\em Proof}. According to (\ref{YW1}), the deterministic
function $Y(\tau)$ satisfies
\be
          dY(\tau)/d\tau=F(Y,\tau),\ \ \ \ \ \ \ \ Y(0)=1/2
          \label{Y'tau17}
\ee
Now Theorems~\ref{tmdV3} and \ref{tmdV217} imply that for all
$\omega\in\Omega_1^{\ast}$ the random trajectory satisfies
\be
         \partial Y_L(\tau,\omega)/\partial\tau=
         F(Y,\tau)+\chi(\tau,\omega),
         \ \ \ \ \ \ Y_L(0,\omega)=1/2
          \label{Y'tauomega17}
\ee
with some
$$
        |\chi(\tau,\omega)|\leq\frac{C\,\ln L}{L^{1/7}}
$$
Recall that $|\partial F(Y,\tau)/\partial Y|\leq \kappa$, see
(\ref{dFdY}). Therefore, the difference
$Z_L(\tau,\omega):=Y_L(\tau,\omega)-Y(\tau)$ satisfies
$$
          |Z_L'(\tau,\omega)|\leq \kappa|Z_L(\tau,\omega)|+
          \frac{C\,\ln L}{L^{1/7}}
$$
and $Z_L(0,\omega)=0$. Using the standard Gronwall inequality in
differential equations, see, e.g., Lemma~2.1 in \cite{TVS}, gives
$$
          |Z_L(\tau,\omega)|\leq
          \frac{C\,\ln L}{\kappa L^{1/7}}\,\Big (e^{\kappa\tau}-1\Big )
$$
and
$$
          |Z_L'(\tau,\omega)|\leq
          \frac{C\,\ln L}{L^{1/7}}\, e^{\kappa\tau}
$$
for all $\tau < S_2/L$, which imply (\ref{YLY017}) and
(\ref{WLW017}).

Next, (\ref{WLW017}) enables us to apply Proposition~\ref{prS1}
with $\Delta=CL^{-1/7}\ln L$ and thus prove (\ref{T1S10}). Now we
employ the same argument as in the proof of
Theorem~\ref{tmzerorec}. By (\ref{WLW017}), random fluctuations of
the piston velocity are bounded by $CL^{-1/7}\ln L$. Hence, random
fluctuations of the velocities of particles that have had one or
two collisions with the piston are bounded by $4CL^{-1/7}\ln L$.
The random fluctuations of the positions of both piston and
particles at every moment of time $t<\min\{\tau_2L,S_2\}$ are
bounded by the same quantities (with, possibly, a different value
of $C$) in the coordinate $y=x/L$. This implies (\ref{S2fluct}) in
the same way, as (\ref{tau0T0L}) in Theorem~\ref{tmzerorec}. Now
we can apply Proposition~\ref{prS1} to the deterministic dynamics
constructed in Section~\ref{secHE} with the same result, and thus
prove (\ref{tau1S1}). $\Box$\medskip

The bound (\ref{S2fluct}) shows that the limit
$$
       \tau_{\ast\ast}:=\lim_{L\to\infty} S_2(\omega)/L
$$
does not depend on $\omega\in\Omega_1^{\ast}$. We have
$|\tau_{\ast\ast} -2/v_{\max}|\leq \,{\rm const} \cdot
\varepsilon_0$, due to (\ref{tau1S1}) and the estimates in
Lemma~\ref{lm29}.

The convergence claimed in Theorem~\ref{tmmain} is now proved on
the interval $(0,\tau_{\ast\ast})$, but, generally,
$\tau_{\ast\ast} <\tau_{\ast}$ with $\tau_{\ast}-\tau_{\ast\ast}
=O(\varepsilon_0)$, so we may still be $O(\varepsilon_0)$ short of
our target value $\tau_{\ast}$. To extend our results all the way
to $\tau_{\ast}$ we need to redefine $S_2$ and the neighborhood
${\cal X}_1$ introduced by (\ref{calX1}) more accurately. We need
to set
\begin{eqnarray}
     {\cal X}_1(t)=\{(x,v)\in G^+(t):\, x=X(t)+0,\ v<0\}
       \nonumber\\
       \cup \{(x,v)\in G^+(t):\, x=X(t)-0,\ v>0\}
        \label{calX1a}
\end{eqnarray}
The domain $G^+(t)=F^t(G^+)$ is random (it depends of $\omega$),
i.e.\ we should write $G^+(t)= G^+(t,\omega)$, and this is why we
could not adopt the above definition of ${\cal X}_1(t)$ earlier
and opted for a cruder one (\ref{calX1}). But now, as we have just
shown in the proof of Theorem~\ref{tmonerec}, random fluctuations
of the velocities and positions (in the $y$ coordinate) of the
particles and the piston are bounded by ${\rm const}\cdot
L^{-1/7}\ln L$. Hence, at every time moment
$t<\min\{\tau_2L,S_2\}$ all the domains $G^+(t,\omega)$, are close
to each other -- the distance between $G^+(t,\omega)$ and
$G^+(t,\omega')$ for $\omega,\omega' \in\Omega_1^{\ast}$ in the
Hausdorff metric on the coordinate plane $y,v$ is bounded by ${\rm
const}\cdot L^{-1/7}\ln L$. By the same reason, every domain
$G^+(t,\omega)$ will be $O(L^{-1/7}\ln L)$-close in the Hausdorff
metric to the deterministic domain ${\cal G}^+(t/L)$ defined in
Section~\ref{secHE}. Hence we can easily make $G^+(t)$ in
(\ref{calX1a}) independent of $\omega$ by, say, taking the union
\be
    G^+(t)=\cup_{\omega\in\Omega_1^{\ast}} G^+(t,\omega)
       \label{GtGt}
\ee
It is important to note that all the gas particles colliding with
the piston at time $t$ for every $\omega\in\Omega_1^{\ast}$ will
be in ${\cal X}_1$ defined by (\ref{calX1a})--(\ref{GtGt}). With
this new definition of ${\cal X}_1$ replacing (\ref{calX1}) and
with $S_2$ changing accordingly, we have
$$
      \lim_{L\to\infty} |S_2/L-\tau_{\ast}|=0
$$
which follows from the $O(L^{-1/7}\ln L)$-closeness of the random
dynamics to the deterministic dynamics on the $y,v$ plane. Thus we
extend all our results to the interval $(0,\tau_{\ast})$.

\section{Beyond the second recollision}
\label{secBORI} \setcounter{equation}{0}

The main goal of our analysis in Sections~\ref{secZRI} and
\ref{secORI} is to prove that under suitable initial conditions
random fluctuations in the motion of a massive piston in a closed
container filled with an ideal gas are small and vanish in the
thermodynamic limit. We are, however, able to control those
fluctuations effectively only as long as the surrounding gas of
particles can be described by a Poisson process, i.e.\ during the
zero-recollision interval $0<\tau<\tau_1$. In that case the random
fluctuations are bounded by const$\cdot L^{-1}\, \ln L$, see
Theorem~\ref{tmzerorec}. Up to the logarithmic factor, this bound
is optimal, according to Theorem~\ref{tmH} by Holley.

During the one-recollision interval $\tau_1<\tau<\tau_{\ast}$, the
situation is different. The probability distribution of gas
particles that have experienced one collision with the piston is
no longer a Poisson process, it has intricate correlations. We are
only able to show that random fluctuations remain bounded by
$L^{-1/7}$, see Theorem~\ref{tmonerec}. Perhaps, our bound is far
from optimal, but our numerical experiments reported below
demonstrate that random fluctuations indeed grow during the
one-recollision interval and beyond.

At present, we do not know if our methods or results can be
extended beyond the critical time $\tau_{\ast}$, this remains an
open question. We emphasize, however, that our
Theorem~\ref{tmonerec} is the first rigorous treatment of the
evolution of a piston in an ideal gas where most or all of the
particles experience more than one collision with the piston.

In order to understand what is going on beyond the critical time
${\tau}_{\ast}$, and in particular whether random fluctuations
grow or remain small, we undertook experimental and heuristic
studies of the piston dynamics on a large time scale. Below we
describe our findings and discuss further research in this
direction. A detailed account of our work can be found in
\cite{CL}.

We set the initial density of the gas to
\be
    \pi_0(y,v)=\pi_0(|v|)=\left\{\begin{array}{ll}
    1 & {\rm if}\ \ 0.5\leq |v|\leq 1\\
    0 & {\rm elsewhere}
    \end{array}\right .
        \label{pini}
\ee
It satisfies our requirements (P1)--(P5), in particular
$v_{\min}=0.5$ and $v_{\max}=1$, and most importantly
$\varepsilon_0=0$. Therefore, by Corollary~\ref{crprop}, the
solution of the hydrodynamical equations is trivial:
$Y(\tau)\equiv 0.5$, $W(\tau)\equiv 0$, and
$\pi(y,v,\tau)\equiv\pi_0(y,v)$ for all $\tau>0$.

To generate an initial configuration of particles, we used a
random number generator described in \cite{MN}. For our density
(\ref{pini}), the $x$ and $v$ coordinates of all the particles are
independent random variables uniformly distributed in their ranges
$0<x<L$ and $v_{\min}\leq |v|\leq v_{\max}$. Our computer program
first selects the number of particles $N$ according to the Poisson
law with mean $L^3$, and then generates all $(x_i,v_i)$, $1\leq
i\leq N$, independently according to their uniform distributions.
The parameter $L$ changed in our simulations from $L=30$ to
$L=300$. For $L=300$ the system contains $\approx L^3=27,000,000$
particles.

Once the initial data is generated randomly, the program computes
the dynamics by using the elastic collision rules
(\ref{V'})--(\ref{v'}). All calculations were performed in double
precision, with coordinates and velocities of all particles stored
and computed individually.

Figure 6 presents a typical trajectory of the piston. Here
$L=100$. The position and time are measured in hydrodynamic
variables $Y=X/L$, $0<Y<1$, and $\tau=t/L$.

\begin{figure}[h]
\centering \epsfig{figure=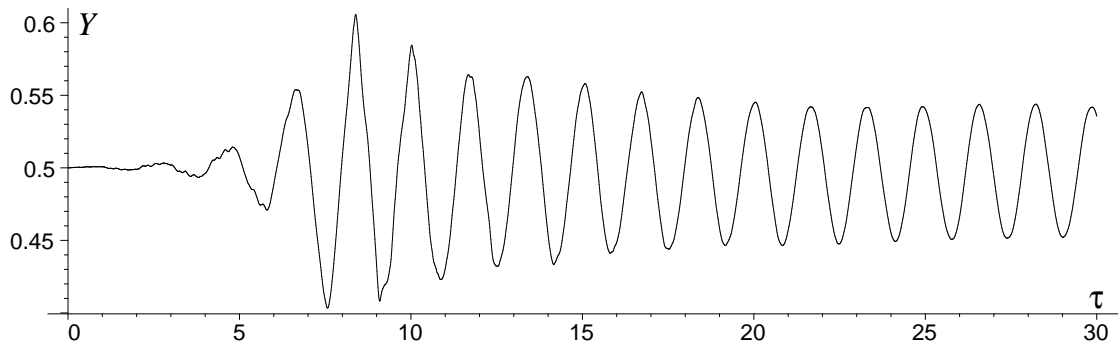}\caption{The piston coordinate
$Y$ as a function of time $\tau$. Here $L=100$, $N=1000229$.}
\end{figure}

Initially, the piston barely moves about its stationary point $y=
0.5$. Then, at times $\tau$ between $3$ and $5$, the random
vibrations of the piston grow and become quite visible on the
$y$-scale, but for a short while they look random, as a trajectory
of a Brownian motion. After that the piston starts travelling back
and forth along the $y$ axis in a more regular manner, making
excursions farther and farther away from the stationary point
$y=0.5$. Very soon, at $\tau = \tau_{\max} \approx 8$, the
swinging motion of the piston reaches its maximum, $(\Delta
Y)_{\max}=\max|Y_L(\tau)-0.5|\approx 0.1$. Then the oscillations
of the piston dampen in size and seem to stabilize at an amplitude
$A\approx 0.04$. At the same time the trajectory of the piston
smoothes out and enters an oscillatory mode with a period
$\tau_{\rm per}\approx 1.63$. The velocity of the piston $W(\tau)$
follows similar patterns, see Figure~7.

\begin{figure}[h]
\centering \epsfig{figure=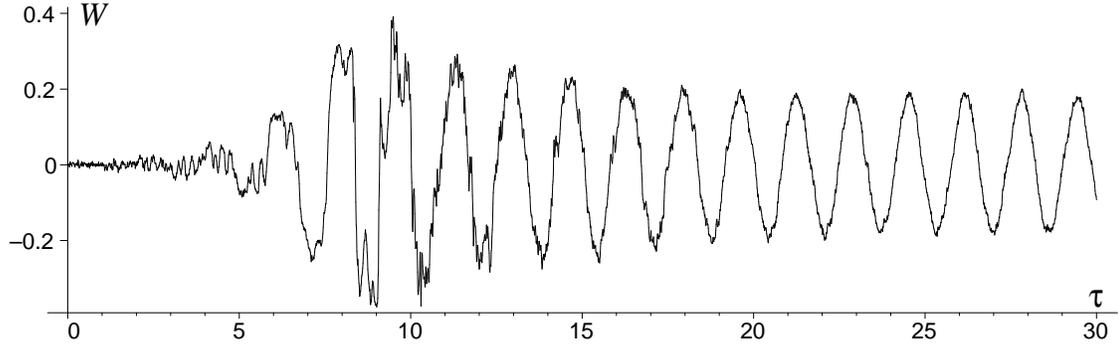}\caption{The piston velocity
$W$ as a function of time $\tau$. The same run as in Fig.~6.}
\end{figure}

Both the coordinate and velocity of the piston continue almost
perfect harmonic oscillations for a long time with the same period
$\tau_{\rm per} \simeq 1.63$ (this is independent of $L$) but the
amplitudes of both $Y(\tau)$ and $W(\tau)$ are slowly decreasing,
see Figure~8.

\begin{figure}[h]
\centering \epsfig{figure=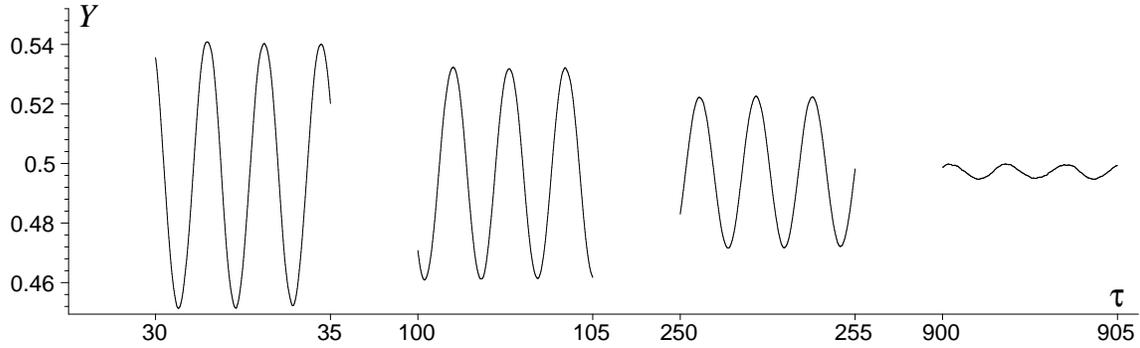}\caption{The piston coordinate
$Y$ during the intervals $(30,35)$, $(100,105)$, $(250,255)$, and
$(900,905)$. The same run as in Fig.~6 and 7.}
\end{figure}

The oscillations of the piston with decaying amplitude can be
described, in the interval $20 <\tau<1000$, approximately by
\be
      Y(\tau)\simeq Ae^{-\lambda(\tau-20)}\sin\omega
      (\tau-\alpha)
         \label{Y1}
\ee
with $A = 0.046$ and some constant $\lambda>0$. Correspondingly,
$W=dY/d\tau$ in the same interval $20<\tau<1000$ is
\begin{eqnarray}
      W(\tau) &\simeq & -\lambda Y(\tau)+Ae^{-\lambda(\tau-20)}
      \omega\cos\omega (\tau-\alpha)\nonumber\\
      &=& Ae^{-\lambda(\tau-20)}
      [-\lambda\sin\omega (\tau-\alpha)+\omega \cos\omega (\tau-\alpha)]
      \nonumber\\
      &=&
      A_1e^{-\lambda (\tau-20)}\sin\omega (\tau-\beta)
         \label{W1}
\end{eqnarray}
with $A_1=A\sqrt{\omega^2+\lambda^2}$ and some $\beta$ related to
$\alpha$.

To check how well our prediction (\ref{Y1}) agrees with the
experimental data, we computed the amplitude $A(\tau)$ as a
function of time $\tau$, by fitting a sine function
$Y(\tau)=A\sin\omega(\tau-\alpha)$ ``locally'', on the interval
$(\tau-5,\tau+5)$ for each $\tau$. Fig.~9 shows $A(\tau)$ on the
logarithmic scale, which looks almost linear on the interval
$30<\tau<800$.

\begin{figure}[h]
\centering \epsfig{figure=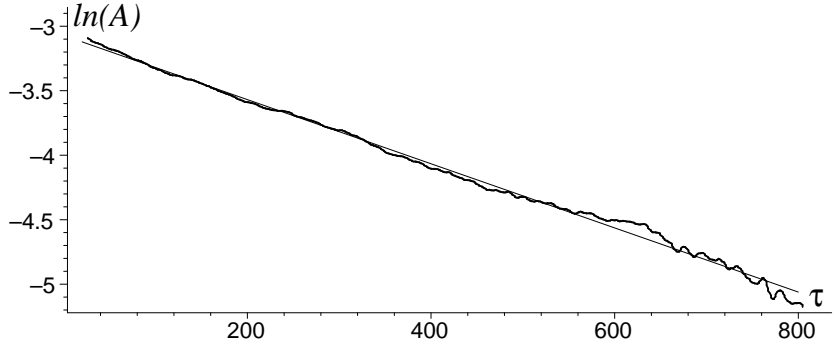}\caption{The amplitude
$A(\tau)$ on the logarithmic scale: experimental curve (bold) and
a linear fit (thin). The same run as the one shown in Fig.~6, 7
and 8.}
\end{figure}

We used the least squares fit to estimate $\lambda=0.00264$ for
the run shown on Figs.~6-9. Since $\lambda$ is small, the
oscillations indeed die out very slowly. The ``half-life'' time
(the time it takes to reduce the amplitude by a factor of two) is
$\tau_{1/2}=\lambda^{-1}\ln 2 \approx 263$.  The parameter
$\lambda$ and hence $\tau_{1/2}$ depend on the system size $L$. We
estimated numerically that $\tau_{1/2} \sim L^{1.3}$, hence
$\lambda\sim L^{-1.3}$.

The key characteristics of the piston trajectory described above,
in particular, $(\Delta Y)_{\max}$, $W_{\max}$, $A$, $\tau_{\rm
per}$, appear to be independent of $L$. Even for $L=300$ (the
largest system tested experimentally) the piston experiences large
oscillations very similar to the ones shown on Fig.~6 and 7. Some
other quantities, such as $\tau_{1/2}$ and the related $\lambda$,
depend in a systematic way on $L$.

But most importantly, the time of the largest oscillations
$\tau_{\max}$ and the related time of the onset of the instability
$\tau_c$, see below, seem to slowly grow with $L$, very likely as
$\log L$. To understand this fact, we looked into the mechanism of
the build-up of random fluctuations of the piston position and
velocity displayed on Figures~6 and 7. To this end we plotted the
histogram of the (empirical) density of gas particles in the $y,v$
plane at various times $0<\tau <30$, see samples in Figure~10. The
initial density (at time zero) is almost uniform over the domain
$0<x<L$ and $v_{\min}\leq |v|\leq v_{\max}$ (variations in the
initial configuration always exist, because it is generated
randomly). Then, for $0<\tau<1$, the piston experiences random
collisions with particles and acquires a speed of order
$M^{-1/2}=O(1/L)$, see Theorem~\ref{tmH}. These small fluctuations
of the piston velocity result in bigger changes of the velocities
of the particles which leave the piston after collisions due to
the rule (\ref{v'}). In particular, the outgoing particles on the
right hand side of the piston have velocities in the interval
$(v_{\min}+2W(\tau),v_{\max}+2W(\tau))$ while those on the left
hand side of the piston have velocities in the interval
$(-v_{\min}+2W(\tau),-v_{\max}+2W(\tau))$. Hence, the region in
the $y,v$ plane where the density of the particles is positive is
no longer a rectangle with straight sides, now its boundaries are
curves whose shape nearly repeats the graph of the randomly
evolving piston velocity $W(\tau)$. While the variations of
$O(1/L)$ of these boundary curves may seem small, it is crucial
that on opposite sides of the piston they go in opposite
directions. Indeed, when $W(\tau)>0$, then the outgoing particles
on the right hand side accelerate and those on the left hand side
slow down. When $W(\tau)<0$ the opposite happens.

\begin{figure}[h]
\centering \epsfig{figure=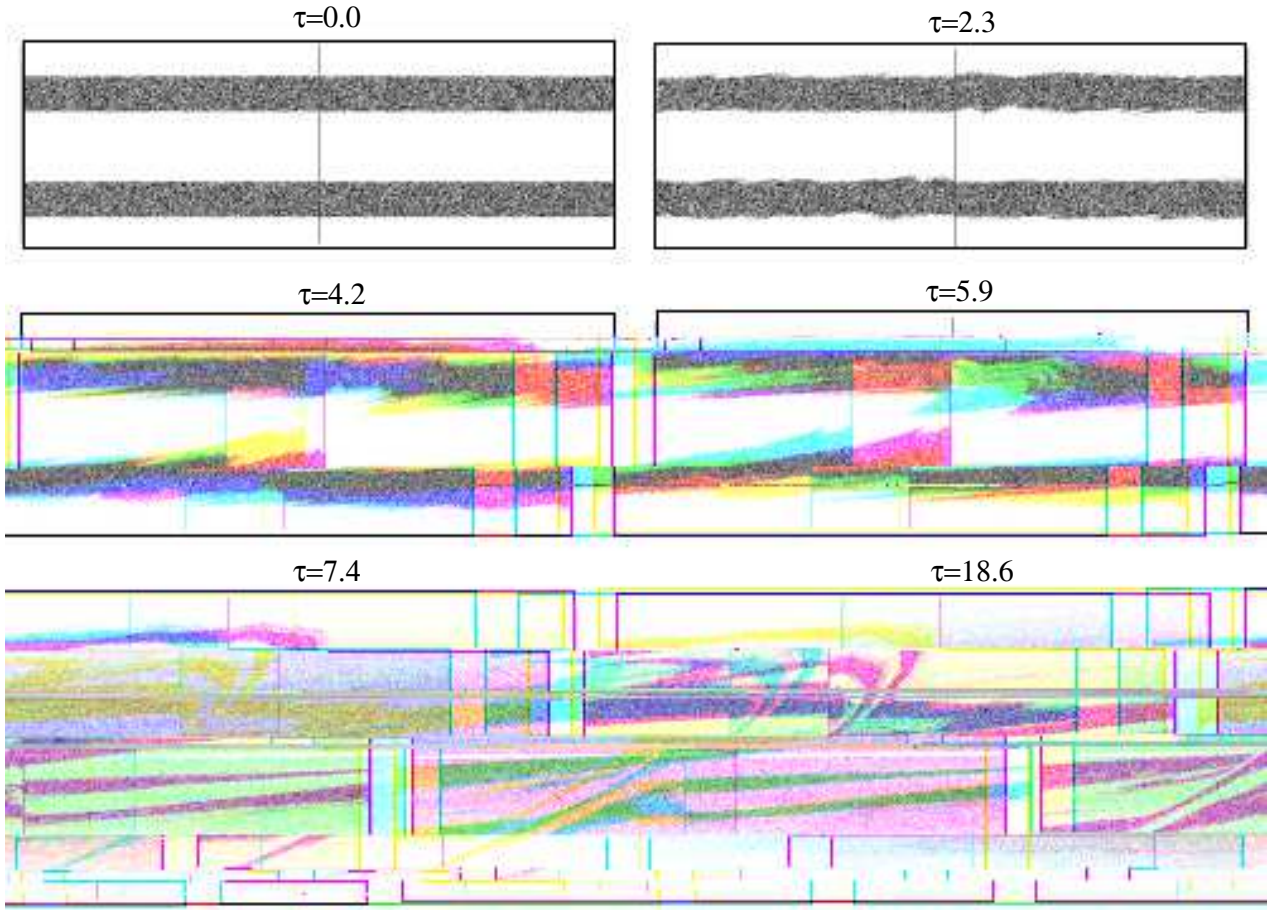}\caption{Six snapshots of the
empirical gas density (in the $x,v$ plane) at times $\tau=0$, 2.3,
4.2, 5.9, 7.4 and 18.6.}
\end{figure}

Next, the particles that have collided with the piston travel to
the wall and come back to the piston. Now their densities are less
regular than they were initially -- the regions in the $x,v$ plane
where the density is positive, are curvilinear domains. When they
hit the piston, they shake it back and forth more forcefully than
before, because the velocities of the incoming particles on the
opposite sides of the piston are now negatively correlated. When
particles on the right hand side are fast, those on the left hand
side are slow, and vice versa. The fluctuations of the gas
densities thus ``cooperate'' to push the piston harder, with a
``double'' force. This produces a resonance-type effect
destabilizing the piston dramatically and the velocity of the
piston $W(\tau)$ experiences larger fluctuations than before. The
velocities of the newly outgoing particles will again go up and
down in opposite direction, on a greater scale than before.

As time goes on, the above phenomenon repeats over and over, with
larger and larger fluctuations of the gas and piston velocities,
until the distribution of gas particles completely breaks down.
For $L=100$, at times $\tau\sim 10$, two large clusters of
particles are formed, one on each side of the piston. When one
cluster bombards the piston, the other moves away from it and hits
the wall, then they exchange their roles. The clusters have sizes
of about 0.3--0.5 in the $y$ direction and the particle velocities
range from about 0.2 to just over 1. The average velocity is about
0.5--0.6 and so the clusters hammer the piston periodically with
period 1.6--2.0, which is close to the experimentally determined
period of piston oscillations, see above.

Fig.~10 shows six snapshots of the empirical density of gas
particles taken at different times. At $\tau=0$ the gas fills
(almost uniformly) two rectangles $\{(y,v):\ 0.5<|v|<1,\ 0<y<1\}$.
At $\tau=2.3$ one can see some ripples on the boundaries of these
rectangles. At time $\tau=4.2$ the irregularities grow and at
$\tau=5.9$ the rectangular shape is broken down. Two large
clusters of particles are formed, both appear in the upper
half-plane $v>0$, i.e.\ at that time both clusters move to the
right (one toward the piston, the other away from it). Later the
density undergoes strange formations ($\tau=7.4$) but eventually
smoothes out and enters a slow process of convergence to
Maxwellian ($\tau=18.6$) described below.

The above analysis suggests that the fluctuations of the piston
velocity roughly increase by a constant factor during each time
interval of length one. Indeed, initial random fluctuations
$W_a\sim O(1/L)$ result in additional changes of velocities of
outgoing particles by $2W_a$. When those particles come back to
the piston (in time $\Delta \tau\approx 1$), they kick its
velocity to the level of $2W_a$. Then the newly outgoing particles
acquire an additional velocity $4W_a$, etc. Over each time
interval of length one the fluctuations double in size. This is an
obvious oversimplification of the real dynamics, but it leads to a
reasonable conjecture
\be
    W_a(\tau) \approx \frac{C\, R^{\tau}}{L}
      \label{Wa}
\ee
where $W_a(\tau)$ are typical fluctuations of the piston velocity
at time $\tau$ and $C,R>0$ are constants.

\begin{figure}[h]
\centering \epsfig{figure=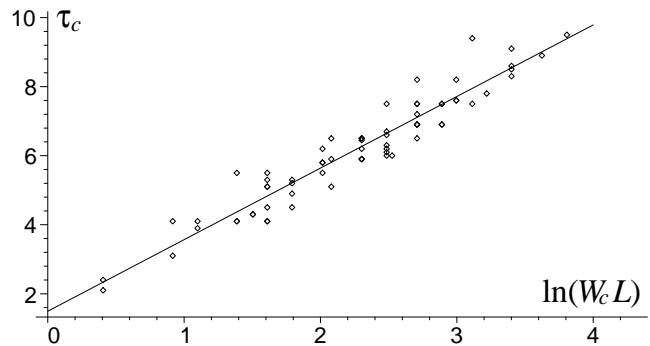}\caption{The value $\tau_c$ as
a function of $\ln (W_cL)$: experimental points and a linear fit.}
\end{figure}

We tested the above formula numerically as follows. Let $W_c>0$ be
some preset critical value of the piston speed and
$\tau_c=\inf\{\tau>0:\, |W(\tau)| > W_c\}$ the (random) time when
$W_c$ is first exceeded. This time plays the role of the ``onset''
of large fluctuations of the piston velocity. One would expect,
based on (\ref{Wa}) that
\be
      \tau_c \approx \ln (W_cL/C)/\ln R
        \label{tauc}
\ee
i.e. $\tau_c$ grows as $\ln L$ when $L$ increases.

We found $\tau_c$ experimentally for $W_c=0.1$ and $W_c=0.15$ and
checked that (\ref{tauc}) agreed well with the data, see Fig.~11.
By the least squares fit we estimated $C=0.45$ and $R=1.6$.

To summarize our experimental observations, we conclude that the
random fluctuations of the density function $p_L(y,v,\tau)$ and
the piston coordinate $Y_L(\tau)$ grow exponentially in time
$\tau$, and at times $\tau\sim\log L$ they become large even on a
macroscopic scale. At that point the evolution of the system
deviates far from the solution of the hydrodynamical equations
(H1)--(H4), and they become completely separated afterwards.

Interestingly, our observations do not indicate that the
convergence (\ref{YY})--(\ref{WW}) claimed in Theorem~\ref{tmmain}
fails on any interval of time. In fact, if the random fluctuations
behave as $CR^{\tau}/L$, as predicted by (\ref{Wa}), then
(\ref{YY})--(\ref{WW}) should hold as $L\to\infty$ on {\em every}
finite interval $(0,\tau_{\ast})$. However, a rigorous proof of
this fact would be a very challenging task.

Next, we also examined numerically and heuristically how the
system behaves asymptotically, as $\tau\to\infty$. On physical
grounds \cite{B}, we expect the system to approach thermal
equilibrium, see Section~\ref{secI}, i.e.\ the velocity
distribution of gas particles should converge to a Maxwellian.

We used the Kolmogorov-Smirnov statistical test to verify the
convergence of the velocity distribution to a normal law. At any
given time $\tau>0$, let
$$
   F_{\tau}(u)=\#\{i:\, v_i<u\}/N
$$
be the empirical (cumulative) distribution function of particle
velocities. For the corresponding normal distribution function
$\Phi(x)$, we compute
$$
           D_{\tau}=\sup_{-\infty<u<\infty}|F_{\tau}(u)-\Phi(u)|
$$
Initially, $D_0\approx 0.245$ for our choice of $\pi_0(v)$ in
(\ref{pini}). If the velocities $v_i$ were independent normal
random variables, then $D_{\tau}$ would be of order
$O(1/\sqrt{N})$ and the product $D_{\tau}\sqrt{N}$ would have a
certain limit distribution, see, e.g.\ \cite{Lu}. In particular,
it is known that the probability $P(D_{\tau}\sqrt{N}>1)\approx
0.2$. Based on this, we opted to define the time of convergence to
equilibrium by
\be
           \tau_{\rm eq}=\inf\{\tau>0:\, D_{\tau}\sqrt{N}<1\}
              \label{taueq}
\ee
We estimated $\tau_{\rm eq}$ for various $L$'s and found that
$\tau_{\rm eq}\approx aL^b$ with some constants $a,b>0$. By a
least squares fit to experimental points we found $a=0.18$ and
$b=2.47$.

The plot of the product $S=D_{\tau}\sqrt{N}$ versus $\tau$ is
given on Fig.~12 (for a particular run with $L=40$). It shows
that, after an initial sharp drop over the period $0<\tau<20$, the
statistic $S$ decreases exponentially in $\tau$. Another commonly
used (and popular among experimentalists) statistic to measure
closeness to a normal distribution is
$$
           S'=3-\frac{M_4}{M_2^2}\
$$
where $M_2$ and $M_4$ are the second and the fourth sample moments
of the empirical velocity distribution, respectively. Fig.~12
shows that $S'$ converges to zero in a similar manner (for the
same run with $L=40$).\\

\begin{figure}[h]
\centering \epsfig{figure=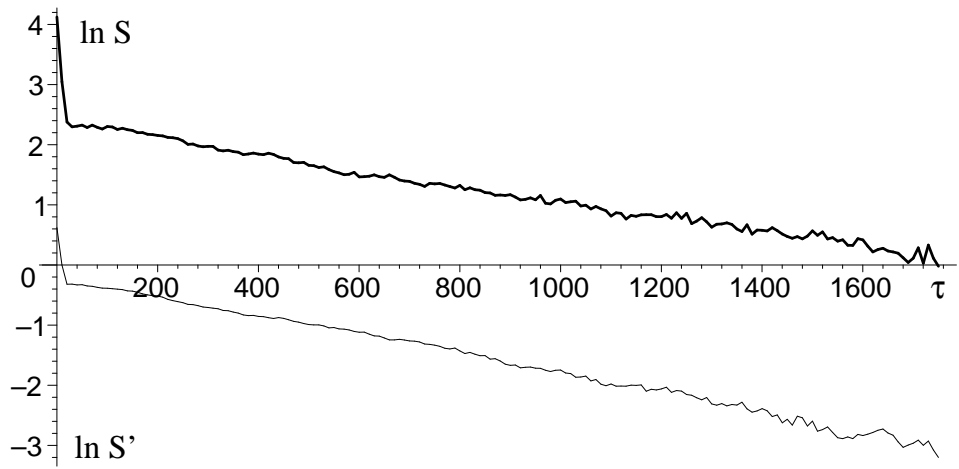}\caption{$\ln S$ (thick line)
and $\ln S'$ (thin line) as functions of $\tau$.}
\end{figure}

The convergence to a thermal equilibrium can also be justified
mathematically. If one fixes three integrals of motion -- the
total energy $E$ and the numbers of particles in the left and
right compartments $N_L$ and $N_R$ -- the dynamics can be reduced
to a billiard system in a high-dimensional polyhedron by standard
techniques, as we show next.

Let $\{x_i\}$, $i=1,\ldots,N_R$, denote the $x$-coordinates of the
particles to the right of the piston, and  $\{x_i\}$,
$i=-1,\ldots,-N_L$, those to the left of it (ordered arbitrarily).
Put $x_0=X\sqrt{M}$, where $X$ is the coordinate of the piston and
$M$ is its mass. Then the configuration space of the system (in
the coordinates $x_i$, $-N_L\leq i\leq N_R$) is a polyhedron
$Q\subset \IR^{N+1}$ (recall that $N=N_L+N_R$) defined by
inequalities
$$
   0\leq x_{-N_L},\ldots,x_{-1}\leq x_0/\sqrt{M} \leq x_1,\ldots,x_{N_R}\leq L
$$
It is known that the dynamics of our mechanical system corresponds
to the billiard dynamics in $Q$, see \cite{CFS}. That is, the
configuration point ${\bf q}\in Q$ moves freely and experiences
specular reflections at the boundary $\partial Q$. The velocity
vector
$$
    {\bf p}=\dot{\bf q}=\{v_{-N_L},\ldots,v_{-1},V\sqrt{M},v_1,\ldots, v_{N_R}\}
$$
has constant length, since $\|{\bf p}\|^2=2E=\,$const. Therefore,
the phase space of the billiard system is ${\cal M}=Q\times
S^N_{\rho}$ where $S^N_{\rho}$ is the $N$-dimensional sphere of
radius $\rho=\sqrt{2E}$.

The billiard system has the Liouville invariant measure $\mu$ on
$\cal M$, which is the product of a uniform measure on the
polyhedron $Q$ and a uniform (Lebesgue) measure on the sphere
$S^N_{\rho}$, i.e. $d\mu=dq\, dp$. The properties of billiard
dynamics depend heavily on the curvature of the boundary $\partial
Q$. In our case $Q$ is a polyhedron, hence its boundary consists
of flat sides with zero curvature. A prototype of such systems is
billiard in a polygon. It is well known that (see, e.g., \cite{C})

\medskip
\noindent{\bf Fact}. For billiards in polygons and polyhedra (and
hence, for our mechanical model of a piston in the ideal gas) all
Lyapunov exponents vanish, and so does the measure-theoretic
(Kolmogorov-Sinai) entropy.
\medskip

Systems with zero Lyapunov exponents and zero entropy are not
regarded as truly chaotic, but they still may be ergodic. In fact,
billiards in generic polygons {\em are} ergodic \cite{KMS}.
Moreover, for many nonergodic polygons, the phase space is
foliated by invariant subsurfaces on which the dynamics is
ergodic.

Even though there are no similar results, to our knowledge, for
billiards in high-dimensional polyhedra, one can expect that they,
too, have similar properties. That is, they are generically
ergodic or become ergodic after trivial reductions. In our case,
the billiard in $Q$ is, perhaps, ergodic for typical values of
$M$, or else the phase space is foliated by invariant submanifolds
on which the dynamics is ergodic, and that those submanifolds fill
${\cal M}$ pretty densely. In the latter case, one would hardly
distinguish experimentally between such a nonergodic system and a
truly ergodic one.

Hence, we can assume that our system is ergodic or very close to
ergodic in the above sense. Then almost every trajectory
eventually behaves according to the invariant measure $\mu$,
independently of the initial state. In particular, for any initial
gas density and velocity distribution (given by the function
$\pi_0(y,v)$, see Section~\ref{secI}) the hydrodynamic regime for
a finite $L$ is only valid on a finite interval of time --
eventually the system will relax to a thermal equilibrium.  We
expect in fact that in terms of the ``macroscopic'' variables,
say, the one particle distribution function, the system will relax
to an effective equilibrium, as defined by (\ref{taueq}) in terms
of $\tau_{\rm eq}$, which is much smaller than the exponentially
long time (in $L$) required for the ergodic theorem. So the real
question is how does this time depend on $L$. According to our
earlier discussion $\tau_c \sim \log L$ and $\tau_{\rm eq} \sim
L^{5/2}$.

At equilibrium, the distribution of coordinates $x_i$ and
velocities $v_i$ are determined by the Liouville measure $\mu$,
which is uniform in the phase space. Physically interesting (and
only observable) are its marginal measures, i.e. projections, on
lower-dimensional subspaces. The marginal measures of the
velocities are approximately normal for large $N$.

In particular, each individual velocity $v_i$ converges in
distribution to a Maxwellian (i.e., normal) random variable with
zero mean and variance $2E/N=\,$const. The same holds for the
``piston'' component of the velocity, $\dot{x}_0=V\sqrt{M}$, hence
the piston velocity $V$ will be normally distributed with zero
mean and standard deviation const$/\sqrt{M}=\,$const$/L$, as
$L\to\infty$. In our case $V$ has standard deviation
$\sqrt{7/24}/L\approx 0.5/L$. This conclusion agrees well with
Holley's theorem~\ref{tmH} and our numerical data.

The equilibrium distribution of the piston coordinate $X$ is also
determined by the projection of the uniform measure $dq$ on $Q$
onto the $x_0$ axis. Before we do that, let us get rid of $M$ in
the definition of both $Q$ and $x_0$. A simple change of variable
$X=x_0/\sqrt{M}$ allows us to redefine $Q$ by
$$
   0\leq x_{-N_L},\ldots,x_{-1}\leq X \leq x_1,\ldots,x_{N_R}\leq L
$$
Furthermore, rescaling $Y=X/L$ and $y_i=x_i/L$ gives a new,
simpler, definition of $Q$:
$$
   0\leq y_{-N_L},\ldots,y_{-1}\leq Y \leq y_1,\ldots,y_{N_R}\leq 1
$$
This is a variation of the so called Brownian bridge.
``Integrating away'' the variables $y_i$ yields the following
equilibrium density for $Y$:
\be
            f(Y)=c\, Y^{N_L}(1-Y)^{N_R}
             \label{fY}
\ee
for $0<Y<1$, where $c$ is  the normalizing factor that can be
computed explicitly:
\be
      c^{-1}=\int_0^1Y^{N_L}(1-Y)^{N_R}\,dY
      =\frac{N_L!N_R!}{(N_L+N_R+1)!}
         \label{c}
\ee
Asymptotically, as $L\to\infty$, we have $N_L\sim L^3/2$ and
$N_R\sim L^3/2$. Assume, for simplicity, that $N_L=N_R=N/2$ and
denote $K=N/2$, then
$$
        c=\frac{(2K+1)!}{(K!)^2}\simeq
        \frac{2\cdot 4^K\sqrt{K}}{\sqrt{\pi}}
$$
Put $z=(Y-0.5)\sqrt{8K}$, then the density of $z$ is given
asymptotically by
\begin{eqnarray*}
     f(z) &=& \frac{c}{\sqrt{8K}}
           \left (0.5+\frac{z}{\sqrt{8K}}\right)^K
       \left(0.5-\frac{z}{\sqrt{8K}}\right)^K\\
           &=& \frac{c}{4^K\sqrt{8K}}\left (1-\frac{z^2}{2K}\right )^K\\
       &\approx & \frac{1}{\sqrt{2\pi}}\, e^{-z^2/2}
\end{eqnarray*}
Hence, $Y$ is asymptotically gaussian with mean $0.5$ and variance
$(4N)^{-1}=(4L^3)^{-1}$. Therefore, in equilibrium
$$
         |Y-0.5|\sim \frac{1}{2L\sqrt{L}}\sim\frac{1}{2\sqrt{N}}
$$
Note that this estimate is independent of the piston mass.

Next, by using (\ref{fY}) one can easily compute the probability
that the piston coordinate $Y$ deviates from its mean value $0.5$
by a fixed amount $d>0$, say $d=0.1$: this probability is $<\,{\rm
const}\cdot e^{-aN}$ with some $a=a(d)>0$. Therefore, we have
observed experimentally a very rare event whose probability was
exponentially small in $N$. If we started with an equilibrium
state, such an observation would be practically impossible. But we
started with a state described by the density (\ref{pini}), which
itself has probability less than ${\rm const}\cdot e^{-bN}$,
$b>0$, with respect to the equilibrium measure $\mu$. So we only
observed how one highly improbable initial state evolved to
another highly improbable state along its (very slow)
transformation to an equilibrium state. It would be interesting to
understand why the system ``chooses'' such a strange evolution to
equilibrium, i.e.\ why starting with a state described by a
``double-peaked'' distribution (\ref{pini}) the system behaves as
a damped harmonic oscillator over an extended time interval, with
initial oscillations as large as 1/10 of the system size.

We are currently working on this problem and will report results in a
separate paper \cite{CCL}. Below we outline our program and mention
some preliminary findings. For simplicity, we assume that
$\pi_0(y,v)=\pi_0(|v|)$ and $X(0)=L/2$, $V(0)=0$.
Corollary~\ref{crprop} ensures that the hydrodynamical equations have
the trivial solution $Y(\tau)\equiv 0.5$, $W(\tau)\equiv 0$, and
$\pi(y,v,t)\equiv \pi_0(y,v)$ for all $\tau>0$. Now, since the initial
configuration of particles is generated randomly from a Poisson process
with the density $\pi_0(y,v)$, the ``actual'' (empirical) density of
the particles, such as the one shown on Fig.~10 at $\tau=0$, does not
exactly coincide with $\pi_0(y,v)$. Random fluctuations of the
empirical density are typically of order $O(1/L)$. Hence, the
``actual'' initial distribution of particles can be thought of as a
small perturbation of the function $\pi_0(y,v)=\pi_0(|v|)$ and can be
written as $\pi_0(|v|)+\varepsilon \pi_1(y,v)$ with $\varepsilon=1/L$
and some (random) function $\pi_1(y,v)$ of order one.

Now, we conjecture that the evolution of the mechanical system
closely follows the solutions of the hydrodynamical equations
(H1)--(H4) with a perturbed initial density $\pi_0(|v|)
+\varepsilon \pi_1(y,v)$, rather than the stationary solution
corresponding to the unperturbed density $\pi_0(|v|)$. This
accounts for significant differences between the behavior of the
mechanical system and the stationary solution, if the latter is
unstable. In particular, two trajectories which are initially
$\varepsilon$-close (in our case $\varepsilon=1/L$) can deviate
from each other exponentially fast in time $\tau$, and at times
$\tau\sim -\log \varepsilon=\log L$ will look completely
different. This would be in agreement with our experimental
observations and the estimate (\ref{tauc}) of the time
$\tau_c\sim\log L$ of the onset of ``instability''.

To test our conjecture, we solved the hydrodynamical equations
(H1)--(H4) numerically starting with a perturbed initial density
obtained by adding to (\ref{pini}) a function $\varepsilon$-small
in the $L^1$ metric (with $\varepsilon\simeq 0.01$). We found that
the corresponding solution resembled strikingly well the evolution
of the mechanical system described above. In particular, the
coordinate and velocity of the piston followed large nearly
harmonic oscillations during the interval $10<\tau<30$. The
corresponding plots of the piston position and velocity along the
perturbed solutions of (H1)--(H4) were almost indistinguishable
from our Figures~6 and 7. Hence, the behavior of the mechanical
system can be traced to that of the perturbed solutions of the
hydrodynamical equations, and the instability of the latter
becomes an important issue.

When we were finishing the present paper, we received a message
from E.~Caglioti and E.~Presutti who (a) proved that the
hydrodynamical equations (H1)--(H4) are stable when $\pi_0(|v|)$
is monotonically nonincreasing in $|v|$, i.e.\ $\pi_0'(|v|)\leq
0$, and (b) suggested that they might be unstable for our class on
non-monotone $\pi_0(|v|)$. We checked the suggestion (b) for our
particular density (\ref{pini}) and found that it was indeed
correct; we proved that small perturbations grow exponentially in
$\tau$.

Conversely, when we simulated a particle dynamics with a
nonincreasing initial density $\pi_0(|v|)$ the oscillations
essentially disappeared (to this end we tried a uniform ``flat''
function given by $\pi_0(|v|)=1$ for $|v|\leq v_{\max}$ and a
triangular one $\pi_0(|v|)=1-|v|/v_{\max}$.). On the other hand,
the particle velocity distribution still approached a Maxwellian,
albeit at a somewhat slower pace.

\medskip Finally, we describe some other open problems related to
the piston dynamics.

\medskip\noindent
1. It is clear that recollisions of gas particles with the piston
have a very ``destructive'' effect on the dynamics in the system.
However, we need to distinguish between two types of recollisions.

We say that a recollision of a gas particle with the piston is
{\em long} if the particle hits a wall $x=0$ or $x=L$ between the
two consecutive collisions with the piston. Otherwise a
recollision is said to be {\em short}. Long recollisions require
some time, as the particle has to travel all the way to a wall,
bounce off it, and then travel back to the piston before it hits
it again. Short recollisions can occur in rapid succession.

We have imposed the velocity cut-off (P4) in order to avoid any
recollisions for at least some initial period of time (which we
called the zero-recollision interval). More precisely, the upper
bound $v_{\max}$ guarantees the absence of long recollisions.
Without it, we would have to deal with arbitrarily fast particles
that dash between the piston and the walls many times in any
interval $(0,\tau)$. On the other hand, the lower bound $v_{\min}$
was assumed to exclude short recollisions.

There are good reasons to believe, though, that short recollisions
may not be so destructive for the piston dynamics. Indeed, let a
particle experience two or more collisions with the piston in
rapid succession (i.e. without hitting a wall in between). This
can occur in two cases: (i) the particle's velocity is very close
to that of the piston, or (ii) the piston's velocity changes very
rapidly. The latter should be very unlikely, since the
deterministic acceleration of the piston is very small, cf.\
Theorem~\ref{tmprop}c. In case (i), the recollisions should have
very little effect on the velocity of the piston according to the
rule (\ref{V'}), so that they may be safely ignored, as it was
done already in some earlier studies \cite{H,DGL}.

We therefore expect that our results can be extended to velocity
distributions without a cut-off from zero, i.e. allowing
$v_{\min}=0$.

\medskip\noindent
2. In our paper, $L$ plays a dual role: it parameterizes the mass
of the piston ($M\sim L^2$), and it represents the length of the
container ($0\leq x\leq L$). This duality comes from our
assumption that the container is a cube.

However, our model is essentially one-dimensional, and the mass of
the piston $M$ and the length of the interval $0\leq x\leq L$ can
be treated as two independent parameters. In particular, we can
assume that the container is infinitely long in the $x$ direction
(i.e., {\em that} $L$ is infinite), but the mass of the piston is
still finite and given by $M\sim L^2$ ({\em this} $L$ is the size
of the container in the $y$ and $z$ directions). In this case
there are no recollisions with the piston, as long as its velocity
remains small. Hence, our zero-recollision interval is effectively
infinite. As a result, Theorem~\ref{tmmain} can be extended to
arbitrarily large times. Precisely, for any $T>0$ we can prove the
convergence in probability:
$$
   P\left (\sup_{0\leq\tau\leq T}
        |Y_L(\tau,\omega) - Y(\tau)| \leq C_T\ln L/L\right )\to 1
$$
and
$$
   P\left (\sup_{0\leq\tau\leq T}
        |W_L(\tau,\omega) - W(\tau)| \leq C_T\ln L/L\right )\to 1
$$
as $L\to\infty$, where $C_T>0$ is a constant and $Y(\tau)$ and
$W(\tau)=\dot{Y}(\tau)$ are the solutions of the hydrodynamical
equations described in Section~2.

\medskip\noindent
3. Along the same lines as above, we can assume that the container
is $d$-dimensional with $d\geq 4$. Then the mass of the piston and
the density of the particles are proportional to $L^{d-1}$ rather
than $L^2$.

When $d$ is large, the gas particles are very dense on the $x,v$
plane. This leads to a much better control over fluctuations of
the particle distribution and the piston trajectory. During the
zero-recollision interval, for example, the piston trajectory is
$L^{-(d-1)/2}$-close to its deterministic trajectory. This is an
easy modification of the results of our Section~\ref{secZRI}.
During the one-recollision interval, the piston trajectory is
$L^{-(2d-5)/7}$-close to its deterministic trajectory. This can be
shown with the methods developed in Section~\ref{secORI} but
requires some extra work. Moreover, the methods and results of
that section can be extended to the $k$-recollision interval
$(\tau_{k},\tau_{k+1})$ for any $k\geq 1$. It can be shown that
there is a $d_k\geq 3$ such that for all $d\geq d_k$ we have
$$
     P\left (
     \sup_{\tau_{k}<\tau<\tau_{k+1}} |W_L(\tau,\omega)-W(\tau)|
        \leq L^{-b}\right )\to 1
$$
as $L\to\infty$, here $b>0$ depends on $k$ and $d$. This
extension, however, requires quite substantial work, which is
beyond the scope of this article. The upshot is that a high
dimensional piston is more stable than a lower dimensional one.

It would be interesting to investigate other modifications of our
model that lead to more stable regimes. For example, let the
initial density $\pi_0(y,v)$ of the gas depend on the factor
$a=\varepsilon L^2$ in such a way that
$\pi_{0}(y,v)=a^{-1}\rho(y,v)$, where $\rho(y,v)$ is a fixed
function. Then the particle density grows as $a\to 0$. This is
another way to increase the density of the particles, but without
changing the dimension. One may expect a better control over
random fluctuations in this case, too.

\medskip\noindent
{\bf Acknowledgements}.  We thank J.~Piasecki, Ch.~Gruber, E.~Lieb,
M.~Mansour, V.~Yahot, N.~Simanyi for many illuminating discussions and
valuable suggestions. N.~Chernov was partially supported by NSF grant
DMS-0098788. J.~Lebowitz was partially supported by NSF grant
DMR-9813268 and by Air Force grant F49620-01-0154. Ya.~Sinai was
partially supported by NSF grant DMS-9706794. This work was completed
when N.~C. and J.~L. stayed at the Institute for Advanced Study with
partial support by NSF grant DMS-9729992.

\section*{Appendix}
\addcontentsline{toc}{section}{Appendix}

\setcounter{section}{1} \setcounter{theorem}{0}

\renewcommand{\thetheorem}{\Alph{section}.\arabic{theorem}}
\renewcommand{\theequation}{\Alph{section}.\arabic{equation}}
\setcounter{equation}{0}

In this section, we derive various probabilistic estimates on the
distribution of gas particles and their velocities. The number of
particles $K=N_D$ in any domain $D$ on the $x,v$ plane at time
$t=0$ is a Poisson random variable, and we need bounds on its
large deviations. Perhaps, some of our estimates are known in
probability theory, but we include proofs for the sake of
completeness.

\begin{lemma}
Let $K$ be a Poisson random variable with parameter $\lambda>0$.
Then for any $A>\lambda$ we have
$$
    P(K>A)\leq e^{A-\lambda-A\ln(A/\lambda)}
$$
and for any $A<\lambda$
$$
    P(K<A)\leq e^{A-\lambda-A\ln(A/\lambda)}
$$
\label{lmPoi1}
\end{lemma}

\noindent{\em Proof}. The moment generating function of $K$ is
$$
     \varphi_K(t)=E(e^{tK})=e^{\lambda(e^t-1)}
$$
First, let $A>\lambda$. Then, obviously, for all $t>0$
$$
     \varphi_K(t)\geq e^{At}\cdot P(K>A)
$$
Hence for all $t>0$
$$
     P(K>A)\leq e^{\lambda(e^t-1)-At}
$$
The expression on the right hand side takes minimum at
$$
           t=\ln(A/\lambda)>0
$$
This proves the first part of the lemma.

Now let $A<\lambda$. Then for all $t<0$
$$
     \varphi_K(t)\geq e^{At}\cdot P(K<A)
$$
Hence for all $t<0$
$$
     P(K<A)\leq e^{\lambda(e^t-1)-At}
$$
The expression on the right hand side takes minimum at
$$
           t=\ln(A/\lambda)<0
$$
This proves the second part of the lemma. $\Box$\medskip

\begin{lemma}
Let $K$ be a Poisson random variable with parameter $\lambda>0$.
For any $b>0$ there is a $c>0$ such that for all
$0<B<b\sqrt{\lambda}$ we have
$$
    P(|K-\lambda|>B\sqrt{\lambda})\leq 2e^{-cB^2}
$$
\label{lmPoi2}
\end{lemma}

\noindent{\em Proof}. A direct application of the previous lemma
gives
\be
      P(|K-\lambda|>B\sqrt{\lambda})\leq 2e^{-B^2g(q)}
        \label{PKB}
\ee
where
\be
        g(q)=\frac{(1+q)\ln(1+q)-q}{q^2}=
        \frac{\int_0^q\ln(1+s)\, ds}{2\int_0^qs\, ds}
       \label{gq}
\ee
and $q=B/\sqrt{\lambda}$ in (\ref{PKB}). By direct inspection one
can verify that the function $g(q)$ is a positive and strictly
monotonically decreasing function on the interval $0<q<\infty$. We
complete the proof by setting $c=g(b)$. $\Box$
\medskip

\begin{lemma}
Let $\lambda_0>0$ and $a>0$. Then for all sufficiently large $L>0$
and every Poisson random variable with parameters
$\lambda\geq\lambda_0$ we have
$$
    P(|K-\lambda|>a\sqrt{\lambda}\,\ln L)\leq L^{-d\,\ln\ln L}
$$
where $d=a\sqrt{\lambda_0}/2$. \label{lmPoi3}
\end{lemma}

\noindent{\em Proof}. Using (\ref{PKB}) with $B=a\ln L$ gives
$$
    P(|K-\lambda|>a\sqrt{\lambda}\,\ln L)\leq
    2e^{-(a\ln L)^2g(a\ln L/\sqrt{\lambda_0})}
$$
with $g(q)$ defined by (\ref{gq}). Observe that, for large $L$,
$$
    g\left (\frac{a\ln L}{\sqrt{\lambda_0}}\right )\sim
    \frac{\sqrt{\lambda_0}\,\ln\ln L}{a\ln L}
$$
This complete the proof. $\Box$ \medskip

\noindent{\bf Remark}. In most of our applications, $\lambda$ is
large, of order $\lambda\sim L^b$ with some $b>0$. Hence, the
factor $\ln L$ is small compared to $\sqrt{\lambda}$.

\begin{corollary}
Let $\lambda_0>0$ and $a>0$. Let $K$ be a Poisson random variable
with parameter $\lambda\leq\lambda_0$. Then for all sufficiently
large $L>0$ we have
$$
    P(K>a\ln L)\leq L^{-d\,\ln\ln L}
$$
where $d=a/2$. \label{crPoi1}
\end{corollary}

\noindent{\em Proof}. The case $\lambda=\lambda_0$ easily follows
from the previous lemma. Now, if $\lambda<\lambda_0$, then the
event $K>a\ln L$ is even less likely than it is for
$\lambda=\lambda_0$. $\Box$ \medskip

Next, we need to study another random variable related to a
Poisson process. For any domain $D$ on the $x,v$ plane consider
the sum of the velocities
$$
      Z=Z_D=\sum_{(x,v)\in D} v
$$
of the particles in $D$ at time $0$. We assume that
$D\subset\{v_{\min}<v<v_{\max}\}$ (the case
$D\subset\{-v_{\max}<v<-v_{\min}\}$ is completely symmetric and
analogous). By projecting the domain $D$ onto the $v$ axis we
obtain a Poisson process on the interval
$$
         I=(v_{\min},v_{\max})
$$
with density
$$
        \pi(v)=L^2\int_{D\cap\{u=v\}}p_L(x,u)\, dx
$$
Now the random variable $Z$ can be described as follows.

Consider a one-dimensional Poisson process with density $\pi(v)$
on the interval $I$. This means that for any subinterval $J\subset
I$ the number of points in $J$, call it $N_J$, is a Poisson random
variable with mean
$$
     E(N_J)=\int_J\pi(v)\, dv
$$

Each realization $\omega$ of this process is a finite subset of
$I$. Consider a random variable
$$
         Z(\omega)=\sum_{v\in\omega} v
$$
We will call $Z$ an {\em integrated Poisson random variable}.

If we fix a large $n\geq 1$ and partition $I$ into small intervals
$$
       \Delta_i=I\cap \left [\frac in ,\frac{i+1}{n}\right )
$$
$i=0,1,2,\ldots$, then we can obviously bound $Z$ by
\be
     \sum_i\frac{i}{n}\,N_{\Delta_i}\leq
     Z<\sum_i\frac{i+1}{n}\,N_{\Delta_i}
        \label{lZr}
\ee
where $N_{\Delta_i}$ is the number of points of the process in the
interval $\Delta_i$. Note that $N_{\Delta_i}$ are independent
Poisson random variables with parameters
$$
      \lambda_i=E(N_{\Delta_i})=\int_{\Delta_i}\pi(v)\, dv
$$
Therefore, the moment generating function $\varphi_Z(t)=E(e^{tZ})$
of $Z$ is bounded by
$$
     \exp\left [\sum_i\lambda_i(e^{t\frac{i}{n}}-1)\right ]
         \leq E(e^{tZ}) <
     \exp\left [\sum_i\lambda_i(e^{t\frac{i+1}{n}}-1)\right ]
$$
Taking the limit $n\to\infty$ we obtain
\be
        \varphi_Z(t)=E(e^{tZ})=
        \exp\left [\int_{I}(e^{tv}-1)\pi(v)\, dv\right ]
           \label{phiZ}
\ee
By using (\ref{lZr}), it is also easy to find the mean value
\be
     E(Z)=\mu_Z=\int_I v\, \pi(v)\, dv
        \label{EZ}
\ee
and the variance
\be
     {\rm Var}(Z)=\sigma_Z^2=\int_I v^2\pi(v)\, dv
        \label{VarZ}
\ee

Note that $Z$ is related to a Poisson random variable $K=N_I$ with
parameter
$$
           \lambda_Z=\int_I\pi(v)\, dv
$$
In particular, we have
\be
          v_{\min}K\leq Z\leq v_{\max}K
        \label{KZK}
\ee
hence
\be
          v_{\min}\lambda_Z\leq \mu_Z\leq v_{\max}\lambda_Z
        \label{lambdaZmuZ}
\ee
We also have
\be
          v_{\min}^2\lambda_Z\leq \sigma_Z^2\leq v_{\max}^2\lambda_Z
        \label{lambdaZsigmaZ}
\ee

The random variable $Z$ admits bounds on large deviations similar
to the ones we found for Poisson random variables:

\begin{lemma}
For any $b>0$ there is a $c>0$ $($determined by $b$, $v_{\min}$
and $v_{\max})$ such that for any integrated Poisson random
variable $Z$ and all $0<B<b\,\sigma_Z$ we have
$$
    P(|Z-\mu_Z| > B\sigma_Z)\leq 2e^{-c B^2}
$$
\label{lmZ1}
\end{lemma}

\noindent{\em Proof}. Put, for brevity, $\mu=\mu_Z$ and
$\sigma=\sigma_Z$. We will show that
\be
    P(Z > \mu+B\sigma)\leq e^{-c B^2}
       \label{PZB}
\ee
(the same bound for $P(Z < \mu-B\sigma)$ is proved similarly, as
we did that in Lemma~\ref{lmPoi1}). For all $t>0$ we have
$$
     \varphi_Z(t)\geq e^{(\mu+B\sigma)t}\cdot P(Z>\mu+B\sigma)
$$
hence
$$
     P(Z>\mu+B\sigma)\leq \exp
     \left [\int_{I}(e^{tv}-1)\pi(v)\, dv-(\mu+B\sigma)t\right ]
$$
We substitute $t=Bs/\sigma$ with $s>0$ to be chosen later and
expand $e^{tv}$ into a Taylor series:
$$
     P(Z>\mu+B\sigma)\leq \exp
     \left [\int_{I}\left (
     \sum_{n=1}^{\infty}\frac{(Bsv)^n}{\sigma^nn!}\right )
     \pi(v)\, dv-\frac{Bs\mu}{\sigma} - B^2s\right ]
$$
The first two terms with $n=1$ and $n=2$ give
$$
    \int_I\frac{Bsv}{\sigma}\,\pi(v)\,dv=
    \frac{Bs\mu}{\sigma}
$$
by (\ref{EZ}) and
$$
    \int_I\frac{(Bsv)^2}{2\sigma^2}\,\pi(v)\, dv=\frac{B^2s^2}{2}
$$
by (\ref{VarZ}), respectively.

Therefore,
\begin{eqnarray*}
     P(Z>\mu+B\sigma) &\leq & \exp
     \Big [- B^2 \Big (s-s^2/2 - \\
     && \ \ \ \ -s^2 \sum_{n=3}^{\infty}
     \frac{(Bs/\sigma)^{n-2}\int_I v^n\,\pi(v)\, dv}{n!\,\sigma^2}
     \Big ) \Big ]
\end{eqnarray*}
Assuming $B/\sigma<b$ and using (\ref{VarZ}) gives
\begin{eqnarray*}
     \sum_{n=3}^{\infty}
     \frac{(Bs/\sigma)^{n-2}\int_I v^n\,\pi(v)\, dv}{n!\,\sigma^2}
     & < &
     \sum_{k=1}^{\infty}\frac{(sb)^k\, v_{\max}^{k+2}}
     {k!\, v_{\min}^2}\\
     & < &
     \frac{v_{\max}^2}{v_{\min}^2}
     \left (e^{sb\, v_{\max}}-1\right )
\end{eqnarray*}
Now if $s$ is small enough, then we have
$$
      c:=s-\frac{s^2}{2}
      -s^2\,\frac{v_{\max}^2}{v_{\min}^2}\left (e^{sb\, v_{\max}}-1\right )>0
$$
This proves (\ref{PZB}), and hence the lemma. $\Box$\medskip

\begin{lemma}
Let $\lambda_0>0$. For all sufficiently large $L>0$ and any
integrated Poisson random variable $Z$ with $\lambda_Z\geq
\lambda_0$
$$
    P(|Z-\mu_Z|>\sigma_Z\ln L)\leq L^{-d\,\ln\ln L}
$$
where $d>0$ is a constant determined by
$\lambda_0,v_{\min},v_{\max}$. \label{lmZ2}
\end{lemma}

\noindent{\em Proof}. Let
$$
        b=\frac{2v_{\max}}{v_{\min}^2}
$$
First, if $\ln L<b\sigma_Z$, then the result easily follows from
the previous lemma.

Now, assume that
\be
    \ln L\geq b\sigma_Z
       \label{lnLbZ}
\ee
Using the inequalities (\ref{lambdaZmuZ})--(\ref{lambdaZsigmaZ})
gives
$$
        \lambda_Z\leq \frac{\sigma_Z^2}{v_{\min}^2}\leq\frac{(\ln L)^2}{b^2v_{\min}^2}
$$
and hence
\be
        \mu_Z\leq v_{\max}\lambda_Z\leq \frac{v_{\max}(\ln L)^2}{b^2v_{\min}^2}
           \label{muZ1}
\ee
and also
\be
        \mu_Z\leq v_{\max}\lambda_Z\leq \frac{v_{\max}}{v_{\min}^2}\sigma_Z^2
      \label{muZ2}
\ee
Multiplying (\ref{muZ1}) and (\ref{muZ2}) and taking the square
root gives
\be
        \mu_Z\leq \frac{v_{\max}}{bv_{\min}^2}\, \sigma_Z\ln L
      =\frac 12\, \sigma_Z\ln L
           \label{muZ3}
\ee
Therefore, since $Z$ is a positive random variable, we have
\begin{eqnarray*}
      P(|Z-\mu_Z|>\sigma_Z\ln L) & = & P(Z>\mu_Z+\sigma_Z\ln L)\\
        & \leq & P(Z>\sigma_Z\ln L)
\end{eqnarray*}
Moreover, combining (\ref{muZ3}) with (\ref{lnLbZ}) and
(\ref{lambdaZsigmaZ}) gives
$$
   \frac 12\, \sigma_Z\ln L=\frac{v_{\max}}{bv_{\min}^2}\,\sigma_Z\ln L
     \geq \frac{v_{\max}}{v_{\min}^2}\, \sigma_Z^2\geq v_{\max}\lambda_Z
$$
and also, by (\ref{lambdaZsigmaZ})
$$
      \frac 12\, \sigma_Z\ln L\geq \frac 12\, v_{\min}\sqrt{\lambda_Z}\ln L
$$
Now, since $Z\leq v_{\max}K$ by (\ref{KZK}), we have
\begin{eqnarray*}
      P(Z>\sigma_Z\ln L) & \leq &
      P\left ( Z>v_{\max}\lambda_Z+\frac 12\, v_{\min}\sqrt{\lambda_Z}\ln L\right )\\
      & \leq & P\left (K>\lambda_Z+
      \frac{v_{\min}}{2v_{\max}}\sqrt{\lambda_Z}\ln L\right )
\end{eqnarray*}
Now the result follows from Lemma~\ref{lmPoi3}. $\Box$\medskip

\begin{corollary}
Let $\lambda_0>0$. Let $Z$ be an integrated Poisson random
variable with $\lambda_Z\leq\lambda_0$. Then for all sufficiently
large $L>0$ we have
$$
    P(Z>\ln L)\leq L^{-d\,\ln\ln L}
$$
where $d>0$ is determined by $\lambda_0,v_{\min},v_{\max}$.
\label{crZ1}
\end{corollary}

This immediately follows from Corollary~\ref{crPoi1}.

\end{document}